\documentclass{aa}  
\usepackage{graphicx}
\usepackage{txfonts}
\usepackage[hidelinks]{hyperref}
\usepackage{tikz}
\usepackage{geometry}
\usepackage{orcidlink}
\newcommand\asloth{\textsc{a-sloth}\xspace}
\newcommand{\ctp}{Caterpillar\xspace}
\newcommand{\Msun}{\,\ensuremath{\mathrm{M}_\odot}}

\begin{document} 

 \title{Mass distribution of Pop III star clusters: A-SLOTH predictions for JWST observability}

   \author{Veronika Lipatova
  \inst{1}\orcidlink{0000-0002-6111-2570}
   \and 
   Simon C.~O.\ Glover\inst{1}\orcidlink{0000-0001-6708-1317}
   \and
   Ralf S.\ Klessen\inst{1,2}\orcidlink{0000-0002-0560-3172}
   \and
   Boyuan\ Liu\inst{1}\orcidlink{0000-0002-4966-7450}
   }
   
   \institute{Zentrum f\"{u}r Astronomie, Institut f\"{u}r Theoretische Astrophysik, Universit\"{a}t Heidelberg, Albert-Ueberle-Str. 2, D-69120 Heidelberg, Germany
   \and Universit\"at Heidelberg, Interdisziplinäres Zentrum f\"ur Wissenschaftliches Rechnen, Im Neuenheimer Feld 205, 69120 Heidelberg, Germany}

\abstract
    {Detecting Population~III (Pop.\,III) stars remains a major observational challenge. Their Balmer-series recombination line emission, redshifted into the infrared at $z\sim5$-11, has been proposed as a potential tracer. JWST/NIRSpec offers the first opportunity to search for such lines, provided that Pop.\,III star formation occurs in sufficiently massive systems.}
    {This study aims to model the expected luminosities of the first four Balmer-series transitions from Pop.\,III star-forming halos and assess their detectability with JWST/NIRSpec across $5 \le z \le 11$, while testing whether the massive Pop.\,III stellar systems required for detectability are physically expected to form.}
    {We use the semi-analytical code \asloth\ with merger trees constructed from the extended Press-Schechter (EPS) formalism and cosmological $N$-body simulations targeting Milky Way-like halos and the halo population in an 8~Mpc$/h$ box. Predicted line fluxes are compared to JWST detection limits derived from the Exposure Time Calculator (ETC), assuming a 10\,000~s NIRSpec exposure at a signal-to-noise ratio of 5.}
    {For our default model parameters, Pop.\,III H$\alpha$ fluxes peak at $\sim10^{-20}$\,erg\,s$^{-1}$\,cm$^{-2}$, 1--2 orders of magnitude below the JWST detection threshold ($\sim6\times10^{-19}$\,erg\,s$^{-1}$\,cm$^{-2}$). The other Balmer lines are weaker than H$\alpha$ and are likewise undetectable. This is because, in our models, the massive Pop.\,III stellar systems required to generate detectable Balmer emission do not form. Pop.\,III star formation proceeds in short, feedback-regulated episodes that are terminated by radiative and supernova feedback, yielding young Pop.\,III stellar masses of only $\sim10^{1}$--$10^{4}\,\mathrm{M}_\odot$. In contrast, detectable Balmer emission would require Pop.\,III stellar masses of $M_{\star,\mathrm{III}}\gtrsim 10^{5}\,\mathrm{M}_\odot$, depending on the observable redshift.}
    {Balmer-series emission from Pop.\,III stars is therefore unlikely to be detectable by JWST in unlensed fields, primarily because hierarchical structure formation combined with stellar feedback prevents the formation of massive, Pop.\, III-dominated stellar systems. Detectability would require strong gravitational lensing (magnification $\mu \gtrsim 10$) or fundamentally different modes of Pop.\,III star formation than that considered in \asloth.}

   \keywords{Stars: Population III -- Methods: numerical -- Galaxies: halos -- Galaxies: high-redshift -- Cosmology: dark ages, reionisation, first stars -- Infrared: stars }

\maketitle

\section{Introduction}

The emergence of the first generation of stars, known as Population~III (Pop.\ III), represents a defining milestone in cosmic evolution. Their formation brought an end to the Cosmic Dark Ages and initiated the thermal, radiative, and chemical transformation of the Universe \citep{glover05, greif15, haemmerle20}. Pop.\ III stars formed in pristine, metal-free environments at high redshifts ($z \sim 15$--20) and are believed to have been extremely massive and short-lived, emitting intense ultraviolet radiation that ionised and dissociated the surrounding gas \citep{bromm02, yoshida03, jaacks19, schauer21}.

These stars played a crucial role in shaping their host halos through radiative and mechanical feedback processes, including photoionisation, photoheating, and supernova-driven outflows and metal enrichment \citep{klessen19, klessen23}. Their ejecta seeded the interstellar medium with the first heavy elements (``metals''), facilitating the transition to Population~II star formation. Additionally, the remnants of massive Pop.\ III stars are considered promising candidates for the progenitors of supermassive black holes \citep[e.g.,][]{woods19, Liu2024}, and their compact binaries may have contributed to the early population of gravitational wave sources observed by LIGO/Virgo \citep[e.g.,][]{kinugawa14, hartwig16, liu20a, Tanikawa2021, Santoliquido2023, Liu2024gw}.

Despite their cosmological significance, no Pop.\ III stars have yet been observed directly. Their short lifespans, high formation redshifts, and likely low number densities place them beyond the reach of current instrumentation \citep{magg16, kulkarni21, skinner20, hartwig22, hegde23}. While strong gravitational lensing could magnify rare, luminous examples into detectability, the frequency of such occurrences and the clustering behaviour of Pop.\ III star formation remain uncertain \citep{riaz22}.

As a result, indirect observational tracers have become central to probing Pop.\ III populations. Among these, nebular recombination lines powered by ionising photons from massive stars offer a promising window into high-redshift star formation. In particular, hydrogen Balmer lines such as H$\alpha$ and H$\beta$ are commonly used to trace star formation in low- and intermediate-redshift galaxies \citep{kennicutt98}, and could likewise reveal primordial star formation if produced in sufficient strength. However, at $z > 8$, H$\alpha$ is redshifted beyond 6~$\mu$m into the mid-infrared, where observational sensitivity is reduced, and ground-based detection is not feasible. Previous missions like \textit{Hubble} lacked the infrared coverage and sensitivity required, but the \textit{James Webb Space Telescope} (JWST) offers unprecedented infrared capabilities. Instruments such as NIRSpec and MIRI are optimised for detecting redshifted recombination lines from the earliest galaxies \citep{jakobsen2022nirspec, rieke2015miri}.

It has been known for a long time that massive Pop.\ III stars are hotter than their present-day counterparts and hence produce larger fluxes of ionising photons \citep[e.g.][]{Schaerer2002}. This fact, together with the absence of dust absorption, leads one quickly to the idea that Pop.\ III dominated galaxies at high redshift are likely to be strong sources of H and He recombination line emission \citep[see e.g.][]{oh1999observational, oh2001he, cen2003galaxies,zackrisson2011hst}. However, the likely detectability of these systems in the Balmer series lines remains unclear. 

HII regions produced by individual massive Pop.\ III stars produce H$\alpha$ fluxes that are far too faint to be detectable by JWST \citep{Greif2009}. Since we expect the higher Balmer series transitions to be weaker than H$\alpha$, the same should be true for them as well. 
\citet{johnson2009first} modelled the H$\alpha$ emission produced by Pop.\ III clusters with stellar masses of a few times $10^{3} \: {\rm M_{\odot}}$ to a few $10^{4} \: {\rm M_{\odot}}$ and showed that this is also likely to be too faint for JWST to detect. On the other hand, other studies have shown that Pop.\ III clusters with masses of $10^{6}$--$10^{7} \: {\rm M_{\odot}}$ should produce detectable Balmer series emission \citep{oh1999observational, trussler23}.

One problem with all of these previous studies is that they adopt the Pop.\ III cluster mass as an input, and hence do not address whether it is actually possible to form Pop.\ III-dominated star clusters of the chosen mass in the real Universe. In our present study, we aim to remedy this weakness. We employ the \asloth semi-analytical framework \citep{hartwig22, hartwig24} to model the luminosity in the H$\alpha$, H$\beta$, H$\gamma$, and H$\delta$ lines produced by Pop.\ III star-forming halos. We aim to assess whether these lines are detectable with JWST, and under what conditions such detections might be possible. This approach enables us to connect theoretical predictions with the sensitivity thresholds of JWST's spectroscopic instruments, and evaluate the potential for constraining Pop.\ III star formation through its spectral imprint, even in the absence of direct stellar light.
        
\section{Methods} \label{sec:methods}

    \subsection{\asloth} \label{sec:asloth}
    
    \begin{figure}
        \begin{center}
        \includegraphics[width=.39\textwidth]{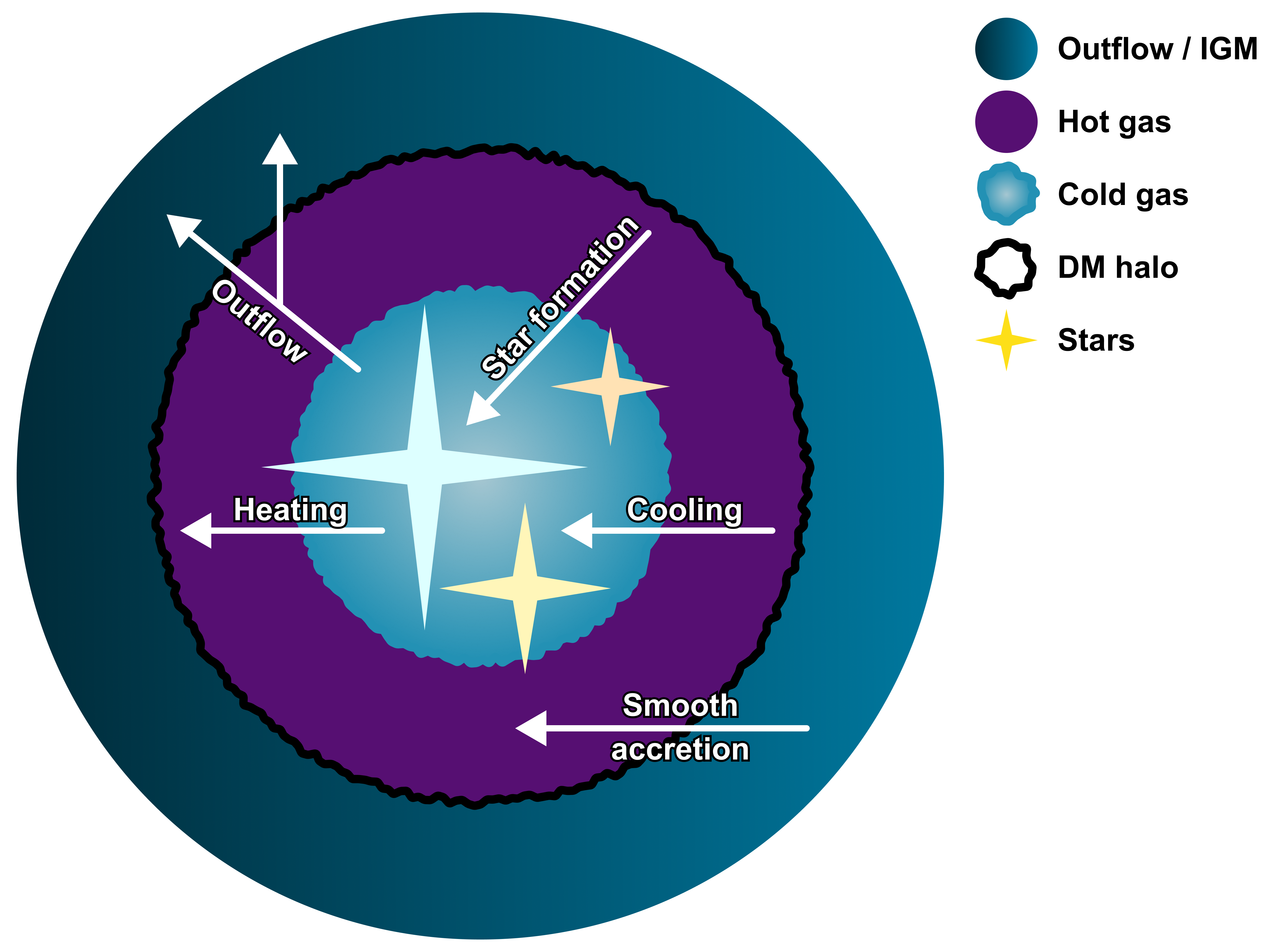}
            \caption{{Schematic overview of the \asloth\ workflow and feedback loops. Dark matter merger trees feed gas reservoirs; star formation episodes (Pop.\,III/Pop.\,II) trigger radiative heating of the halo gas and surrounding intergalactic medium (set by $f_{\rm esc}$), as well as supernova-driven outflows (set by $\alpha_{\rm out}$ and $M_{\rm out,0}$), and chemical enrichment (modulated by $c_{\rm ZIGM}$). These channels regulate the subsequent supply of cold gas, quench Pop.\,III activity, and determine the nebular emission.}
            \end{center}
    \label{fig:aslothdiagram}}
    \end{figure}

    \asloth is a semi-analytical framework for modelling the formation and evolution of stellar populations in the early Universe, with a focus on metal-free (Population~III or Pop.\ III) and metal-poor (Population~II or Pop.\ II) star formation. Originally introduced by \citet{hartwig22}, the model builds baryonic physics atop dark matter merger trees, which can be generated using the Extended Press-Schechter formalism (EPS) \citep{lacey93} or extracted from various cosmological N-body simulations. This approach enables efficient yet physically grounded simulations of primordial halo evolution and feedback-regulated star formation. The \asloth code is publicly available \footnote{\url{https://gitlab.com/thartwig/asloth}} for community use.

    Unlike fully numerical hydrodynamical simulations, which can be computationally expensive and often limited in scale or resolution, \asloth offers a more resource-efficient alternative while retaining predictive power. Its semi-analytical nature makes it especially suitable for large parameter studies and statistical explorations of early star formation processes.
    
    \asloth\ tracks Pop.\,III and Pop.\,II episodes on top of DM merger trees and couples star formation to three feedback channels that regulate subsequent growth and determine quenching pathways \citep{hartwig22, magg22b, hartwig24}:
    
    \begin{enumerate}
        
        \item Radiative feedback. Ionising photons heat and ionise the ISM/CGM. Their local impact is controlled by fixed escape fractions $f_{\mathrm{esc,III}}$ and $f_{\mathrm{esc,II}}$. Inside halos, the retained fraction drives Case~B recombination emission and boosts pressure support; photons that escape contribute to the metagalactic background and global reionisation history. Photodissociating Lyman–Werner (LW) radiation suppresses $\mathrm{H_2}$ cooling in minihalos and raises the star-formation threshold in low-mass systems.
            
        \item Mechanical (SN) feedback. Core–collapse and pair–instability SNe inject energy and momentum into the halo gas. In \asloth, the resulting outflow mass loading is parameterised by a halo-mass–dependent efficiency with slope $\alpha_{\mathrm{out}}$ and normalisation $M_{\mathrm{out},0}$ (cf.\ Eq.~\ref{eq:mass_loading}), which sets the fraction of hot/cold gas ejected during a timestep and thereby regulates fuel supply for subsequent bursts.
            
        \item Chemical feedback. Metals from SNe enrich the ISM/IGM and modify cooling. The transition from Pop.\,III to Pop.\,II is triggered once local gas metallicity exceeds the critical threshold assumed by \asloth\ (see \citealt{hartwig22,magg22b} for defaults). Re-accretion is modelled with an IGM metallicity clumping factor $c_{\mathrm{ZIGM}} > 1$, which captures preferential return of metal-rich clumps and accelerates the Pop.\,III$\rightarrow$Pop.\,II transition.
            
    \end{enumerate}
    
    These channels yield two generic quenching modes for Pop.\,III systems that we will quantify in Sec.~\ref{sec:discussion}: (i) radiative heating quenching, where photoheating suppresses further cold gas supply in minihalos; and (ii) SN–driven blowout quenching, where mechanical feedback ejects a large fraction of the gas such that subsequent star formation is delayed until re-accretion.
    
    \subsection{Modelling Pop.\ III star formation with \asloth} \label{sec:asloth_popiii}

        \subsubsection{Model parameters} \label{sec:model_params}

        \asloth is calibrated to reproduce multiple key observational constraints simultaneously, including (but not limited to) the global reionisation history, the properties of the Milky Way and its satellite galaxies, and the cosmic star formation rate density. In its most recent application, \citet{hartwig24} employed \asloth to constrain the nature of Pop.\ III star formation by optimising a likelihood function based on a total of nine independent observables or sets of observables. This calibration yields best-fit values and uncertainties for eleven critical parameters that govern early star formation physics. This set of parameters is briefly introduced below and summarised in Table \ref{tab:parameters}. The reader is referred to the code release and calibration papers \citep{magg22b,hartwig22,hartwig24} for details of the roles played by these parameters in the galaxy evolution model of \asloth and the calibration process.  

\begin{table*}[h]
    \centering
    \caption{Free parameters used in the \asloth semi-analytical framework.}
    \label{tab:parameters}
    \small
    \begin{tabular}{lcccc}
        \hline
        Parameter & Description & Central 68\% & Best-fit value & Ranges \\
        \hline

        $M_\mathrm{max}$ & Max. mass of Pop.\ III stars, \Msun & $110-313$ & $197$ & $100 - 320$ \\

        $M_\mathrm{min}$ & Min. mass of Pop.\ III stars, \Msun & $6.6-21.1$ & $13.6$ & $3 - 42$ \\

        $\alpha_\mathrm{III}$ & power-law index of the Pop.\ III IMF & $0.23-2.27$ & $1.77$ & $0.2 - 2.3$ \\

        $\eta _{\rm III}$ & Pop.\ III star formation efficiency & $0.60-87.6$ & $8.15$ & $0.3 - 87.6$ \\

        $\eta _{\rm II}$ & Pop.\ II star formation efficiency & $0.099-1.64$ & $0.237$ & $0.1 - 1.7$ \\

        $\alpha_\mathrm{out}$ & slope of outflow efficiency & $1.78-4.05$ & $2.59$ & $1 - 5$ \\

        $M_\mathrm{out0}$ & outflow efficiency normalisation, \Msun & $(6.22-10.92) \times 10^{9}$ & $8.39 \times 10^{9}$ & $(6 - 11) \times 10^{9}$ \\

        $f_\mathrm{esc,II}$ & Pop.\ II ion. photon escape fraction & $0.093-0.279$ & $0.175$ & $0.1 - 0.3$ \\

        $f_\mathrm{esc,III}$ & Pop.\ III ion. photon escape fraction & $0.196-0.865$ & $0.525$ & $0.1 - 0.9$ \\

        $v_\mathrm{sv} / \sigma_\mathrm{sv}$ & MW baryonic streaming velocity & $1.18-2.16$ & $1.75$ & $0.8 - 2.2$ \\

        $c_\mathrm{ZIGM}$ & IGM metallicity clumping factor & $2.87-3.72$ & $3.32$ & $2.5 - 4$ \\

        \hline
    \end{tabular}
    \tablefoot{
        The first nine parameters were originally introduced in \citet{hartwig22}, while the last two were added as new free parameters by \citet{hartwig24}. The values in the ``Central 68\%'' and ``Best-fit value'' columns are taken from \citet{hartwig24}, whereas the ``Ranges'' column reflects the parameter space explored in this paper.}
\end{table*}

    \begin{itemize}

        \item Pop.\ III IMF parameters: The initial mass function for Population~III stars is described by a power-law distribution:

            \begin{equation}
                \frac{dN}{dM_\star} \propto M_\star^{-\alpha_{\mathrm{III}}}\;,
            \end{equation}

        bounded between $M_{\mathrm{min}}$ and $M_{\mathrm{max}}$, which represent the minimum and maximum stellar mass. \citet{hartwig24} proposes best-fit values of $M_{\mathrm{min}} = 13.6\,\mathrm{M}_\odot$, $M_{\mathrm{max}} = 197\,\mathrm{M}_\odot$, and a slope $\alpha_{\mathrm{III}} = 1.77$, indicating a top-heavy IMF compared to present-day star-forming regions \citep{kroupa01, chabrier03}.

        \item Star formation efficiencies: The efficiencies $\eta_{\mathrm{III}}$ and $\eta_{\mathrm{II}}$ represent the fraction of cold gas converted into stars per freefall time. They are defined as:

            \begin{equation} \label{eq:efficiency_def}
                \eta = \frac{\dot{M}_\star}{M_{\mathrm{cold}}/t_{\mathrm{ff}}}\;,
            \end{equation}
    
        where $\dot{M}_\star$ is the star formation rate, defined as the mass of gas converted into stars per unit time. $M_{\mathrm{cold}}$ is the cold gas mass, and $t_{\mathrm{ff}}$ is the freefall time. Since this freefall time is computed for the mean density of the cold gas, $\eta > 1$ is physically allowed and corresponds to rapid star formation in dense regions, where the gas density is much higher than the mean value and where the star-formation timescale $t_{\mathrm{form}} \ll t_{\mathrm{ff}}$.

        \item Outflow parameters: The impact of supernova feedback is modelled using the slope $\alpha_{\mathrm{out}}$ and normalisation $M_{\mathrm{out},0}$ of the outflow efficiency. The fraction of the hot gas in the halo that is ejected by feedback during a given timestep is calculated via:

            \begin{equation} \label{eq:mass_loading}
                \epsilon_{\mathrm{out}} \simeq \frac{f_{\rm hot/cold}E_{\mathrm{SN}}}{E_{\mathrm{bind,hot/cold}}} \cdot \left( \frac{M_h}{M_{\mathrm{out},0}} \right)^{-\alpha_{\mathrm{out}}}\;,
            \end{equation}

        where $f_{\rm hot/cold}$ is the fraction of SN energy that affects hot/cold gas \citep[see sec.~2.3.4 of][]{hartwig22}, \( M_h \) denotes the total mass of the dark matter halo hosting the star-forming region, $E_{\mathrm{SN}}$ is the total supernova energy input during the timestep and $E_{\mathrm{bind,hot/cold}}$ is the gravitational binding energy of hot/cold gas. This formulation regulates how efficiently halos of different mass eject baryons via winds.

        \item Escape fractions: $f_{\mathrm{esc,III}}$ and $f_{\mathrm{esc,II}}$ denote the escape fractions of hydrogen-ionising photons for Pop.\ III and Pop.\ II stars, respectively. These are fixed for each stellar population and affect both internal nebular emission and external reionisation. \citet{hartwig24} find best-fit values of $f_{\mathrm{esc,III}} \approx 0.53$ and $f_{\mathrm{esc,II}} \approx 0.18$. The model uses Pop.\ III escape fraction value only when the halo is not metal-enriched, and Pop.\ II stars have not formed yet. When Pop.\ II stars already formed, the model uses $f_{\mathrm{esc,II}}$ for each type of star. 

        \item Streaming velocity: The parameter $v_{\mathrm{sv}}/\sigma_{\mathrm{sv}}$ represents the baryon-dark matter streaming velocity normalised by its rms value \cite[][]{tseliakhovich10}. High values reduce gas accretion into minihalos and delay early star formation \citep[see e.g.,][]{greif11_stream, stacy11_stream,schauer19}. The most probable cosmological value for the parameter is $v_{\mathrm{sv}}/\sigma_{\mathrm{sv}} = 0.8$, consistent with the mode of the Maxwellian distribution for large cosmological volumes. However, for the specific case of the Milky Way, simulations suggest a significantly higher value of $v_{\mathrm{sv}}/\sigma_{\mathrm{sv}} \approx 1.75$, as inferred from constrained local universe models \citep{uysal23}. Within \asloth, this parameter is treated as free and can be varied to capture both average and environment-specific conditions.

        \item IGM metallicity clumping: $c_{\mathrm{ZIGM}}$ is a clumping factor for re-accreted, metal-enriched gas from outflow bubbles. Instead of assuming homogeneous mixing, \asloth models preferential re-accretion of metal-rich clumps, resulting in an enhanced metallicity of the returning gas by a factor $c_\mathrm{ZIGM} > 1$. Values around $3.3$ provide a better match to measurements of the metallicities of Milky Way satellite galaxies \citep{chen22, chen22a} than no clumping (i.e.\ $c_{\rm ZIGM} = 1$).

    \end{itemize}

    These parameters were jointly calibrated using a Markov Chain Monte Carlo (MCMC) algorithm, ensuring consistency with both Milky Way-specific and cosmologically representative observables as discussed by  \citet{hartwig24}.
    
    \subsubsection{Simulation setup} \label{sec:sim_setup_concrete}

    To model the hierarchical growth of dark matter halos and their associated baryonic physics, we employ three distinct types of merger tree inputs within the \asloth framework. Each approach provides a different level of physical resolution and cosmological representativeness:

    \begin{itemize}
    \item Extended Press-Schechter (EPS) Formalism: 
        The EPS approach is a semi-analytical method for generating dark matter halo merger histories based on excursion set theory \citep{press74,bond91}. It statistically describes the hierarchical growth of structure by computing the probability distribution of halo progenitors across time, using an analytical barrier-crossing condition applied to the linear matter power spectrum. In this study, we generate a single EPS-based merger tree corresponding to a target halo of mass $M_{\mathrm{tot}} = 1.3 \times 10^{12}\,\mathrm{M_\odot}$ at redshift $z = 0$, using a fixed random seed.\footnote{Specifically, we set {\tt RNG\_SEED} = 161803398 in the \asloth configuration file.}
        The tree is constructed with a redshift range from $z = 35$ to $z = 0$, resolved over $n_{\mathrm{lev}} = 311$ time steps. The minimum halo mass resolution is set to $5 \times 10^5\,\mathrm{M_\odot}$, comfortably below the typical threshold for Pop.\ III star formation ($\sim 10^6\,\mathrm{M_\odot}$). Although this method lacks spatial information, it remains computationally efficient and is used for cosmological galaxy/black hole population synthesis models of the early Universe \citep[e.g.,][]{lacey93,Jeon2025}. 

    \item Caterpillar Merger Trees (CTP): 
        The second merger tree is taken from the high-resolution Caterpillar simulation suite \citep{Caterpillar}, which follows the formation of 30 Milky Way-like halos in a fully cosmological $N$-body framework. For this paper, we consider a single representative example. To construct the Caterpillar trees,
        dark matter halos and subhalos were identified using the \textsc{Rockstar} phase-space halo finder \citep{Behroozi13}, and the merger trees were constructed using the \textsc{Consistent-Trees} algorithm \citep{Behroozi13b},
        as described in \citet{hartwig22}.
        This combination allows robust tracking of halo assembly histories and substructure evolution across snapshots. The trees preserve both the mass accretion history and spatial relationships within a well-resolved zoom-in region, making them especially useful for studying halo-to-halo variations and environmental effects.

    \item 8~Mpc Cosmological Volume: 
        Finally, we include a merger tree extracted from a cosmological dark matter simulation of a uniform box with a comoving side length of 8~Mpc/$h$ \citep{ishiyama16}. This box is representative of average cosmic environments and includes a statistically significant range of halo masses and formation pathways. Dark matter halos are identified using the Friends-of-Friends (FoF) algorithm \citep{davis1985} with a standard linking length of $b=0.2$, and a merger tree is constructed following the method described in \citet{Ishiyama15}.

    \end{itemize}

Each of these tree-building techniques feeds into \asloth, which then applies its baryonic prescriptions such as gas cooling, Pop.\ III and Pop.\ II star formation, and radiative feedback self-consistently to the evolving halo populations.

\subsubsection{Stellar mass assembly and star formation rate} \label{sec:mass_sfr_estimates}

For each halo and global timestep $\Delta t$, \asloth\ forms stars according to the population-specific efficiency $\eta$ applied to the available cold-gas reservoir (see Eq.~\ref{eq:efficiency_def} in Sec.~\ref{sec:model_params}). We record: (i) the star formation rate,
$\mathrm{SFR}(t)=\Delta M_\star/\Delta t$ per population, where $\Delta M_\star$ is the mass of new stars in the population in question formed during the timestep; (ii) the peak mass in massive Pop.\,III stars, $M_{\mathrm{mass,III}}^{\mathrm{peak}}$, defined as the maximum total mass of massive main sequence Pop.\,III stars (with $M > 5\,\mathrm{M}_\odot$) reached per halo. This quantity corresponds to the peak population of short-lived massive stars tracked within the star formation sub-cycles of each timestep and therefore more closely traces the maximum instantaneous ionising output of the halo than quantities averaged over the full timestep, 
and (iii) the cumulative Pop.\,III mass, $M_{\star,\mathrm{III}}^{\mathrm{formed}}$, formed within the halo or any of its progenitors. 
Unlike observationally inferred star formation rates, the recorded SFR is directly computed within the model and is fully determined by the mass formed over the timestep $\Delta t$.

\subsection{Modelling the Balmer series luminosities}

The \asloth framework computes the ionising photon luminosities ($L_{\rm ion}$) for Pop.\ II and Pop.\ III stars using stellar evolution tracks from the SEVN code \citep{spera22, spera15}. These tracks provide time-dependent ionising photon emission rates and stellar lifetimes across a range of initial masses and metallicities. We use analytical fits to SEVN stellar evolution models for Pop.\,III and Pop.\,II stars. Individual stellar masses are sampled from the IMF in each star formation episode, and the total ionising luminosity $L_{\rm ion}$ is obtained by summing the contributions of the sampled stars, thereby naturally capturing stochastic fluctuations due to incomplete IMF sampling (see Appendix~\ref{appendix:overline}). This approach captures the evolving luminosity contribution of the stellar population more accurately than fixed photon yield assumptions. In the following, we denote by $N_{\rm ion,II}$ and $N_{\rm ion,III}$ the intrinsic ionising photon production rates of Pop.\,II and Pop.\,III stars, respectively. To estimate the resulting nebular H$\alpha$ emission produced within the halo itself\footnote{Any contribution from the surrounding IGM will have a much lower surface brightness than the halo contribution and hence is unlikely to be detectable.}, we calculate the fraction of ionising photons retained within the halo that contributes to hydrogen recombination.

For a halo containing only Pop.\,III stars, we write the rate at which ionisations occur within the halo as
\begin{equation}
R_{\rm ion} =  N_{\rm ion,III} \left(1 - f_{\rm esc,III}\right).
\end{equation}

For halos containing both Pop.\,II and Pop.\,III stars, which will typically be more massive and hence have lower photon escape fractions, we use instead the expression

\begin{equation}
R_{\rm ion} = \left(N_{\rm ion,II} + N_{\rm ion,III}\right) \left(1 - f_{\rm esc,II}\right).
\end{equation}

Here, $f_{\rm esc,II}$ and $f_{\rm esc,III}$ are the escape fractions of hydrogen-ionising photons for Pop.\,II and Pop.\,III stars, respectively. Thus, the factor $(1-f_{\rm esc})$ gives the fraction of ionising photons absorbed locally and available to power nebular recombination emission. Note that we distinguish between these two expressions based on the mass in Pop.\ II stars and not the ionising photon luminosity, and so for a halo containing massive Pop.\ III stars but only low mass Pop.\ II stars we would nevertheless use the second expression.

The total H$\alpha$ luminosity then follows as:
\begin{equation}
L_{\text{H}\alpha} =  0.45 \cdot 3 \cdot 10^{-12}\ {\rm erg}\cdot R_{\rm ion}.
\end{equation}

The factor of $0.45$ accounts for the average number of H$\alpha$ photons produced per hydrogen recombination under Case B conditions \citep{hummer87}. This coefficient is mildly temperature and density-dependent, but varies by less than 10\% across the range of densities and temperatures likely to be encountered in metal-free HII regions, and so for simplicity we keep this factor constant. The final multiplicative factor, $3 \times 10^{-12}$ erg, represents the energy of a single H$\alpha$ photon, resulting in a total luminosity $L_{\text{H}\alpha}$ in units of erg s$^{-1}$.

Note that although our calculation of $R_{\rm ion}$ accounts for contributions from both Pop.\ II and Pop.\ III stars, in practice we find that in almost all cases either the Pop.\ III or the Pop.\ II contribution dominates, i.e.\ there are very few halos in which Pop.\ III and Pop.\ II stars simultaneously make large contributions to the total ionising luminosity.

To relate the modelled H$\alpha$ luminosities to observable quantities, we follow a two-step procedure:

Line Ratio Conversion. Since  H$\alpha$ is only observable with NIRSpec for redshifts $z \lesssim 7$ (see Fig. \ref{fig:balmer_vs_JWST}), we convert  H$\alpha$ luminosities to those of higher-order Balmer lines that are more accessible at higher redshifts. For a  Pop.\ III HII region with $T = 2 \times 10^4$ K and $n = 100$ cm$^{-3}$, the relative line intensities under Case B recombination are:
\begin{equation}
\label{eq:scaling}
    \frac{\text{H}\beta}{\text{H}\alpha} = 0.36, \quad 
    \frac{\text{H}\gamma}{\text{H}\alpha} = 0.17, \quad 
    \frac{\text{H}\delta}{\text{H}\alpha} = 0.096\;.
\end{equation}
            
The luminosity for each line is obtained by scaling the  H$\alpha$ luminosity with the appropriate factor. These values are obtained from quantum-mechanical calculations of the hydrogen emissivity \citep{hummer87, osterbrock2006astrophysics, storey1995recombination} and are widely adopted in nebular diagnostics. The use of constant ratios is justified by their weak dependence on gas conditions over the typical range found in high-redshift HII regions. (i.e., electron temperatures $T_{\rm e} \sim 10^4$--$2.5 \times 10^4$\,K and densities $n_{\rm e} \sim 10^2$--$10^4$\,cm$^{-3}$; see e.g.\ \citealt{Nakajima23, Curti23}.)

Flux Calculation. To convert intrinsic line luminosities to observed fluxes (in erg cm$^{-2}$ s$^{-1}$), we apply the inverse square law using the luminosity distance $D_L$:
            
        \begin{equation}
            F = \frac{L}{4\pi D_L^2}
        \end{equation}
            
where $L$ is the luminosity of the recombination line and $D_L$ is the luminosity distance corresponding to the source redshift. The luminosity distance is computed using the \texttt{astropy} cosmology module \citep{astropy:2022}, assuming a flat \(\Lambda\)CDM cosmology with parameters from the Planck 2018 results \citep{aghanim20}.

This modelling framework enables us to assess whether the predicted line fluxes from Pop.\ III star-forming regions are detectable with JWST's instruments.
        
\subsection{Detectability criteria}\label{sec:obs}

\begin{figure}[h]
    \begin{center}
    \includegraphics[width=.39\textwidth]{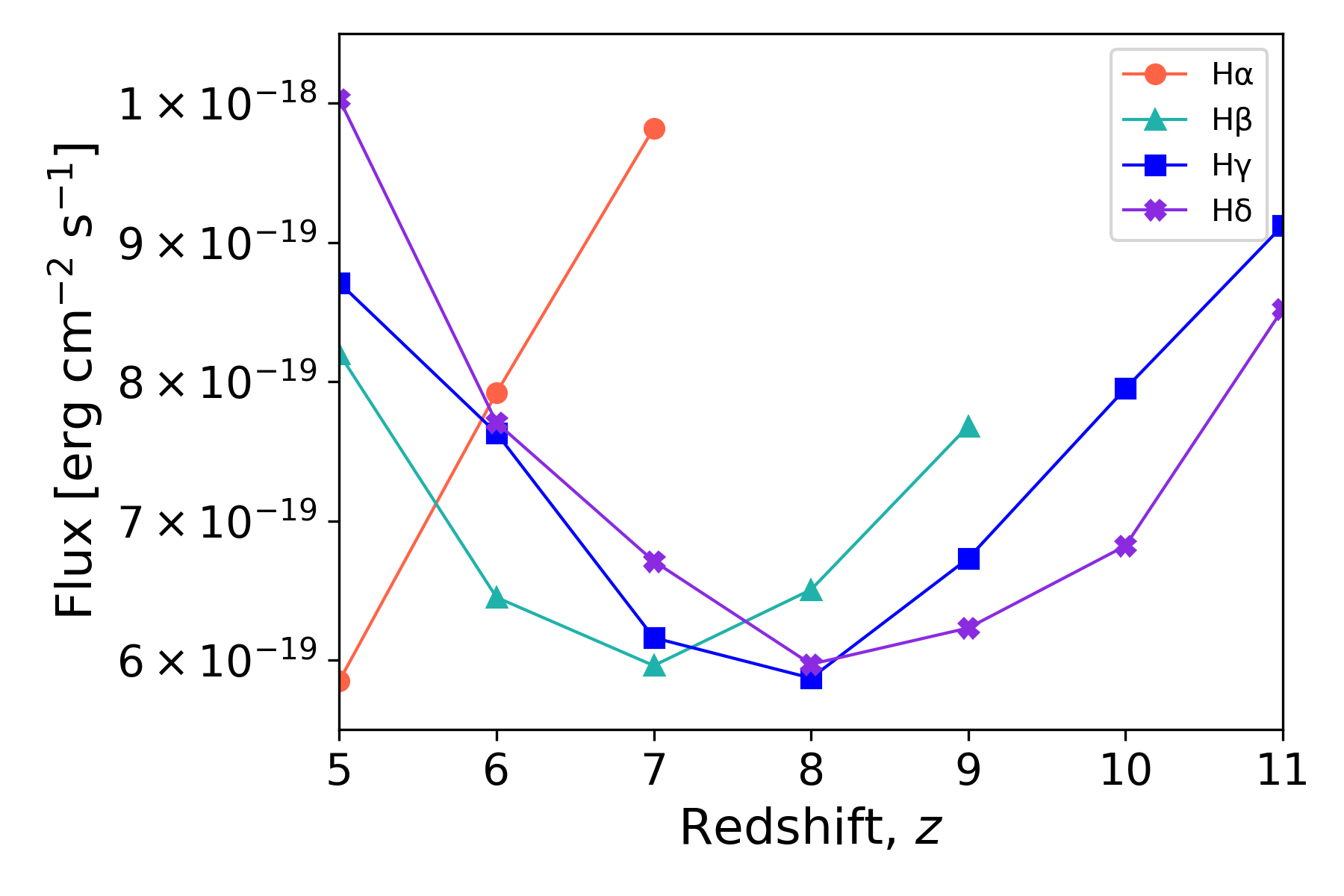}
    \caption{Minimum flux required for a signal-to-noise ratio (S/N) of 5 detection with JWST/NIRSpec under a total exposure time of $\sim 10^4\ \rm s$ as a function of redshift for the first four Balmer lines (H$\alpha$,  H$\beta$,  H$\gamma$,  H$\delta$) based on Exposure Time Calculator (ETC) simulations. The calculations assume a synthetic high-redshift halo emitting these lines. This plot illustrates the sensitivity of NIRSpec across redshifts $z=5-11$, indicating the flux thresholds necessary for line detection. } 
    \label{fig:balmer_flux_ETC}
    \end{center}
\end{figure}
        
To assess the observational feasibility of detecting Balmer emission lines from high-redshift Population~III star-forming regions, we employed the \textit{JWST Exposure Time Calculator} (ETC)\footnote{\url{https://jwst.etc.stsci.edu}}. The ETC is an official web-based tool developed by the Space Telescope Science Institute (STScI) to simulate JWST instrument performance under realistic astrophysical conditions. It supports all JWST observing modes and models three-dimensional astronomical scenes, including both point-like and extended sources across spatial and spectral dimensions.

For our simulations, we assumed a \texttt{FULL} subarray with the \texttt{NRS} readout pattern. We designed an observation that would require a total of around $10^{4} \: {\rm s}$ of on-source time, as an example of what JWST could expect to detect in a relatively shallow exposure. Specifically, the configuration included 29 groups per integration, 2 integrations per exposure, and a total of 4 dithers, yielding 8 total integrations. This setup resulted in a total exposure time of 10,049.62~s ($\approx$\,2\,h\,47\,min\,30\,s), which served as the baseline for our flux sensitivity estimates and detectability analysis. Further details are given in Appendix~\ref{appendix:obs}

The ETC output provides the minimum flux required for each Balmer line to be detectable at the specified threshold, effectively defining the sensitivity limit of NIRSpec under these conditions. These flux thresholds, as a function of redshift, are presented in Fig.~\ref{fig:balmer_flux_ETC}. We see that for a $\sim 10^{4}$ second observation and an S/N of 5, the minimum detectable flux for all four of the lines lies in the range 0.6--$1.0 \times 10^{-18} \: {\rm erg \, cm^{-2} \, s^{-1}}$ for redshifts $z = 5$--11. Although we can improve on this limit by observing for a longer time, even in the ideal case where we are limited by the photon noise from the source, we expect the sensitivity at fixed S/N to scale with the exposure time $t$ as $t^{-1/2}$, meaning that improving on this limit by more than a factor of a few would require an impractically larger amount of JWST time. In practice, even achieving an improvement of a factor of a few may be difficult, owing to the non-negligible contribution of detector noise \citep{jakobsen2022nirspec}.

\section{Results} \label{sec:results}

    \begin{figure}[h]
        \begin{center}
        \includegraphics[width=.39\textwidth]{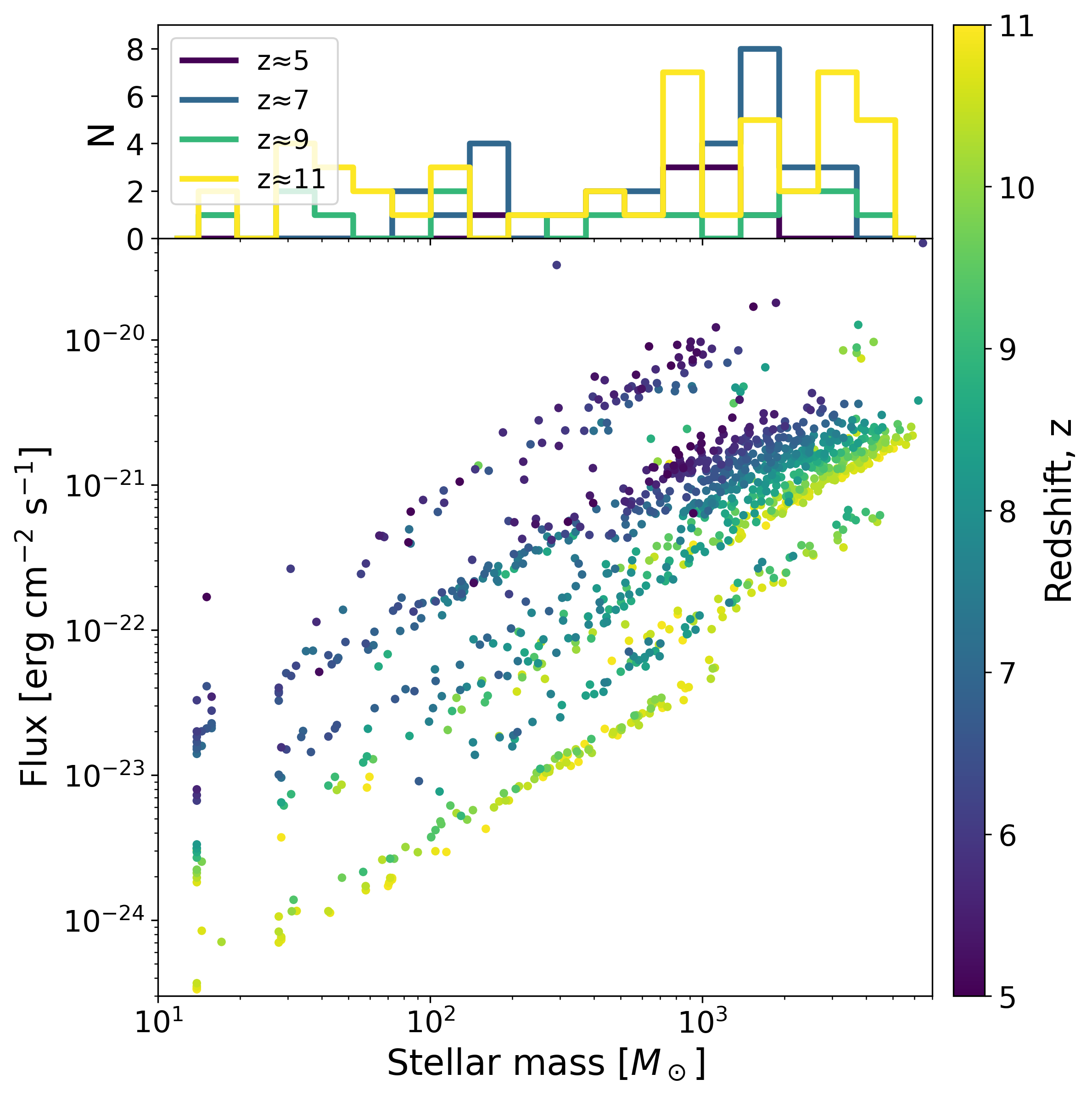}
        \caption{Observed flux in the H$\alpha$ line as a function of peak Pop.\ III stellar mass for Pop.\ III star-forming halos, derived from the \asloth model using the EPS merger tree. Each point represents one halo, colour-coded by redshift in the range $5 \leq z \leq 11$. The fluxes are computed from the corresponding peak ionising-photon luminosities reported by \asloth\ and converted to H$\alpha$ using Case~B recombination (Sec.~\ref{sec:methods}). While the most massive Pop.\ III systems ($M_{\star,\mathrm{III}}^{\rm peak}\sim10^4\,\Msun$) approach flux levels of $\sim10^{-20}\,\mathrm{erg\,cm^{-2}\,s^{-1}}$, even these remain far below the JWST/NIRSpec detection threshold ($\sim 6 \times 10^{-19}\,\mathrm{erg\,cm^{-2}\,s^{-1}}$ at $z \sim 5$) . The corresponding values for H$\beta$, H$\gamma$ and H$\delta$ are factors of 3-10 smaller.} 
        \label{fig:flux_mass}
        \end{center}
    \end{figure}

    \begin{figure*}[h]
        \centering
        \includegraphics[width=.75\textwidth]{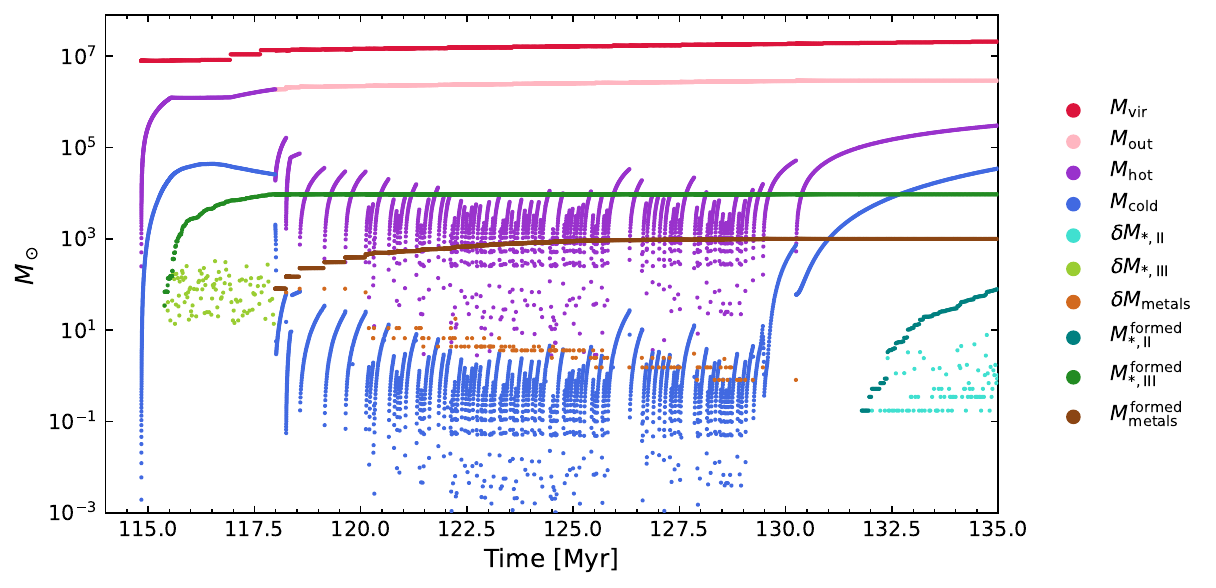}
        \caption{Time evolution of baryonic quantities for a single halo branch.
        Shown are the total halo mass ($M_{\mathrm{vir}}$), outflowing gas ($M_{\mathrm{out}}$), hot and cold gas reservoirs ($M_{\mathrm{hot}}$, $M_{\mathrm{cold}}$), the stellar mass formed per substep ($\delta M_{*}$), and the corresponding metal production ($\delta M_{\mathrm{metals}}$).
        We also show the cumulative stellar mass formed in Pop.\,III and Pop.\,II stars ($M^{\mathrm{formed}}_{*,\mathrm{III}}$, $M^{\mathrm{formed}}_{*,\mathrm{II}}$).
        Supernova events are associated with metal ejections and indicated by orange markers.
        The episodic star-formation cycle reflects both feedback-driven heating/outflows and the inefficient cooling of the post-burst hot gas, which delays the re-establishment of a substantial cold-gas reservoir.}
        \label{fig:baryonicbranch}
    \end{figure*} 

    \begin{figure*}[h]
        \begin{center}
        \includegraphics[width=.32\textwidth]{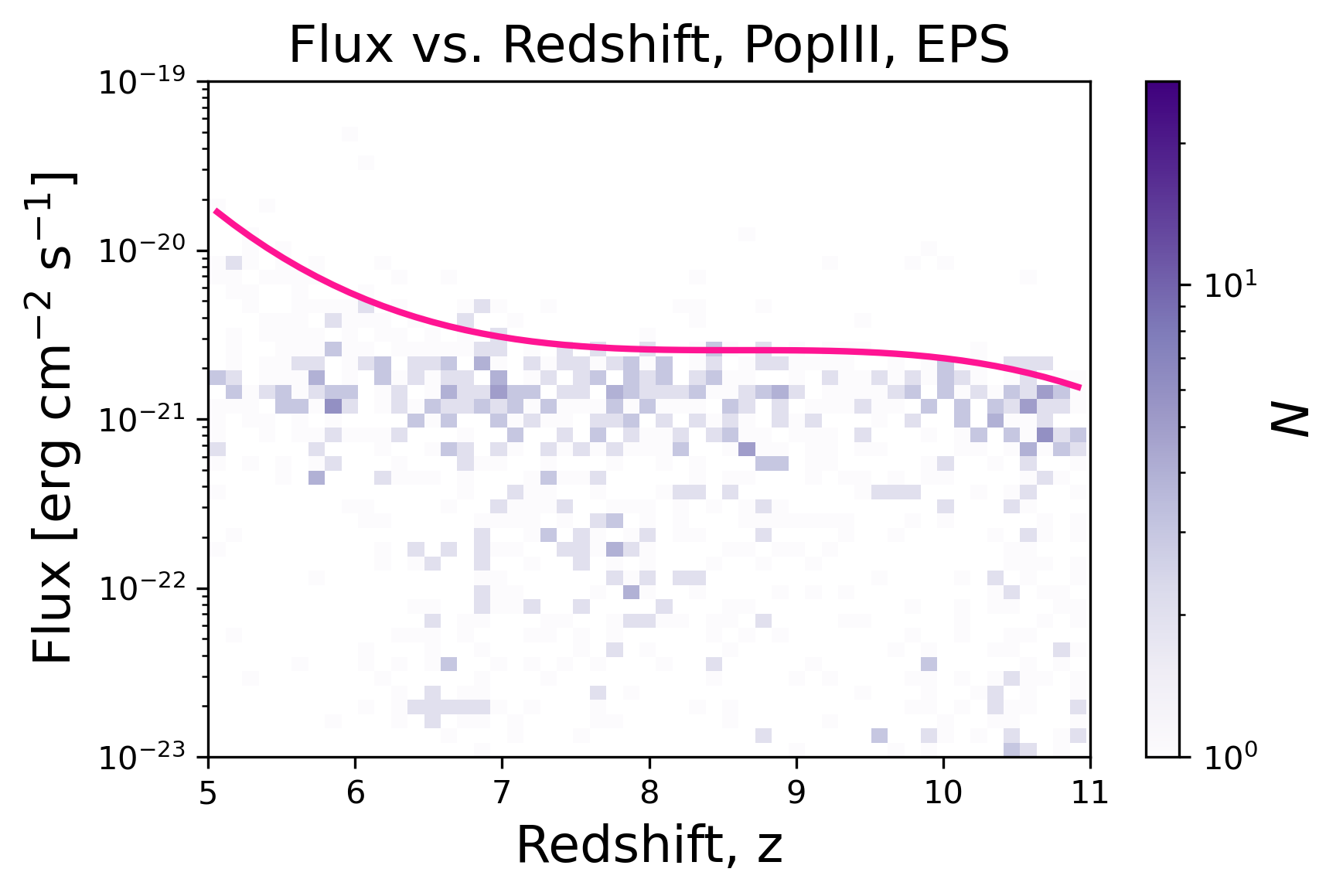}
        \includegraphics[width=.32\textwidth]{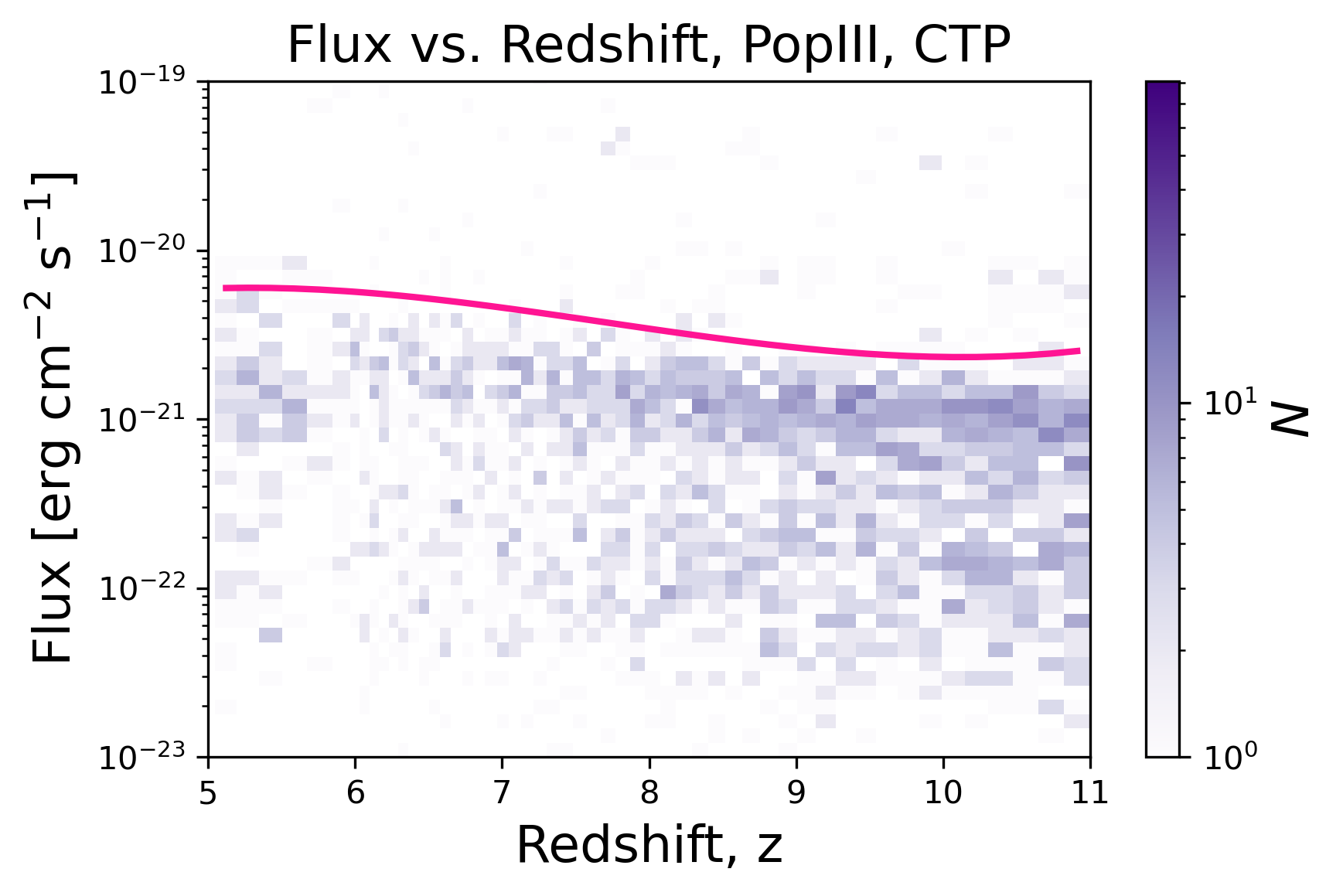}
        \includegraphics[width=.32\textwidth]{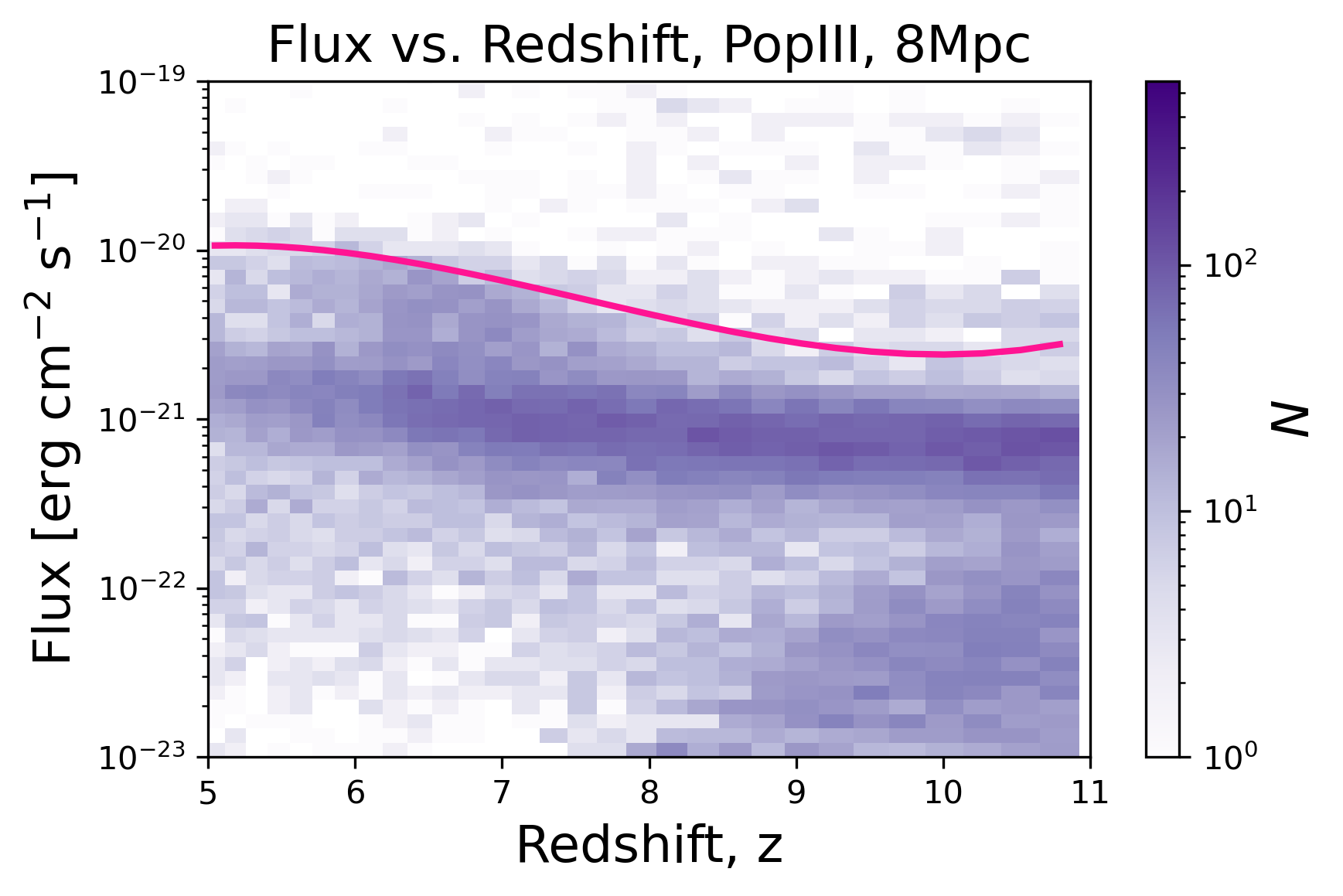}
        \includegraphics[width=.32\textwidth]{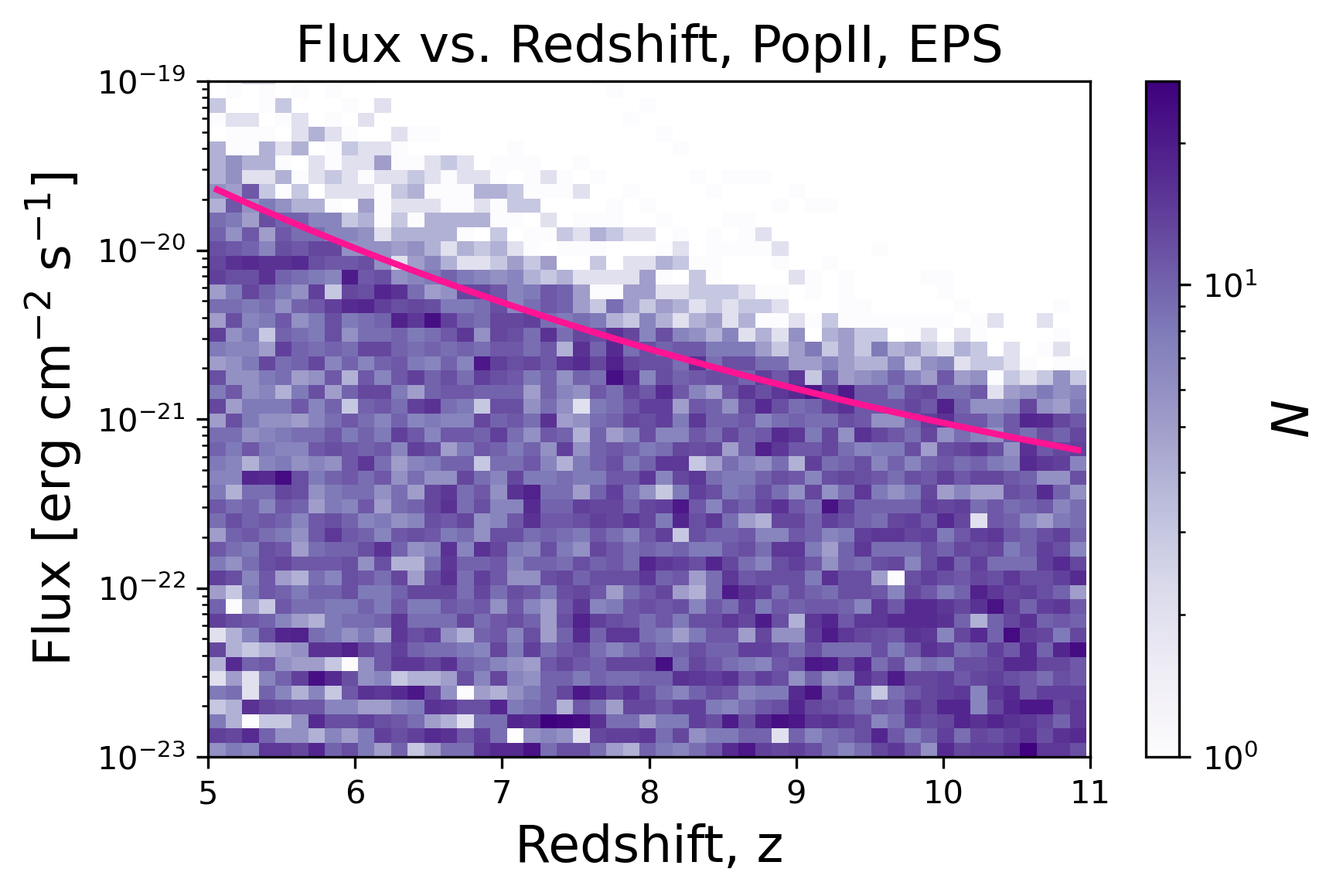}
        \includegraphics[width=.32\textwidth]{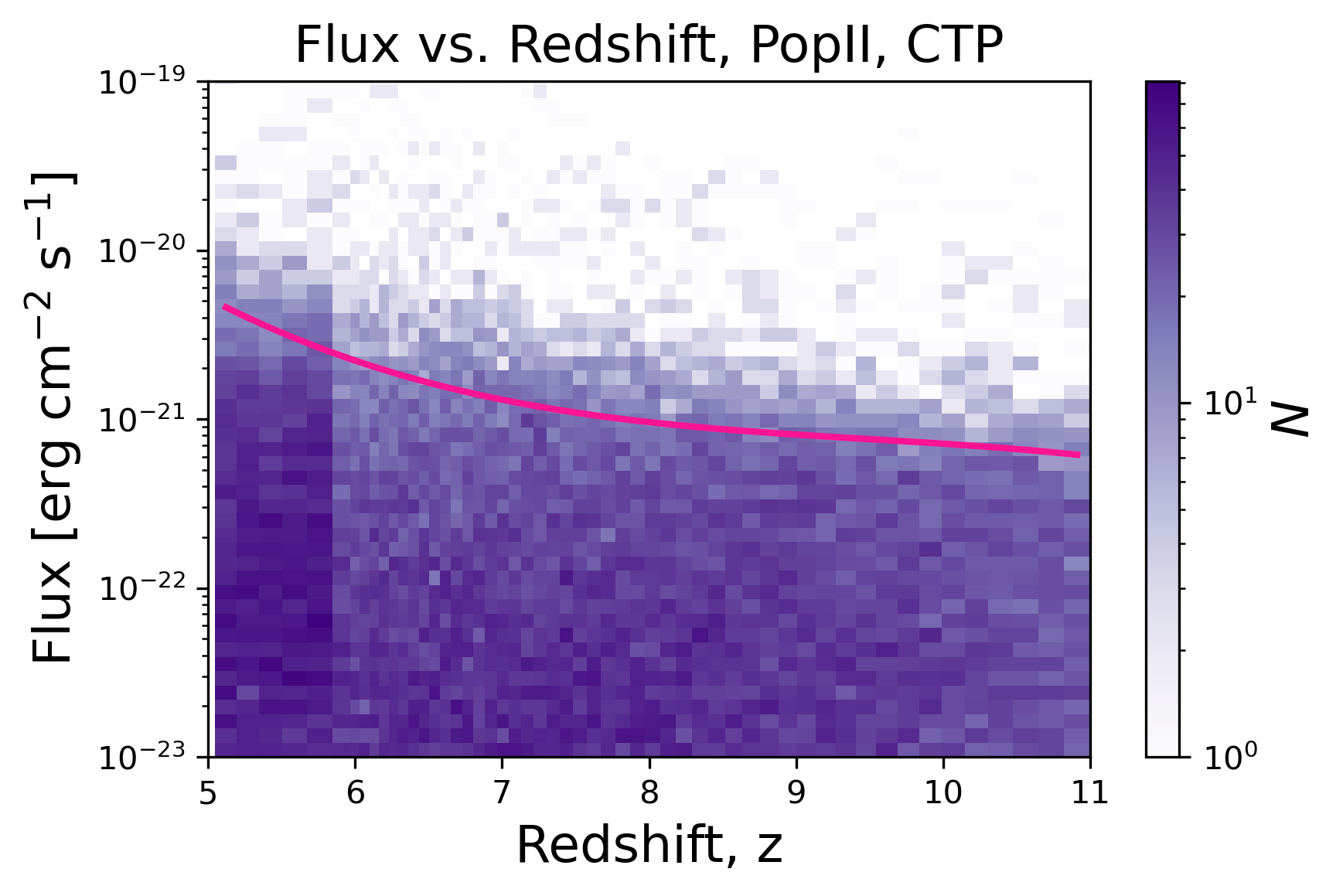}
        \includegraphics[width=.32\textwidth]{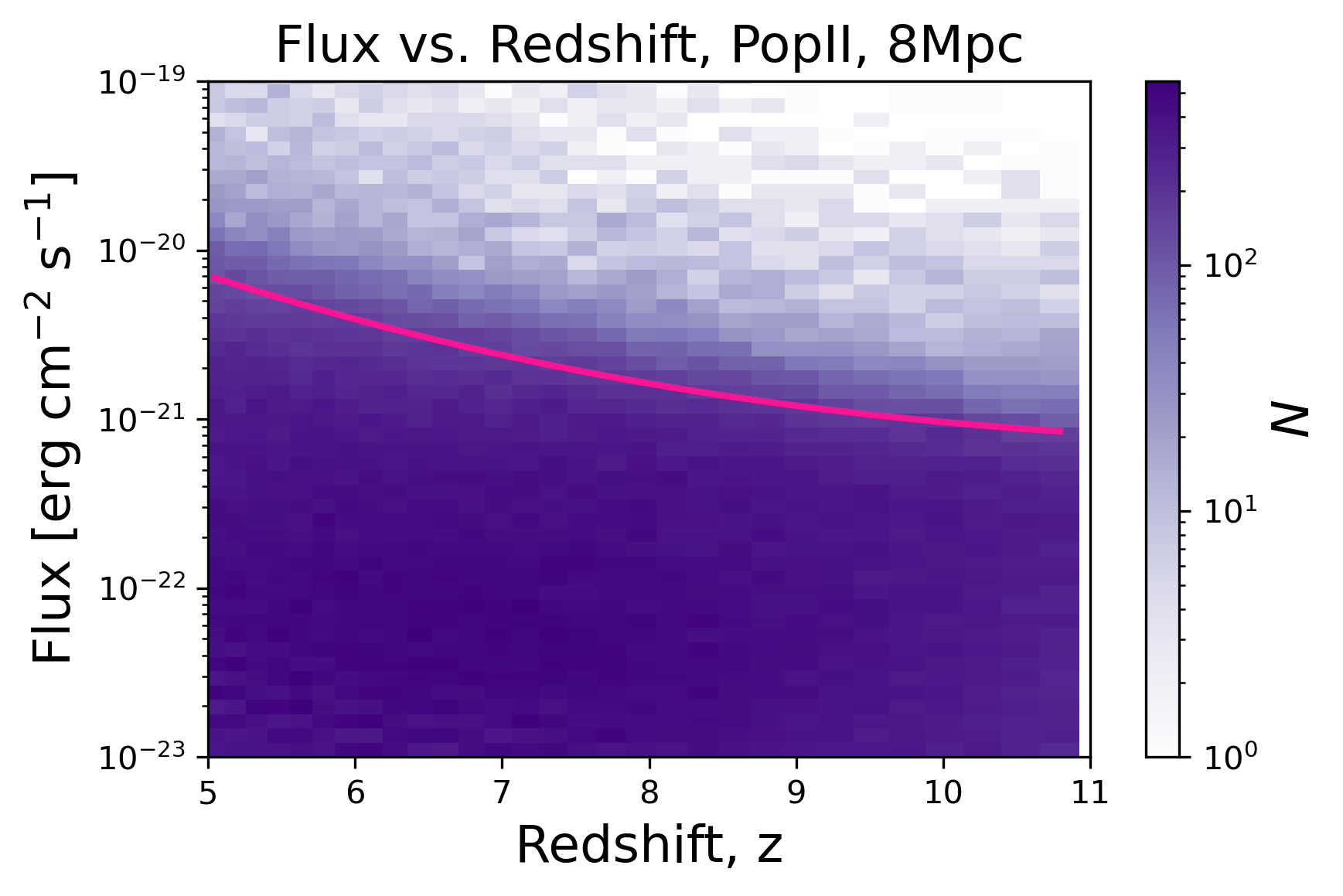}
        \caption{{H$\alpha$ flux as a function of redshift for Pop.\,III (top row) and Pop.\,II (bottom row) stellar populations, modelled with \asloth\ using three different merger-tree inputs. Left: Extended Press--Schechter (EPS) tree. Middle: Caterpillar merger tree representative of a Milky Way progenitor. Right: cosmologically representative merger tree extracted from an 8~Mpc/$h$ dark-matter simulation box. Each panel shows a two-dimensional histogram of model outputs, with fluxes binned logarithmically and colour-coded by $\log_{10}(N)$. The pink curve marks the smoothed upper envelope, defined as the 97th percentile of the flux distribution in each redshift bin.}}
        \label{fig:alpha_flux_vs_redshift}
        \end{center}
    \end{figure*}

    \begin{figure}[h]
        \begin{center}
        \includegraphics[width=.39\textwidth]{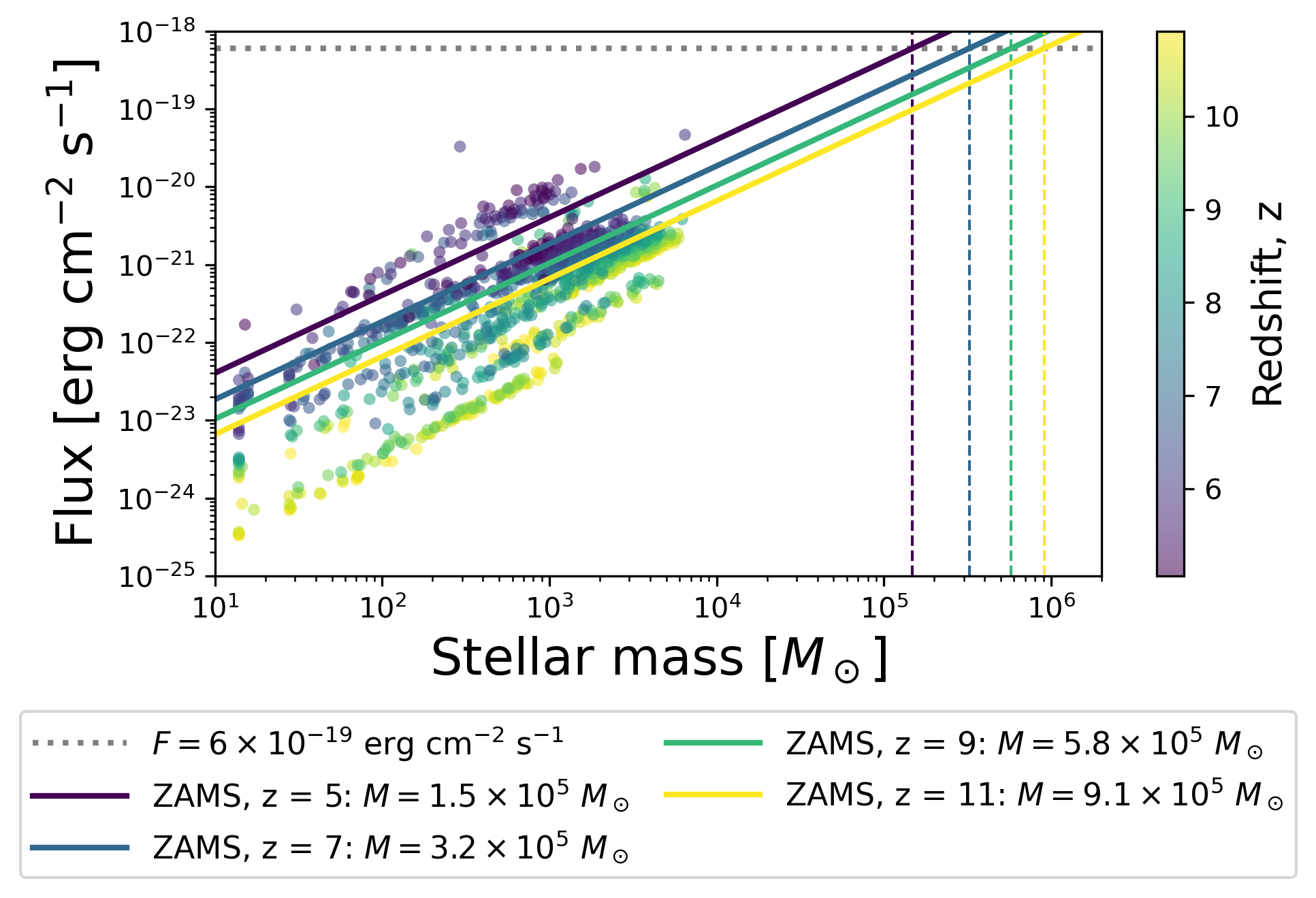}
        \caption{H$\alpha$ flux as a function of peak Pop.\ III stellar mass for Pop.\ III star-forming halos, derived from the \asloth model using the EPS merger tree. Each point represents one halo, colour-coded by redshift in the range $5 \le z \le 11$. The fluxes are computed from the corresponding peak ionising-photon luminosities reported by \asloth and converted to H$\alpha$ using Case B recombination (Sec.~\ref{sec:methods}). While the most massive Pop.\ III systems ($M^{\rm peak}_{\star,\rm III}\sim10^4\,M_\odot$) approach flux levels of $\sim10^{-20}$ erg cm$^{-2}$ s$^{-1}$, even these remain far below the JWST/NIRSpec detection threshold ($\sim6\times10^{-19}$ erg cm$^{-2}$ s$^{-1}$ at $z\sim5$). Possible deviations from the analytic ZAMS expectation are discussed in Appendix~\ref{appendix:overline}.}
        \label{fig:mass_cross}
        \end{center}
    \end{figure}
    
    \subsection{Masses, SFRs, and baryonic evolution of Pop.\ III systems}

    For our fiducial choice of parameters, Pop.\ III star formation occurs in halos with virial masses of $10^{6}$–$10^{8}\,\Msun$.
    However, these halos form only a few Pop.\ III stars. Young Pop.\ III stellar populations ($t_{\mathrm{age}}<10\,\mathrm{Myr}$) have typical masses $M_{\star,\mathrm{III}}^{\mathrm{young}}\sim10^{1}$-$10^{4}\,\Msun$, and the peak mass of Pop.\ III stars formed in a halo or any of its progenitors does not exceed $10^{4}\,\Msun$. Here, the peak mass refers to the maximum instantaneous total mass of massive Pop.\ III main-sequence stars (with $M>5\,\mathrm{M}_\odot$) reached in a halo, as tracked within the star-formation sub-cycles of the \asloth\ model. Such systems form through brief, feedback-regulated bursts with instantaneous SFRs of $10^{-5}$–$10^{-3}\,\Msun\,\mathrm{yr^{-1}}$. No halos reach peak Pop.\ III stellar masses $\gtrsim10^{5}\,\Msun$, the characteristic scale required to produce observable Balmer-line emission based on the analytic ZAMS flux–mass relations shown in Fig.~\ref{fig:mass_cross} (see also Sec.~\ref{sec:line_flux})

    The absence of very massive Pop.\,III stellar systems in our models can be traced to the self-regulating nature of baryonic feedback. As illustrated in Fig.~\ref{fig:baryonicbranch}, the star formation cycle in a typical primordial halo is highly episodic: a short burst of Pop.\,III star formation rapidly heats and ionises the gas, while subsequent supernova explosions enrich the surrounding medium with metals and drive strong mechanical feedback. Once the cold-gas reservoir is depleted, star formation is temporarily suppressed until gas is reaccreted or cools again, after which the halo transitions to Pop.\,II formation after a few Myr. This feedback loop prevents uninterrupted growth of the Pop.\,III stellar component and confines the total Pop.\,III stellar mass to $\sim10^{3}$--$10^{4}\,\Msun$ in most halos. Note also that aside from feedback, the main factor limiting the formation of Pop.\,III stars in these systems is not the poorly-constrained efficiency with which these stars form from cold gas. Rather, it is the lengthy cooling time of the gas itself, which limits the amount of cold gas available at the time that Pop.\,III stars start to form. For example, in the case illustrated in Fig.~\ref{fig:baryonicbranch}, the total mass in cold gas at the onset of star formation is only a few times $10^{4} \: {\rm M_{\odot}}$, which is only a few percent of the total baryonic mass associated with the halo. Even in the unrealistic case in which all of this cold gas forms stars before the first supernova explosion, we would still only form a total Pop.\,III stellar population mass of a few times $10^{4} \: {\rm M_{\odot}}$.

    As shown in Fig.~\ref{fig:baryonicbranch}, the sharp decline of the cold-gas mass and the rise of hot and outflowing components occur immediately after the first supernova event (red circle). This indicates that mechanical feedback dominates the quenching of Pop.\ III star formation in this branch. Radiative heating before the explosion mildly regulates gas accretion but does not remove the cold reservoir. The supernova-driven outflow expels most of the gas and enriches the halo, marking the transition to Pop.\ II formation.

    As will be discussed below, the lack of massive Pop.\,III stellar systems above a few $10^{4}\,\Msun$ is a common feature across all merger-tree realisations (EPS, CTP, and 8~Mpc box). This upper limit is set by the balance between the short Pop.\,III lifetime ($\lesssim2$\,Myr), efficient photoheating, and supernova–driven gas blowout, which halt further accretion before large Pop.\,III clusters can assemble. As a result, \asloth\ does not predict the formation of massive, Pop.\,III–dominated galaxies: radiative and mechanical feedback ensure that each halo experiences only a brief, low–mass Pop.\,III episode before transitioning to metal–enriched star formation.

    \subsection{Balmer-line fluxes and dependence on model parameters}\label{sec:line_flux}

    Figure~\ref{fig:alpha_flux_vs_redshift} shows the H$\alpha$ fluxes of all simulated halos over $5\le z\le11$. Pop.\ III systems produce $F_{\mathrm{H}\alpha}\sim10^{-22}$–$10^{-20}\,\mathrm{erg\,cm^{-2}\,s^{-1}}$, one to three orders of magnitude below the JWST/NIRSpec sensitivity limit ($\sim6\times10^{-19}\,\mathrm{erg\,cm^{-2}\,s^{-1}}$). Higher Balmer transitions are weaker by factors of 3--10 (see Eq.~\ref{eq:scaling}) and are also undetectable.

    The flux–mass relation in Fig.~\ref{fig:flux_mass} shows that even the most massive Pop.\ III systems ($M_{\star,\mathrm{III}}\sim10^{4}\,\Msun$) remain far below the detection threshold.   Extrapolation in Fig.~\ref{fig:mass_cross} indicates that observable fluxes would require $M_{\star,\mathrm{III}}\gtrsim(1$–$9)\times10^{5}\,\Msun$.
    Across all formation modes (EPS, \ctp, and 8~Mpc/h box), the absence of such massive halos persists, confirming that this result is independent of environment or resolution.

    We have also made use of the flexibility and speed of the \asloth model to explore the impact of varying the input parameters within the ranges allowed by the current observational constraints used to calibrate \asloth. Full results of this exploration are shown in Appendix~\ref{appendix:param}, but the main takeaway is that although some of the parameter variations have an appreciable impact on the predicted fluxes, the largest increases are around a factor of 3--5, far below the factor of $\sim 60$ needed to change the conclusions of this paper.

    Among the explored parameters, the most noticeable effects arise from the Pop.\,III IMF parameters and the escape fractions. In particular, increasing the minimum Pop.\,III stellar mass $M_{\mathrm{min}}$ shifts the upper flux envelope upward by up to $\sim0.5$--$0.7$\,dex across the redshift range considered, reflecting the higher ionising photon yield of a more top-heavy IMF. The IMF slope $\alpha_{\mathrm{III}}$ also affects the flux envelope at the level of $\sim0.3$--$0.5$\,dex, primarily at lower redshifts. However, there is a substantial degeneracy between these two parameters, since flattening the IMF slope and increasing the minimum mass both act to make the IMF more top heavy. Consequently, varying both parameters simultaneously does not give a significantly larger boost to the Balmer line fluxes than varying them individually. Changes in the escape fractions $f_{\mathrm{esc,II}}$ and $f_{\mathrm{esc,III}}$ alter the normalisation of the recombination luminosity by a comparable amount ($\lesssim0.5$\,dex), as expected because the retained ionising-photon fraction directly controls the nebular emission. 

    In summary, Pop.\ III halos in \asloth\ remain small ($M_{\star,\mathrm{III}}\lesssim10^{4}\,\Msun$) and short-lived.  
    Feedback rapidly quenches star formation, preventing the formation of massive Pop.\ III galaxies and leaving Balmer-line fluxes one to three orders of magnitude below current JWST sensitivity.
    
\section{Discussion} \label{sec:discussion}

In Fig. \ref{fig:alpha_flux_vs_redshift}, we show the H$\alpha$ flux produced by each Pop.\ III and Pop.\ II star-forming halo in our A-SLOTH models in the redshift range $5 \leq z \leq 11$. Results are shown for all three A-SLOTH structure formation modes (EPS, CTP, and 8 Mpc box; see Sec.~\ref{sec:asloth}), using the best-fit input parameters from Table~\ref{tab:parameters}. We see that in all three cases, 

Pop.\ III stellar populations produce H$\alpha$ fluxes of around 
$10^{-20}$ erg cm$^{-2}$ s$^{-1}$ or below, declining further with increasing redshift. Most Pop.\ II systems also produce fluxes below this level, but there is a small population of Pop.\ II systems that have much higher H$\alpha$ fluxes, reaching $10^{-19}$ erg cm$^{-2}$ s$^{-1}$ or above. Pop.\ II systems also reach much lower flux values than the majority of Pop.\ III systems, which is a consequence of the different IMFs assumed for the two stellar populations.

Comparing these results with the sensitivity limits inferred from our ETC calculations reveals a significant disparity: the vast majority of both Pop.\ II and Pop.\ III fluxes predicted by \asloth lie well below the ETC-determined detectability limits for JWST/NIRSpec. In particular, Pop.\ III halos consistently fall short by one to three orders of magnitude.

Population\ II halos, while closer to the detection boundary at lower redshifts, still do not reach the threshold required for confident spectroscopic identification beyond $z \sim 7$ in unlensed fields. The results for the other Balmer series lines (not shown) are very similar, but with the fluxes decreased by the scaling factors given in Equation~\ref{eq:scaling}.

The reason why we obtain this result becomes clear if we plot the H$\alpha$ flux as a function of the Pop.\ III stellar mass formed in each halo (Figure~\ref{fig:flux_mass}). We see that the vast majority of Pop.\ III star-forming halos are associated with total Pop.\ III stellar masses of $10^{1}$--$10^{4} \: {\rm M_{\odot}}$, and that even the youngest systems with this mass do not produce enough H$\alpha$ photons to be detectable by JWST, in agreement with the earlier results of \citet{johnson2009first}. Pop.\ III systems with total masses in the range necessary to produce detectable emission do not form in these models. Extrapolating from our results here (see Figure~\ref{fig:mass_cross}), we estimate that halos would have to form a total Pop.\ III stellar mass of a few times $10^{5} \: {\rm M_{\odot}}$ to produce H$\alpha$ fluxes that exceed our adopted detection threshold.

This conclusion is broadly consistent with the majority of hydrodynamical simulations of Pop.\,III star formation \citep[e.g.,][]{hirano14,Xu16a,liu2020,skinner20,Storck2026}, which generally find that metal-free star formation is self-limiting because radiative feedback, supernova feedback, and rapid metal enrichment terminate the Pop.\,III phase after relatively small stellar masses have formed. A possible exception is provided by rare atomic-cooling halos exposed to strong Lyman--Werner backgrounds, where Pop.\,III star formation can be delayed until larger gas reservoirs have assembled. For example, \citet{Jeong2026} used cosmological radiation-hydrodynamical zoom-in simulations of atomic-cooling halos at $7\lesssim z\lesssim8$ with high Lyman--Werner backgrounds and high star formation efficiencies, and found that intense Pop.\,III starbursts can occur for sufficiently strong Lyman--Werner radiation fields. Even in this favourable case, however, the Pop.\,III stellar mass remains limited to $M_{\star,\mathrm{Pop\,III}}<10^6\,\mathrm{M}_\odot$, because the first Pop.\,III supernovae rapidly enrich the gas and trigger the transition to Pop.\,II star formation. This supports our conclusion that very massive Pop.\,III-dominated systems should be rare, although it also highlights a possible formation channel for the brightest and most exceptional Pop.\,III candidates \citep[e.g., GN-z11][]{Maiolino2024,Maiolino2026,Jeon2026}.

This analysis confirms that JWST, without the aid of gravitational lensing, is unlikely to detect typical high-redshift Pop.\ III star-forming halos via their Balmer emission. While strong gravitational lensing can, in principle, amplify fluxes by an order of magnitude or more and thus bring some Pop.\ III sources within the JWST detection regime, such events are rare and cover a small effective survey volume \cite[see, e.g.,][]{rydberg13, rydberg20, zackrisson24}. As such, the direct detection of pristine stellar populations remains observationally challenging \cite[for further discussions, see e.g.][]{klessen23}. 

Nevertheless, there are two Pop.\ III candidates (with $Z\lesssim 0.004\ \rm Z_\odot$) detected via extreme lensing magnification by a factor of $\sim100$ at $z=6.6$, LAP1 \citep{Vanzella2023} and LAP1-B \citep{Nakajima2025}, with inferred stellar masses $\lesssim 10^4\ \rm M_\odot$.

These systems are consistent with our model predictions and support the need for extreme lensing magnification to detect typical Pop.\,III systems. This interpretation is also supported by \citet{Visbal2025}, who argued that LAP1-B is consistent with several theoretical expectations for Pop.\,III sources, including formation in a low-metallicity halo, a top-heavy IMF, and a low mass of massive Pop.\,III stars of only a few $\times10^3\,\mathrm{M}_\odot$. They further found that the expected abundance of such highly magnified Pop.\,III sources behind MACS~J0416 is of order unity at $z\simeq6$--7, consistent with the detection of LAP1-B.

On the other hand, a few metal-poor ($Z<0.01\rm\ Z_\odot$) systems detected without such lensing magnification at $z\sim 3-11$ are also speculated to be Pop.\ III based on their strong nebular emission, with inferred stellar masses of a few $10^{5}\ \rm M_\odot$, including the HeII emitting region in GNz-11 \citep{Maiolino2024,Maiolino2026,Ubler2026}, AMORE6 \citep{Morishita2025}, and CR3 \citep{Cai2025}. Our results suggest that these systems are unlikely to be dominated by Pop.\ III stars, because of the difficulty of forming such massive clusters of ordinary Pop.\ III stars given the combined effects of inefficient cooling and strong stellar feedback. Based on our results, we conclude that they are more likely to be Pop.\ II systems. In fact, \asloth typically predicts that most metal-poor ($Z<0.01\rm\ Z_\odot$) stars born at $z \sim 5-10$ are Pop~II, and that less than 1\% are Pop.\  III \citep{Liu2025}. If these systems are later confirmed to be dominated by Pop.\ III stars, it would indicate either that our model is under-estimating the mass of cold metal-free gas that can be accumulated in a halo prior to the onset of Pop.\ III star formation, or that it is over-estimating the effectiveness of stellar feedback at removing this gas from the halo.

\section{Summary}

Our comparison between the predicted Balmer-line luminosities from Pop.\,III star-forming halos, as modelled with \asloth, and the flux thresholds required for detection with JWST (based on ETC estimates for a signal-to-noise ratio of 5) shows that detection of Pop.\,III stars via this channel is unlikely with current observational capabilities. For an assumed exposure time of $10^{4}\,\mathrm{s}$, the intrinsic Balmer-line fluxes fall short of the JWST/NIRSpec detection limits by two to three orders of magnitude across all redshifts considered.

Crucially, this shortfall is not primarily an observational limitation that could be overcome with deeper integrations. Instead, it reflects a physical outcome of hierarchical structure formation regulated by stellar feedback: in our models, the massive, Pop.\,III-dominated stellar systems required to produce detectable Balmer emission simply do not form. Pop.\,III star formation occurs in brief, self-regulated episodes that are terminated by radiative and supernova feedback, yielding typical Pop.\,III stellar masses of only $\sim10^{1}$--$10^{4}\,\mathrm{M}_\odot$, far below the $\gtrsim10^{5}\,\mathrm{M}_\odot$ required for detectability.

While strong gravitational lensing could, in principle, boost the fluxes of rare Pop.\,III systems into the observable regime, such configurations are expected to be uncommon and probe only a limited effective volume. Consequently, despite JWST’s unprecedented infrared sensitivity, the direct detection of Balmer-series emission from typical Pop.\,III star-forming halos is unlikely in unlensed fields. Our results, therefore, suggest that the absence of detectable Pop.\,III Balmer emission with JWST would reflect the underlying physics of early star formation rather than insufficient observational depth, and that alternative tracers or fundamentally different Pop.\,III formation pathways would be required to make such systems observable.

\begin{acknowledgements}

We thank the anonymous referee for a constructive report that helped us improve the clarity of the manuscript.

We acknowledge financial support from the German Excellence Strategy via the Heidelberg Cluster of Excellence (EXC 2181 - 390900948) ``STRUCTURES''. RSK is also grateful for support from the European Research Council via the ERC Synergy Grant ``ECOGAL'' (project ID 855130) and from the German Ministry for Economic Affairs and Climate Action in the project ``MAINN'' (funding ID 50OO2206). Furthermore, RSK thanks the 2024/25 Class of Radcliffe Fellows for highly interesting and stimulating discussions. This work relied on computing resources provided by the Ministry of Science, Research and the Arts (MWK) of the State of Baden-W\"{u}rttemberg through bwHPC and the German Science Foundation (DFG) through grants INST 35/1134-1 FUGG and 35/1597-1 FUGG, and also for data storage at SDS@hd funded through grants INST 35/1314-1 FUGG and INST 35/1503-1 FUGG.
\end{acknowledgements}

\bibliographystyle{aa} 
\bibliography{bibliography} 

@ARTICLE{kroupa01,
       author = {{Kroupa}, Pavel},
        title = "{On the variation of the initial mass function}",
      journal = {\mnras},
         year = 2001,
        month = apr,
       volume = {322},
       number = {2},
        pages = {231-246},
          doi = {10.1046/j.1365-8711.2001.04022.x},
archivePrefix = {arXiv},
       eprint = {astro-ph/0009005},
 primaryClass = {astro-ph},
       adsurl = {https://ui.adsabs.harvard.edu/abs/2001MNRAS.322..231K}
}

@ARTICLE{chabrier03,
       author = {{Chabrier}, Gilles},
        title = "{Galactic Stellar and Substellar Initial Mass Function}",
      journal = {\pasp},
         year = 2003,
        month = jul,
       volume = {115},
       number = {809},
        pages = {763-795},
          doi = {10.1086/376392},
archivePrefix = {arXiv},
       eprint = {astro-ph/0304382},
 primaryClass = {astro-ph},
       adsurl = {https://ui.adsabs.harvard.edu/abs/2003PASP..115..763C}
}

@ARTICLE{schauer19,
       author = {{Schauer}, Anna T.~P. and {Glover}, Simon C.~O. and {Klessen}, Ralf S. and {Ceverino}, Daniel},
        title = "{The influence of streaming velocities on the formation of the first stars}",
      journal = {\mnras},
         year = 2019,
        month = apr,
       volume = {484},
       number = {3},
        pages = {3510-3521},
          doi = {10.1093/mnras/stz013},
archivePrefix = {arXiv},
       eprint = {1811.12920},
 primaryClass = {astro-ph.GA},
       adsurl = {https://ui.adsabs.harvard.edu/abs/2019MNRAS.484.3510S}
}

@ARTICLE{tseliakhovich10,
       author = {{Tseliakhovich}, Dmitriy and {Hirata}, Christopher},
        title = "{Relative velocity of dark matter and baryonic fluids and the formation of the first structures}",
      journal = {\prd},
         year = 2010,
        month = oct,
       volume = {82},
       number = {8},
          eid = {083520},
        pages = {083520},
          doi = {10.1103/PhysRevD.82.083520},
archivePrefix = {arXiv},
       eprint = {1005.2416},
 primaryClass = {astro-ph.CO},
       adsurl = {https://ui.adsabs.harvard.edu/abs/2010PhRvD..82h3520T}
}

@ARTICLE{press74,
       author = {{Press}, William H. and {Schechter}, Paul},
        title = "{Formation of Galaxies and Clusters of Galaxies by Self-Similar Gravitational Condensation}",
      journal = {\apj},
         year = 1974,
        month = feb,
       volume = {187},
        pages = {425-438},
          doi = {10.1086/152650},
       adsurl = {https://ui.adsabs.harvard.edu/abs/1974ApJ...187..425P}
}

@ARTICLE{bond91,
       author = {{Bond}, J.~R. and {Cole}, S. and {Efstathiou}, G. and {Kaiser}, N.},
        title = "{Excursion Set Mass Functions for Hierarchical Gaussian Fluctuations}",
      journal = {\apj},
         year = 1991,
        month = oct,
       volume = {379},
        pages = {440},
          doi = {10.1086/170520},
       adsurl = {https://ui.adsabs.harvard.edu/abs/1991ApJ...379..440B}
}

@ARTICLE{rydberg13,
       author = {{Rydberg}, Claes-Erik and {Zackrisson}, Erik and {Lundqvist}, Peter and {Scott}, Pat},
        title = "{Detection of isolated Population III stars with the James Webb Space Telescope}",
      journal = {\mnras},
         year = 2013,
        month = mar,
       volume = {429},
       number = {4},
        pages = {3658-3664},
          doi = {10.1093/mnras/sts653},
archivePrefix = {arXiv},
       eprint = {1206.0007},
 primaryClass = {astro-ph.CO},
       adsurl = {https://ui.adsabs.harvard.edu/abs/2013MNRAS.429.3658R}
}

@ARTICLE{rydberg20,
       author = {{Rydberg}, Claes-Erik and {Whalen}, Daniel J. and {Maturi}, Matteo and {Collett}, Thomas and {Carrasco}, Mauricio and {Magg}, Mattis and {Klessen}, Ralf S.},
        title = "{Detecting strongly lensed supernovae at z {\ensuremath{\sim}} 5-7 with LSST}",
      journal = {\mnras},
         year = 2020,
        month = jan,
       volume = {491},
       number = {2},
        pages = {2447-2459},
          doi = {10.1093/mnras/stz3203},
archivePrefix = {arXiv},
       eprint = {1805.02662},
 primaryClass = {astro-ph.GA},
       adsurl = {https://ui.adsabs.harvard.edu/abs/2020MNRAS.491.2447R}
}

@ARTICLE{zackrisson24,
       author = {{Zackrisson}, Erik and {Hultquist}, Adam and {Kordt}, Aron and {Diego}, Jose M. and {Nabizadeh}, Armin and {Vikaeus}, Anton and {Meena}, Ashish Kumar and {Zitrin}, Adi and {Volpato}, Guglielmo and {Lundqvist}, Emma and {Welch}, Brian and {Costa}, Guglielmo and {Windhorst}, Rogier A.},
        title = "{The detection and characterization of highly magnified stars with JWST: prospects of finding Population III}",
      journal = {\mnras},
         year = 2024,
        month = sep,
       volume = {533},
       number = {3},
        pages = {2727-2746},
          doi = {10.1093/mnras/stae1881},
archivePrefix = {arXiv},
       eprint = {2312.09289},
 primaryClass = {astro-ph.GA},
       adsurl = {https://ui.adsabs.harvard.edu/abs/2024MNRAS.533.2727Z}
}

@ARTICLE{hartwig24,
       author = {{Hartwig}, Tilman and {Lipatova}, Veronika and {Glover}, Simon C.~O. and {Klessen}, Ralf S.},
        title = "{A-SLOTH reveals the nature of the first stars}",
      journal = {\href{https://academic.oup.com/mnras/article/535/1/516/7815909}{\mnras}},
         year = 2024,
        month = nov,
       volume = {535},
       number = {1},
        pages = {516-530},
          doi = {10.1093/mnras/stae2318},
archivePrefix = {arXiv},
       eprint = {2410.05393},
 primaryClass = {astro-ph.GA},
       adsurl = {https://ui.adsabs.harvard.edu/abs/2024MNRAS.535..516H}
}

@ARTICLE{hartwig22,
       author = {{Hartwig}, Tilman and {Magg}, Mattis and {Chen}, Li-Hsin and {Tarumi}, Yuta and {Bromm}, Volker and {Glover}, Simon C.~O. and {Ji}, Alexander P. and {Klessen}, Ralf S. and {Latif}, Muhammad A. and {Volonteri}, Marta and {Yoshida}, Naoki},
        title = "{Public Release of A-SLOTH: Ancient Stars and Local Observables by Tracing Halos}",
      journal = {\href{https://iopscience.iop.org/article/10.3847/1538-4357/ac7150}{\apj}},
         year = 2022,
        month = sep,
       volume = {936},
       number = {1},
          eid = {45},
        pages = {45},
          doi = {10.3847/1538-4357/ac7150},
archivePrefix = {arXiv},
       eprint = {2206.00223},
 primaryClass = {astro-ph.GA},
       adsurl = {https://ui.adsabs.harvard.edu/abs/2022ApJ...936...45H}
}

@ARTICLE{hegde23,
       author = {{Hegde}, Sahil and {Furlanetto}, Steven R.},
        title = "{A self-consistent semi-analytic model for Population III star formation in minihaloes}",
      journal = {\mnras},
         year = 2023,
        month = oct,
       volume = {525},
       number = {1},
        pages = {428-447},
          doi = {10.1093/mnras/stad2308},
archivePrefix = {arXiv},
       eprint = {2304.03358},
 primaryClass = {astro-ph.CO},
       adsurl = {https://ui.adsabs.harvard.edu/abs/2023MNRAS.525..428H}
}

@ARTICLE{klessen23,
       author = {{Klessen}, Ralf S. and {Glover}, Simon C.~O.},
        title = "{The First Stars: Formation, Properties, and Impact}",
      journal = {\araa},
         year = 2023,
        month = aug,
       volume = {61},
        pages = {65-130},
          doi = {10.1146/annurev-astro-071221-053453},
archivePrefix = {arXiv},
       eprint = {2303.12500},
 primaryClass = {astro-ph.CO},
       adsurl = {https://ui.adsabs.harvard.edu/abs/2023ARA&A..61...65K}
}

@ARTICLE{skinner20,
       author = {{Skinner}, Danielle and {Wise}, John H.},
        title = "{Cradles of the first stars: self-shielding, halo masses, and multiplicity}",
      journal = {\mnras},
         year = 2020,
        month = mar,
       volume = {492},
       number = {3},
        pages = {4386-4397},
          doi = {10.1093/mnras/staa139},
archivePrefix = {arXiv},
       eprint = {2001.04480},
 primaryClass = {astro-ph.GA},
       adsurl = {https://ui.adsabs.harvard.edu/abs/2020MNRAS.492.4386S}
}

@ARTICLE{greif15,
       author = {{Greif}, Thomas H.},
        title = "{The numerical frontier of the high-redshift Universe}",
      journal = {Computational Astrophysics and Cosmology},
         year = 2015,
        month = mar,
       volume = {2},
          eid = {3},
        pages = {3},
          doi = {10.1186/s40668-014-0006-2},
archivePrefix = {arXiv},
       eprint = {1410.3482},
 primaryClass = {astro-ph.GA},
       adsurl = {https://ui.adsabs.harvard.edu/abs/2015ComAC...2....3G}
}

@incollection{klessen19,
	Adsurl = {https://ui.adsabs.harvard.edu/abs/2019ffbh.book...67K},
	Author = {{Klessen}, Ralf},
	Booktitle = {Formation of the First Black Holes},
	Doi = {10.1142/9789813227958\_0004},
	Editor = {{Latif}, Muhammad and {Schleicher}, Dominik},
	Pages = {67-97},
	Publisher = {World Scientific Publishing},
	Title = {{Formation of the first stars}},
	Year = 2019}

@article{jaacks19,
	Adsurl = {https://ui.adsabs.harvard.edu/abs/2019MNRAS.488.2202J},
	Archiveprefix = {arXiv},
	Author = {{Jaacks}, Jason and {Finkelstein}, Steven L. and {Bromm}, Volker},
	Doi = {10.1093/mnras/stz1529},
	Eprint = {1804.07372},
	Journal = {\mnras},
	Month = sep,
	Number = {2},
	Pages = {2202-2221},
	Primaryclass = {astro-ph.GA},
	Title = {{Legacy of star formation in the pre-reionization universe}},
	Volume = {488},
	Year = 2019}

@ARTICLE{magg16,
	Adsurl = {http://adsabs.harvard.edu/abs/2016MNRAS.462.3591M},
	Archiveprefix = {arXiv},
	Author = {{Magg}, M. and {Hartwig}, T. and {Glover}, S.~C.~O. and {Klessen}, R.~S. and {Whalen}, D.~J.},
	Doi = {10.1093/mnras/stw1882},
	Eprint = {1606.06294},
	Journal = {\mnras},
	Month = nov,
	Pages = {3591-3601},
	Title = {{A new statistical model for Population III supernova rates: discriminating between {$\Lambda$}CDM and WDM cosmologies}},
	Volume = {462},
	Year = {2016}}

@ARTICLE{woods19,
	Adsurl = {https://ui.adsabs.harvard.edu/abs/2019PASA...36...27W},
	Archiveprefix = {arXiv},
	Author = {{Woods}, Tyrone E. and {Agarwal}, Bhaskar and {Bromm}, Volker and {Bunker}, Andrew and {Chen}, Ke-Jung and {Chon}, Sunmyon and {Ferrara}, Andrea and {Glover}, Simon C.~O. and {Haemmerl{\'e}}, Lionel and {Haiman}, Zolt{\'a}n and {Hartwig}, Tilman and {Heger}, Alexander and {Hirano}, Shingo and {Hosokawa}, Takashi and {Inayoshi}, Kohei and {Klessen}, Ralf S. and {Kobayashi}, Chiaki and {Koliopanos}, Filippos and {Latif}, Muhammad A. and {Li}, Yuexing and {Mayer}, Lucio and {Mezcua}, Mar and {Natarajan}, Priyamvada and {Pacucci}, Fabio and {Rees}, Martin J. and {Regan}, John A. and {Sakurai}, Yuya and {Salvadori}, Stefania and {Schneider}, Raffaella and {Surace}, Marco and {Tanaka}, Takamitsu L. and {Whalen}, Daniel J. and {Yoshida}, Naoki},
	Doi = {10.1017/pasa.2019.14},
	Eid = {e027},
	Eprint = {1810.12310},
	Journal = {\pasa},
	Month = aug,
	Pages = {e027},
	Primaryclass = {astro-ph.GA},
	Title = {{Titans of the early Universe: The Prato statement on the origin of the first supermassive black holes}},
	Volume = {36},
	Year = 2019
}

@ARTICLE{magg22b,
       author = {{Magg}, Mattis and {Hartwig}, Tilman and {Chen}, Li-Hsin and {Tarumi}, Yuta},
        title = "{A-SLOTH: Ancient Stars and Local Observables by Tracing Halos}",
      journal = {The Journal of Open Source Software},
         year = 2022,
        month = jun,
       volume = {7},
       number = {74},
          eid = {4417},
        pages = {4417},
          doi = {10.21105/joss.04417},
archivePrefix = {arXiv},
       eprint = {2209.07339},
 primaryClass = {astro-ph.IM},
       adsurl = {https://ui.adsabs.harvard.edu/abs/2022JOSS....7.4417M}
}

@ARTICLE{chen22a,
       author = {{Chen}, Li-Hsin and {Magg}, Mattis and {Hartwig}, Tilman and {Glover}, Simon C.~O. and {Ji}, Alexander P. and {Klessen}, Ralf S.},
        title = "{Tracing stars in Milky Way satellites with A-SLOTH}",
      journal = {\mnras},
         year = 2022,
        month = jun,
       volume = {513},
       number = {1},
        pages = {934-950},
          doi = {10.1093/mnras/stac933},
archivePrefix = {arXiv},
       eprint = {2202.01220},
 primaryClass = {astro-ph.GA},
       adsurl = {https://ui.adsabs.harvard.edu/abs/2022MNRAS.513..934C}
}

@ARTICLE{riaz22,
       author = {{Riaz}, Shafqat and {Hartwig}, Tilman and {Latif}, Muhammad A.},
        title = "{Unveiling the Contribution of Population III Stars in Primeval Galaxies at Redshift {\ensuremath{\geq}}6}",
      journal = {\apjl},
         year = 2022,
        month = sep,
       volume = {937},
       number = {1},
          eid = {L6},
        pages = {L6},
          doi = {10.3847/2041-8213/ac8ea6},
archivePrefix = {arXiv},
       eprint = {2208.01673},
 primaryClass = {astro-ph.GA},
       adsurl = {https://ui.adsabs.harvard.edu/abs/2022ApJ...937L...6R}
}

@ARTICLE{glover05,
       author = {{Glover}, Simon},
        title = "{The Formation Of The First Stars In The Universe}",
      journal = {\ssr},
         year = 2005,
        month = apr,
       volume = {117},
       number = {3-4},
        pages = {445-508},
          doi = {10.1007/s11214-005-5821-y},
archivePrefix = {arXiv},
       eprint = {astro-ph/0409737},
 primaryClass = {astro-ph},
       adsurl = {https://ui.adsabs.harvard.edu/abs/2005SSRv..117..445G}
}

@ARTICLE{haemmerle20,
       author = {{Haemmerl{\'e}}, L. and {Mayer}, L. and {Klessen}, R.~S. and {Hosokawa}, T. and {Madau}, P. and {Bromm}, V.},
        title = "{Formation of the First Stars and Black Holes}",
      journal = {\ssr},
         year = 2020,
        month = apr,
       volume = {216},
       number = {4},
          eid = {48},
        pages = {48},
          doi = {10.1007/s11214-020-00673-y},
archivePrefix = {arXiv},
       eprint = {2003.10533},
 primaryClass = {astro-ph.GA},
       adsurl = {https://ui.adsabs.harvard.edu/abs/2020SSRv..216...48H}
}

@ARTICLE{schauer21,
	Adsurl = {https://ui.adsabs.harvard.edu/abs/2021MNRAS.507.1775S},
	Archiveprefix = {arXiv},
	Author = {{Schauer}, Anna T.~P. and {Glover}, Simon C.~O. and {Klessen}, Ralf S. and {Clark}, Paul},
	Doi = {10.1093/mnras/stab1953},
	Eprint = {2008.05663},
	Journal = {\mnras},
	Month = oct,
	Number = {2},
	Pages = {1775-1787},
	Primaryclass = {astro-ph.GA},
	Title = {{The influence of streaming velocities and Lyman-Werner radiation on the formation of the first stars}},
	Volume = {507},
	Year = 2021
}

@ARTICLE{kulkarni21,
	Adsurl = {https://ui.adsabs.harvard.edu/abs/2021ApJ...917...40K},
	Archiveprefix = {arXiv},
	Author = {{Kulkarni}, Mihir and {Visbal}, Eli and {Bryan}, Greg L.},
	Doi = {10.3847/1538-4357/ac08a3},
	Eid = {40},
	Eprint = {2010.04169},
	Journal = {\apj},
	Month = aug,
	Number = {1},
	Pages = {40},
	Primaryclass = {astro-ph.GA},
	Title = {{The Critical Dark Matter Halo Mass for Population III Star Formation: Dependence on Lyman-Werner Radiation, Baryon-dark Matter Streaming Velocity, and Redshift}},
	Volume = {917},
	Year = 2021
}

@ARTICLE{xu16a,
	Adsurl = {http://adsabs.harvard.edu/abs/2016ApJ...823..140X},
	Archiveprefix = {arXiv},
	Author = {{Xu}, H. and {Norman}, M.~L. and {O'Shea}, B.~W. and {Wise}, J.~H.},
	Doi = {10.3847/0004-637X/823/2/140},
	Eid = {140},
	Eprint = {1604.03586},
	Journal = {\apj},
	Month = jun,
	Pages = {140},
	Title = {{Late Pop III Star Formation During the Epoch of Reionization: Results from the Renaissance Simulations}},
	Volume = {823},
	Year = {2016}
}

@book{osterbrock2006astrophysics,
  title={Astrophysics Of Gas Nebulae and Active Galactic Nuclei},
  author={Osterbrock, Donald E and Ferland, Gary J},
  year={2006},
  publisher={University science books}
}

@article{storey1995recombination,
  title={Recombination line intensities for hydrogenic ions-IV. Total recombination coefficients and machine-readable tables for Z= 1 to 8},
  author={Storey, PJ and Hummer, DG},
  journal={\mnras},
  volume={272},
  number={1},
  pages={41--48},
  year={1995},
  publisher={Oxford University Press Oxford, UK}
}

@article{jakobsen2022nirspec,
  author       = {Jakobsen, Peter and Ferruit, Pierre and Alves de Oliveira, Catarina and others},
  title        = {The Near Infrared Spectrograph (NIRSpec) on the James Webb Space Telescope: instrument design and capabilities},
  journal      = {\aap},
  year         = {2022},
  volume       = {661},
  pages        = {A80},
  doi          = {10.1051/0004-6361/202142784}
}

@article{rieke2005nircam,
  author       = {Rieke, Marcia J. and Kelly, Douglas M. and Horner, Scott and others},
  title        = {The Near-Infrared Camera (NIRCam) on the James Webb Space Telescope},
  journal      = {Proceedings of the SPIE},
  volume       = {5904},
  pages        = {59040J},
  year         = {2005},
  doi          = {10.1117/12.617540}
}

@article{rieke2015miri,
  author       = {Rieke, George H. and Wright, G. S. and B{\"o}ker, T. and others},
  title        = {The Mid-Infrared Instrument for the James Webb Space Telescope, I: Introduction},
  journal      = {\pasp},
  year         = {2015},
  volume       = {127},
  number       = {953},
  pages        = {584--594},
  doi          = {10.1086/682252}
}

@ARTICLE{hirano14,
       author = {{Hirano}, Shingo and {Hosokawa}, Takashi and {Yoshida}, Naoki and {Umeda}, Hideyuki and {Omukai}, Kazuyuki and {Chiaki}, Gen and {Yorke}, Harold W.},
        title = "{One Hundred First Stars: Protostellar Evolution and the Final Masses}",
      journal = {\href{https://iopscience.iop.org/article/10.1088/0004-637X/781/2/60}{\apj}},
         year = 2014,
        month = feb,
       volume = {781},
       number = {2},
          eid = {60},
        pages = {60},
          doi = {10.1088/0004-637X/781/2/60},
archivePrefix = {arXiv},
       eprint = {1308.4456},
 primaryClass = {astro-ph.CO},
       adsurl = {https://ui.adsabs.harvard.edu/abs/2014ApJ...781...60H}
}

@ARTICLE{hartwig16,
       author = {{Hartwig}, Tilman and {Volonteri}, Marta and {Bromm}, Volker and {Klessen}, Ralf S. and {Barausse}, Enrico and {Magg}, Mattis and {Stacy}, Athena},
        title = "{Gravitational waves from the remnants of the first stars}",
      journal = {\href{https://academic.oup.com/mnrasl/article/460/1/L74/2589670}{\mnras}},
         year = 2016,
        month = jul,
       volume = {460},
       number = {1},
        pages = {L74-L78},
          doi = {10.1093/mnrasl/slw074},
archivePrefix = {arXiv},
       eprint = {1603.05655},
 primaryClass = {astro-ph.GA},
       adsurl = {https://ui.adsabs.harvard.edu/abs/2016MNRAS.460L..74H}
}

@ARTICLE{kinugawa14,
       author = {{Kinugawa}, Tomoya and {Inayoshi}, Kohei and {Hotokezaka}, Kenta and {Nakauchi}, Daisuke and {Nakamura}, Takashi},
        title = "{Possible indirect confirmation of the existence of Pop III massive stars by gravitational wave}",
      journal = {\href{https://academic.oup.com/mnras/article/442/4/2963/1338478}{\mnras}},
         year = 2014,
        month = aug,
       volume = {442},
       number = {4},
        pages = {2963-2992},
          doi = {10.1093/mnras/stu1022},
archivePrefix = {arXiv},
       eprint = {1402.6672},
 primaryClass = {astro-ph.HE},
       adsurl = {https://ui.adsabs.harvard.edu/abs/2014MNRAS.442.2963K}
}

@article{chen22,
    author = {{Chen}, Li-Hsin and Hartwig, Tilman and Klessen, Ralf S and Glover, Simon C O},
    title = {Comparing simulated Milky Way satellite galaxies with observations using unsupervised clustering},
    journal = {\mnras},
    volume = {517},
    number = {4},
    pages = {6140-6149},
    year = {2022},
    month = {11},
    issn = {0035-8711},
    doi = {10.1093/mnras/stac2897},
    url = {https://doi.org/10.1093/mnras/stac2897},
    eprint = {https://academic.oup.com/mnras/article-pdf/517/4/6140/47014792/stac2897.pdf},
}

@article{spera22,
  title={SEVN: Stellar EVolution for N-body},
  author={Spera, Mario and Mapelli, Michela and Bressan, Alessandro},
  journal={Astrophysics Source Code Library},
  pages={ascl:2206.019},
  year={2022}
}

@article{spera15,
    author = {Spera, Mario and Mapelli, Michela and Bressan, Alessandro},
    title = {The mass spectrum of compact remnants from the parsec stellar evolution tracks},
    journal = {\mnras},
    volume = {451},
    number = {4},
    pages = {4086-4103},
    year = {2015},
    month = {06},
    issn = {0035-8711},
    doi = {10.1093/mnras/stv1161},
    url = {https://doi.org/10.1093/mnras/stv1161},
    eprint = {https://academic.oup.com/mnras/article-pdf/451/4/4086/3864018/stv1161.pdf}}

@ARTICLE{aghanim20,
       author = {{Planck Collaboration} and {Aghanim}, N. and {Akrami}, Y. and {Ashdown}, M. and {Aumont}, J. and {Baccigalupi}, C. and {Ballardini}, M. and {Banday}, A.~J. and {Barreiro}, R.~B. and {Bartolo}, N. and {Basak}, S. and {Battye}, R. and {Benabed}, K. and {Bernard}, J. -P. and {Bersanelli}, M. and {Bielewicz}, P. and {Bock}, J.~J. and {Bond}, J.~R. and {Borrill}, J. and {Bouchet}, F.~R. and {Boulanger}, F. and {Bucher}, M. and {Burigana}, C. and {Butler}, R.~C. and {Calabrese}, E. and {Cardoso}, J. -F. and {Carron}, J. and {Challinor}, A. and {Chiang}, H.~C. and {Chluba}, J. and {Colombo}, L.~P.~L. and {Combet}, C. and {Contreras}, D. and {Crill}, B.~P. and {Cuttaia}, F. and {de Bernardis}, P. and {de Zotti}, G. and {Delabrouille}, J. and {Delouis}, J. -M. and {Di Valentino}, E. and {Diego}, J.~M. and {Dor{\'e}}, O. and {Douspis}, M. and {Ducout}, A. and {Dupac}, X. and {Dusini}, S. and {Efstathiou}, G. and {Elsner}, F. and {En{\ss}lin}, T.~A. and {Eriksen}, H.~K. and {Fantaye}, Y. and {Farhang}, M. and {Fergusson}, J. and {Fernandez-Cobos}, R. and {Finelli}, F. and {Forastieri}, F. and {Frailis}, M. and {Fraisse}, A.~A. and {Franceschi}, E. and {Frolov}, A. and {Galeotta}, S. and {Galli}, S. and {Ganga}, K. and {G{\'e}nova-Santos}, R.~T. and {Gerbino}, M. and {Ghosh}, T. and {Gonz{\'a}lez-Nuevo}, J. and {G{\'o}rski}, K.~M. and {Gratton}, S. and {Gruppuso}, A. and {Gudmundsson}, J.~E. and {Hamann}, J. and {Handley}, W. and {Hansen}, F.~K. and {Herranz}, D. and {Hildebrandt}, S.~R. and {Hivon}, E. and {Huang}, Z. and {Jaffe}, A.~H. and {Jones}, W.~C. and {Karakci}, A. and {Keih{\"a}nen}, E. and {Keskitalo}, R. and {Kiiveri}, K. and {Kim}, J. and {Kisner}, T.~S. and {Knox}, L. and {Krachmalnicoff}, N. and {Kunz}, M. and {Kurki-Suonio}, H. and {Lagache}, G. and {Lamarre}, J. -M. and {Lasenby}, A. and {Lattanzi}, M. and {Lawrence}, C.~R. and {Le Jeune}, M. and {Lemos}, P. and {Lesgourgues}, J. and {Levrier}, F. and {Lewis}, A. and {Liguori}, M. and {Lilje}, P.~B. and {Lilley}, M. and {Lindholm}, V. and {L{\'o}pez-Caniego}, M. and {Lubin}, P.~M. and {Ma}, Y. -Z. and {Mac{\'\i}as-P{\'e}rez}, J.~F. and {Maggio}, G. and {Maino}, D. and {Mandolesi}, N. and {Mangilli}, A. and {Marcos-Caballero}, A. and {Maris}, M. and {Martin}, P.~G. and {Martinelli}, M. and {Mart{\'\i}nez-Gonz{\'a}lez}, E. and {Matarrese}, S. and {Mauri}, N. and {McEwen}, J.~D. and {Meinhold}, P.~R. and {Melchiorri}, A. and {Mennella}, A. and {Migliaccio}, M. and {Millea}, M. and {Mitra}, S. and {Miville-Desch{\^e}nes}, M. -A. and {Molinari}, D. and {Montier}, L. and {Morgante}, G. and {Moss}, A. and {Natoli}, P. and {N{\o}rgaard-Nielsen}, H.~U. and {Pagano}, L. and {Paoletti}, D. and {Partridge}, B. and {Patanchon}, G. and {Peiris}, H.~V. and {Perrotta}, F. and {Pettorino}, V. and {Piacentini}, F. and {Polastri}, L. and {Polenta}, G. and {Puget}, J. -L. and {Rachen}, J.~P. and {Reinecke}, M. and {Remazeilles}, M. and {Renzi}, A. and {Rocha}, G. and {Rosset}, C. and {Roudier}, G. and {Rubi{\~n}o-Mart{\'\i}n}, J.~A. and {Ruiz-Granados}, B. and {Salvati}, L. and {Sandri}, M. and {Savelainen}, M. and {Scott}, D. and {Shellard}, E.~P.~S. and {Sirignano}, C. and {Sirri}, G. and {Spencer}, L.~D. and {Sunyaev}, R. and {Suur-Uski}, A. -S. and {Tauber}, J.~A. and {Tavagnacco}, D. and {Tenti}, M. and {Toffolatti}, L. and {Tomasi}, M. and {Trombetti}, T. and {Valenziano}, L. and {Valiviita}, J. and {Van Tent}, B. and {Vibert}, L. and {Vielva}, P. and {Villa}, F. and {Vittorio}, N. and {Wandelt}, B.~D. and {Wehus}, I.~K. and {White}, M. and {White}, S.~D.~M. and {Zacchei}, A. and {Zonca}, A.},
        title = "{Planck 2018 results. VI. Cosmological parameters}",
      journal = {\href{https://www.aanda.org/articles/aa/full_html/2020/09/aa33910-18/aa33910-18.html}{\aap}},
         year = 2020,
        month = sep,
       volume = {641},
          eid = {A6},
        pages = {A6},
          doi = {10.1051/0004-6361/201833910},
archivePrefix = {arXiv},
       eprint = {1807.06209},
 primaryClass = {astro-ph.CO},
       adsurl = {https://ui.adsabs.harvard.edu/abs/2020A&A...641A...6P}
}

@article{Caterpillar,
	Adsurl = {http://adsabs.harvard.edu/abs/2016ApJ...818...10G},
	Archiveprefix = {arXiv},
	Author = {{Griffen}, B.~F. and {Ji}, A.~P. and {Dooley}, G.~A. and {G{\'o}mez}, F.~A. and {Vogelsberger}, M. and {O'Shea}, B.~W. and {Frebel}, A.},
	Doi = {10.3847/0004-637X/818/1/10},
	Eid = {10},
	Eprint = {1509.01255},
	Journal = {\apj},
	Month = feb,
	Pages = {10},
	Title = {{The Caterpillar Project: A Large Suite of Milky Way Sized Halos}},
	Volume = {818},
	Year = {2016}}

@article{ishiyama16,
	Adsurl = {http://adsabs.harvard.edu/abs/2016ApJ...826....9I},
	Archiveprefix = {arXiv},
	Author = {{Ishiyama}, T. and {Sudo}, K. and {Yokoi}, S. and {Hasegawa}, K. and {Tominaga}, N. and {Susa}, H.},
	Doi = {10.3847/0004-637X/826/1/9},
	Eid = {9},
	Eprint = {1602.00465},
	Journal = {\apj},
	Month = jul,
	Pages = {9},
	Title = {{Where are the Low-mass Population III Stars?}},
	Volume = {826},
	Year = {2016}}

@ARTICLE{yoshida03,
       author = {{Yoshida}, Naoki and {Abel}, Tom and {Hernquist}, Lars and {Sugiyama}, Naoshi},
        title = "{Simulations of Early Structure Formation: Primordial Gas Clouds}",
      journal = {\href{https://iopscience.iop.org/article/10.1086/375810}{\apj}},
         year = 2003,
        month = aug,
       volume = {592},
       number = {2},
        pages = {645-663},
          doi = {10.1086/375810},
archivePrefix = {arXiv},
       eprint = {astro-ph/0301645},
 primaryClass = {astro-ph},
       adsurl = {https://ui.adsabs.harvard.edu/abs/2003ApJ...592..645Y}
}

@article{hummer87,
    author = {Hummer, D. G. and Storey, P. J.},
    title = {Recombination-line intensities for hydrogenic ions – I. Case B calculations for H I and He II},
    journal = {\mnras},
    volume = {224},
    number = {3},
    pages = {801-820},
    year = {1987},
    month = {02},
    issn = {0035-8711},
    doi = {10.1093/mnras/224.3.801},
    url = {https://doi.org/10.1093/mnras/224.3.801},
    eprint = {https://academic.oup.com/mnras/article-pdf/224/3/801/3034101/mnras224-0801.pdf},
}

@ARTICLE{bromm02,
       author = {{Bromm}, Volker and {Coppi}, Paolo S. and {Larson}, Richard B.},
        title = "{The Formation of the First Stars. I. The Primordial Star-forming Cloud}",
      journal = {\apj},
         year = "2002",
        month = "Jan",
       volume = {564},
       number = {1},
        pages = {23-51},
          doi = {10.1086/323947},
archivePrefix = {arXiv},
       eprint = {astro-ph/0102503},
 primaryClass = {astro-ph},
       adsurl = {https://ui.adsabs.harvard.edu/abs/2002ApJ...564...23B}
}

@ARTICLE{liu20a,
       author = {{Liu}, Boyuan and {Bromm}, Volker},
        title = "{Gravitational waves from Population III binary black holes formed by dynamical capture}",
      journal = {\mnras},
         year = 2020,
        month = may,
       volume = {495},
       number = {2},
        pages = {2475-2495},
          doi = {10.1093/mnras/staa1362},
archivePrefix = {arXiv},
       eprint = {2003.00065},
 primaryClass = {astro-ph.CO},
       adsurl = {https://ui.adsabs.harvard.edu/abs/2020MNRAS.495.2475L}
}

@ARTICLE{trussler23,
       author = {{Trussler}, James A.~A. and {Conselice}, Christopher J. and {Adams}, Nathan J. and {Maiolino}, Roberto and {Nakajima}, Kimihiko and {Zackrisson}, Erik and {Austin}, Duncan and {Ferreira}, Leonardo and {Harvey}, Tom},
        title = "{On the observability and identification of Population III galaxies with JWST}",
      journal = {\href{https://academic.oup.com/mnras/article/525/4/5328/7251489}{\mnras}},
         year = 2023,
        month = nov,
       volume = {525},
       number = {4},
        pages = {5328-5352},
          doi = {10.1093/mnras/stad2553},
archivePrefix = {arXiv},
       eprint = {2211.02038},
 primaryClass = {astro-ph.GA},
       adsurl = {https://ui.adsabs.harvard.edu/abs/2023MNRAS.525.5328T}
}

@article{kennicutt98,
  author = {Kennicutt, Robert C.},
  title = {Star Formation in Galaxies Along the Hubble Sequence},
  journal = {\araa},
  volume = {36},
  year = {1998},
  pages = {189--232},
  doi = {10.1146/annurev.astro.36.1.189}
}

@article{Ishiyama15,
    author = {Ishiyama, Tomoaki and Enoki, Motohiro and Kobayashi, Masakazu A. R. and Makiya, Ryu and Nagashima, Masahiro and Oogi, Taira},
    title = {The ν2GC simulations: Quantifying the dark side of the universe in the Planck cosmology},
    journal = {\pasj},
    volume = {67},
    number = {4},
    pages = {61},
    year = {2015},
    month = {05},
    issn = {0004-6264},
    doi = {10.1093/pasj/psv021},
    url = {https://doi.org/10.1093/pasj/psv021},
    eprint = {https://academic.oup.com/pasj/article-pdf/67/4/61/54683095/pasj\_67\_4\_61.pdf},
}

@article{lacey93,
  author = {Lacey, C. and Cole, S.},
  title = {Merger rates in hierarchical models of galaxy formation},
  journal = {MNRAS},
  volume = {262},
  pages = {627--649},
  year = {1993},
  doi = {10.1093/mnras/262.3.627}
}

@article{uysal23,
    author = {Uysal, Betül and Hartwig, Tilman},
    title = {First estimate of the local value of the baryonic streaming velocity},
    journal = {\mnras},
    volume = {520},
    number = {3},
    pages = {3229-3237},
    year = {2023},
    month = {02},
    issn = {0035-8711},
    doi = {10.1093/mnras/stad350},
    url = {https://doi.org/10.1093/mnras/stad350},
    eprint = {https://academic.oup.com/mnras/article-pdf/520/3/3229/49199633/stad350.pdf},
}

@ARTICLE{astropy:2022,
       author = {{Astropy Collaboration} and {Price-Whelan}, Adrian M. and {Lim}, Pey Lian and {Earl}, Nicholas and {Starkman}, Nathaniel and {Bradley}, Larry and {Shupe}, David L. and {Patil}, Aarya A. and {Corrales}, Lia and {Brasseur}, C.~E. and {N{"o}the}, Maximilian and {Donath}, Axel and {Tollerud}, Erik and {Morris}, Brett M. and {Ginsburg}, Adam and {Vaher}, Eero and {Weaver}, Benjamin A. and {Tocknell}, James and {Jamieson}, William and {van Kerkwijk}, Marten H. and {Robitaille}, Thomas P. and {Merry}, Bruce and {Bachetti}, Matteo and {G{"u}nther}, H. Moritz and {Aldcroft}, Thomas L. and {Alvarado-Montes}, Jaime A. and {Archibald}, Anne M. and {B{'o}di}, Attila and {Bapat}, Shreyas and {Barentsen}, Geert and {Baz{'a}n}, Juanjo and {Biswas}, Manish and {Boquien}, M{'e}d{'e}ric and {Burke}, D.~J. and {Cara}, Daria and {Cara}, Mihai and {Conroy}, Kyle E. and {Conseil}, Simon and {Craig}, Matthew W. and {Cross}, Robert M. and {Cruz}, Kelle L. and {D'Eugenio}, Francesco and {Dencheva}, Nadia and {Devillepoix}, Hadrien A.~R. and {Dietrich}, J{"o}rg P. and {Eigenbrot}, Arthur Davis and {Erben}, Thomas and {Ferreira}, Leonardo and {Foreman-Mackey}, Daniel and {Fox}, Ryan and {Freij}, Nabil and {Garg}, Suyog and {Geda}, Robel and {Glattly}, Lauren and {Gondhalekar}, Yash and {Gordon}, Karl D. and {Grant}, David and {Greenfield}, Perry and {Groener}, Austen M. and {Guest}, Steve and {Gurovich}, Sebastian and {Handberg}, Rasmus and {Hart}, Akeem and {Hatfield-Dodds}, Zac and {Homeier}, Derek and {Hosseinzadeh}, Griffin and {Jenness}, Tim and {Jones}, Craig K. and {Joseph}, Prajwel and {Kalmbach}, J. Bryce and {Karamehmetoglu}, Emir and {Ka{l}uszy{'n}ski}, Miko{l}aj and {Kelley}, Michael S.~P. and {Kern}, Nicholas and {Kerzendorf}, Wolfgang E. and {Koch}, Eric W. and {Kulumani}, Shankar and {Lee}, Antony and {Ly}, Chun and {Ma}, Zhiyuan and {MacBride}, Conor and {Maljaars}, Jakob M. and {Muna}, Demitri and {Murphy}, N.~A. and {Norman}, Henrik and {O'Steen}, Richard and {Oman}, Kyle A. and {Pacifici}, Camilla and {Pascual}, Sergio and {Pascual-Granado}, J. and {Patil}, Rohit R. and {Perren}, Gabriel I. and {Pickering}, Timothy E. and {Rastogi}, Tanuj and {Roulston}, Benjamin R. and {Ryan}, Daniel F. and {Rykoff}, Eli S. and {Sabater}, Jose and {Sakurikar}, Parikshit and {Salgado}, Jes{'u}s and {Sanghi}, Aniket and {Saunders}, Nicholas and {Savchenko}, Volodymyr and {Schwardt}, Ludwig and {Seifert-Eckert}, Michael and {Shih}, Albert Y. and {Jain}, Anany Shrey and {Shukla}, Gyanendra and {Sick}, Jonathan and {Simpson}, Chris and {Singanamalla}, Sudheesh and {Singer}, Leo P. and {Singhal}, Jaladh and {Sinha}, Manodeep and {Sip{H{o}}cz}, Brigitta M. and {Spitler}, Lee R. and {Stansby}, David and {Streicher}, Ole and {{{S}}umak}, Jani and {Swinbank}, John D. and {Taranu}, Dan S. and {Tewary}, Nikita and {Tremblay}, Grant R. and {Val-Borro}, Miguel de and {Van Kooten}, Samuel J. and {Vasovi{'c}}, Zlatan and {Verma}, Shresth and {de Miranda Cardoso}, Jos{'e} Vin{'i}cius and {Williams}, Peter K.~G. and {Wilson}, Tom J. and {Winkel}, Benjamin and {Wood-Vasey}, W.~M. and {Xue}, Rui and {Yoachim}, Peter and {Zhang}, Chen and {Zonca}, Andrea and {Astropy Project Contributors}},
        title = "{The Astropy Project: Sustaining and Growing a Community-oriented Open-source Project and the Latest Major Release (v5.0) of the Core Package}",
      journal = {\apj},
         year = 2022,
        month = aug,
       volume = {935},
       number = {2},
          eid = {167},
        pages = {167},
          doi = {10.3847/1538-4357/ac7c74},
archivePrefix = {arXiv},
       eprint = {2206.14220},
 primaryClass = {astro-ph.IM},
       adsurl = {https://ui.adsabs.harvard.edu/abs/2022ApJ...935..167A}
}

@article{cen2003galaxies,
  title={Galaxies inside Str{\"o}mgren Spheres of Luminous Quasars at z> 6: Detection of the First Galaxies},
  author={Cen, Renyue},
  journal={ApJ},
  volume={597},
  number={1},
  pages={L13},
  year={2003},
  publisher={IOP Publishing}
}

@article{curti23,
  title={The chemical enrichment in the early Universe as probed by JWST via direct metallicity measurements at z~ 8},
  author={Curti, Mirko and d’Eugenio, Francesco and Carniani, Stefano and Maiolino, Roberto and Sandles, Lester and Witstok, Joris and Baker, William M and Bennett, Jake S and Piotrowska, Joanna M and Tacchella, Sandro and others},
  journal={\mnras},
  volume={518},
  number={1},
  pages={425--438},
  year={2023},
  publisher={Oxford University Press}
}

@article{nakajima23,
  title={JWST census for the mass--metallicity star formation relations at z= 4--10 with self-consistent flux calibration and proper metallicity calibrators},
  author={Nakajima, Kimihiko and Ouchi, Masami and Isobe, Yuki and Harikane, Yuichi and Zhang, Yechi and Ono, Yoshiaki and Umeda, Hiroya and Oguri, Masamune},
  journal={\apjs},
  volume={269},
  number={2},
  pages={33},
  year={2023},
  publisher={IOP Publishing}
}

@article{oh1999observational,
  title={Observational signatures of the first luminous objects},
  author={Oh, Siang Peng},
  journal={\apj},
  volume={527},
  number={1},
  pages={16},
  year={1999},
  publisher={IOP Publishing}
}

@ARTICLE{zackrisson2011hst,
       author = {{Zackrisson}, Erik and {Inoue}, Akio K. and {Rydberg}, Claes-Erik and {Duval}, Florent},
        title = "{The Hubble Space Telescope colours of high-redshift Population III galaxies with strong Ly{\ensuremath{\alpha}} emission}",
      journal = {\mnras},
         year = 2011,
        month = nov,
       volume = {418},
       number = {1},
        pages = {L104-L108},
          doi = {10.1111/j.1745-3933.2011.01153.x},
archivePrefix = {arXiv},
       eprint = {1109.1556},
 primaryClass = {astro-ph.CO},
       adsurl = {https://ui.adsabs.harvard.edu/abs/2011MNRAS.418L.104Z}
}

@article{oh2001he,
  title={He II Recombination Lines from the FirstLuminous Objects},
  author={Oh, S Peng and Haiman, Zolt{\'a}n and Rees, Martin J},
  journal={\apj},
  volume={553},
  number={1},
  pages={73},
  year={2001},
  publisher={IOP Publishing}
}

@article{johnson2009first,
  title={The first galaxies: signatures of the initial starburst},
  author={Johnson, Jarrett L and Greif, Thomas H and Bromm, Volker and Klessen, Ralf S and Ippolito, Joseph},
  journal={\mnras},
  volume={399},
  number={1},
  pages={37--47},
  year={2009},
  publisher={Blackwell Publishing Ltd Oxford, UK}
}

@ARTICLE{Schaerer2002,
       author = {{Schaerer}, D.},
        title = "{On the properties of massive Population III stars and metal-free stellar populations}",
      journal = {\aap},
         year = 2002,
        month = jan,
       volume = {382},
        pages = {28-42},
          doi = {10.1051/0004-6361:20011619},
archivePrefix = {arXiv},
       eprint = {astro-ph/0110697},
 primaryClass = {astro-ph},
       adsurl = {https://ui.adsabs.harvard.edu/abs/2002A&A...382...28S}
}

@article{Behroozi13b,
    author = {Behroozi, Peter and Knebe, Alexander and Pearce, Frazer R. and Elahi, Pascal and Han, Jiaxin and Lux, Hanni and Mao, Yao-Yuan and Muldrew, Stuart I. and Potter, Doug and Srisawat, Chaichalit},
    title = {Major mergers going Notts: challenges for modern halo finders},
    journal = {MNRAS},
    volume = {454},
    number = {3},
    pages = {3020-3029},
    year = {2015},
    month = {10},
    issn = {0035-8711},
    doi = {10.1093/mnras/stv2046},
    url = {https://doi.org/10.1093/mnras/stv2046},
    eprint = {https://academic.oup.com/mnras/article-pdf/454/3/3020/4027299/stv2046.pdf},
}

@article{Behroozi13,
doi = {10.1088/0004-637X/762/2/109},
url = {https://dx.doi.org/10.1088/0004-637X/762/2/109},
year = {2012},
month = {dec},
publisher = {The American Astronomical Society},
volume = {762},
number = {2},
pages = {109},
author = {Behroozi, Peter S. and Wechsler, Risa H. and Wu, Hao-Yi},
title = {THE ROCKSTAR PHASE-SPACE TEMPORAL HALO FINDER AND THE VELOCITY OFFSETS OF CLUSTER CORES},
journal = {ApJ}
}

@article{davis1985,
  title={The evolution of large-scale structure in a universe dominated by cold dark matter},
  author={Davis, Marc and Efstathiou, George and Frenk, Carlos S and White, Simon DM},
  journal={\apj},
  volume={292},
  pages={371--394},
  year={1985}
}

@ARTICLE{greif11_stream,
       author = {{Greif}, Thomas H. and {White}, Simon D.~M. and {Klessen}, Ralf S. and {Springel}, Volker},
        title = "{The Delay of Population III Star Formation by Supersonic Streaming Velocities}",
      journal = {\apj},
         year = 2011,
        month = aug,
       volume = {736},
       number = {2},
          eid = {147},
        pages = {147},
          doi = {10.1088/0004-637X/736/2/147},
archivePrefix = {arXiv},
       eprint = {1101.5493},
 primaryClass = {astro-ph.CO},
       adsurl = {https://ui.adsabs.harvard.edu/abs/2011ApJ...736..147G}
}

@ARTICLE{stacy11_stream,
       author = {{Stacy}, Athena and {Bromm}, Volker and {Loeb}, Abraham},
        title = "{Effect of Streaming Motion of Baryons Relative to Dark Matter on the Formation of the First Stars}",
      journal = {\apjl},
         year = 2011,
        month = mar,
       volume = {730},
       number = {1},
          eid = {L1},
        pages = {L1},
          doi = {10.1088/2041-8205/730/1/L1},
archivePrefix = {arXiv},
       eprint = {1011.4512},
 primaryClass = {astro-ph.CO},
       adsurl = {https://ui.adsabs.harvard.edu/abs/2011ApJ...730L...1S}
}

@ARTICLE{Liu2024,
       author = {{Liu}, Boyuan and {Gurian}, James and {Inayoshi}, Kohei and {Hirano}, Shingo and {Hosokawa}, Takashi and {Bromm}, Volker and {Yoshida}, Naoki},
        title = "{Towards a universal analytical model for Population III star formation: interplay between feedback and fragmentation}",
      journal = {\mnras},
         year = 2024,
        month = oct,
       volume = {534},
       number = {1},
        pages = {290-312},
          doi = {10.1093/mnras/stae2066},
archivePrefix = {arXiv},
       eprint = {2407.14294},
 primaryClass = {astro-ph.GA},
       adsurl = {https://ui.adsabs.harvard.edu/abs/2024MNRAS.534..290L}
}

@ARTICLE{Liu2024gw,
       author = {{Liu}, Boyuan and {Hartwig}, Tilman and {Sartorio}, Nina S. and {Dvorkin}, Irina and {Costa}, Guglielmo and {Santoliquido}, Filippo and {Fialkov}, Anastasia and {Klessen}, Ralf S. and {Bromm}, Volker},
        title = "{Gravitational waves from mergers of Population III binary black holes: roles played by two evolution channels}",
      journal = {\mnras},
         year = 2024,
        month = nov,
       volume = {534},
       number = {3},
        pages = {1634-1667},
          doi = {10.1093/mnras/stae2120},
archivePrefix = {arXiv},
       eprint = {2406.17397},
 primaryClass = {astro-ph.GA},
       adsurl = {https://ui.adsabs.harvard.edu/abs/2024MNRAS.534.1634L}
}

@ARTICLE{Santoliquido2023,
       author = {{Santoliquido}, Filippo and {Mapelli}, Michela and {Iorio}, Giuliano and {Costa}, Guglielmo and {Glover}, Simon C.~O. and {Hartwig}, Tilman and {Klessen}, Ralf S. and {Merli}, Lorenzo},
        title = "{Binary black hole mergers from population III stars: uncertainties from star formation and binary star properties}",
      journal = {\mnras},
         year = 2023,
        month = sep,
       volume = {524},
       number = {1},
        pages = {307-324},
          doi = {10.1093/mnras/stad1860},
archivePrefix = {arXiv},
       eprint = {2303.15515},
 primaryClass = {astro-ph.GA},
       adsurl = {https://ui.adsabs.harvard.edu/abs/2023MNRAS.524..307S}
}

@ARTICLE{Tanikawa2021,
       author = {{Tanikawa}, Ataru and {Susa}, Hajime and {Yoshida}, Takashi and {Trani}, Alessandro A. and {Kinugawa}, Tomoya},
        title = "{Merger Rate Density of Population III Binary Black Holes Below, Above, and in the Pair-instability Mass Gap}",
      journal = {\apj},
         year = 2021,
        month = mar,
       volume = {910},
       number = {1},
          eid = {30},
        pages = {30},
          doi = {10.3847/1538-4357/abe40d},
archivePrefix = {arXiv},
       eprint = {2008.01890},
 primaryClass = {astro-ph.HE},
       adsurl = {https://ui.adsabs.harvard.edu/abs/2021ApJ...910...30T}
}

@ARTICLE{Jeon2025,
       author = {{Jeon}, Junehyoung and {Liu}, Boyuan and {Taylor}, Anthony J. and {Kokorev}, Vasily and {Chisholm}, John and {Kocevski}, Dale D. and {Finkelstein}, Steven L. and {Bromm}, Volker},
        title = "{The Emerging Black Hole Mass Function in the High-redshift Universe}",
      journal = {\apj},
         year = 2025,
        month = jul,
       volume = {988},
       number = {1},
          eid = {110},
        pages = {110},
          doi = {10.3847/1538-4357/ade2e1},
archivePrefix = {arXiv},
       eprint = {2503.14703},
 primaryClass = {astro-ph.GA},
       adsurl = {https://ui.adsabs.harvard.edu/abs/2025ApJ...988..110J}
}

@ARTICLE{Greif2009,
       author = {{Greif}, Thomas H. and {Johnson}, Jarrett L. and {Klessen}, Ralf S. and {Bromm}, Volker},
        title = "{The observational signature of the first HII regions}",
      journal = {\mnras},
         year = 2009,
        month = oct,
       volume = {399},
       number = {2},
        pages = {639-649},
          doi = {10.1111/j.1365-2966.2009.15336.x},
archivePrefix = {arXiv},
       eprint = {0905.1717},
 primaryClass = {astro-ph.CO},
       adsurl = {https://ui.adsabs.harvard.edu/abs/2009MNRAS.399..639G}
}

@ARTICLE{Nakajima2025,
author = {{Nakajima}, Kimihiko and {Ouchi}, Masami and {Harikane}, Yuichi and {Vanzella}, Eros and {Ono}, Yoshiaki and {Isobe}, Yuki and {Nishigaki}, Moka and {Tsujimoto}, Takuji and {Nakamura}, Fumitaka and {Xu}, Yi and {Umeda}, Hiroya and {Zhang}, Yechi},
title = "{An ultra-faint, chemically primitive galaxy forming in the reionization era}",
journal = {\nat},
year = 2026,
month = may,
volume = {653},
number = {8114},
pages = {363-367},
doi = {10.1038/s41586-026-10374-1},
archivePrefix = {arXiv},
eprint = {2506.11846},
primaryClass = {astro-ph.GA},
adsurl = {https://ui.adsabs.harvard.edu/abs/2026Natur.653..363N}
}

@ARTICLE{Vanzella2023,
       author = {{Vanzella}, E. and {Loiacono}, F. and {Bergamini}, P. and {Me{\v{s}}tri{\'c}}, U. and {Castellano}, M. and {Rosati}, P. and {Meneghetti}, M. and {Grillo}, C. and {Calura}, F. and {Mignoli}, M. and {Brada{\v{c}}}, M. and {Adamo}, A. and {Rihtar{\v{s}}i{\v{c}}}, G. and {Dickinson}, M. and {Gronke}, M. and {Zanella}, A. and {Annibali}, F. and {Willott}, C. and {Messa}, M. and {Sani}, E. and {Acebron}, A. and {Bolamperti}, A. and {Comastri}, A. and {Gilli}, R. and {Caputi}, K.~I. and {Ricotti}, M. and {Gruppioni}, C. and {Ravindranath}, S. and {Mercurio}, A. and {Strait}, V. and {Martis}, N. and {Pascale}, R. and {Caminha}, G.~B. and {Annunziatella}, M. and {Nonino}, M.},
        title = "{An extremely metal-poor star complex in the reionization era: Approaching Population III stars with JWST}",
      journal = {\aap},
         year = 2023,
        month = oct,
       volume = {678},
          eid = {A173},
        pages = {A173},
          doi = {10.1051/0004-6361/202346981},
archivePrefix = {arXiv},
       eprint = {2305.14413},
 primaryClass = {astro-ph.GA},
       adsurl = {https://ui.adsabs.harvard.edu/abs/2023A&A...678A.173V}
}

@ARTICLE{Maiolino2024,
       author = {{Maiolino}, Roberto and {{\"U}bler}, Hannah and {Perna}, Michele and {Scholtz}, Jan and {D'Eugenio}, Francesco and {Witten}, Callum and {Laporte}, Nicolas and {Witstok}, Joris and {Carniani}, Stefano and {Tacchella}, Sandro and {Baker}, William M. and {Arribas}, Santiago and {Nakajima}, Kimihiko and {Eisenstein}, Daniel J. and {Bunker}, Andrew J. and {Charlot}, St{\'e}phane and {Cresci}, Giovanni and {Curti}, Mirko and {Curtis-Lake}, Emma and {de Graaff}, Anna and {Egami}, Eiichi and {Ji}, Zhiyuan and {Johnson}, Benjamin D. and {Kumari}, Nimisha and {Looser}, Tobias J. and {Maseda}, Michael and {Nelson}, Erica and {Robertson}, Brant and {Rodr{\'\i}guez Del Pino}, Bruno and {Sandles}, Lester and {Simmonds}, Charlotte and {Smit}, Renske and {Sun}, Fengwu and {Venturi}, Giacomo and {Williams}, Christina C. and {Willmer}, Christopher N.~A.},
        title = "{JADES. Possible Population III signatures at z = 10.6 in the halo of GN-z11}",
      journal = {\aap},
         year = 2024,
        month = jul,
       volume = {687},
          eid = {A67},
        pages = {A67},
          doi = {10.1051/0004-6361/202347087},
archivePrefix = {arXiv},
       eprint = {2306.00953},
 primaryClass = {astro-ph.GA},
       adsurl = {https://ui.adsabs.harvard.edu/abs/2024A&A...687A..67M}
}

@ARTICLE{Maiolino2026,
author = {{Maiolino}, Roberto and {{"U}bler}, Hannah and {Perna}, Michele and {Witstok}, Joris and {Jones}, Gareth C. and {Perez-Gonzalez}, Pablo G. and {Nakajima}, Kimihiko and {Rusta}, Elka and {Salvadori}, Stefania and {Tacchella}, Sandro and {Madau}, Piero and {Trussler}, James A. A. and {D'Eugenio}, Francesco and {Ji}, Xihan and {Scholtz}, Jan and {Carniani}, Stefano and {Isobe}, Yuki and {Katz}, Harley and {Arribas}, Santiago and {Baker}, William M. and {B{"o}ker}, Torsten and {Bromm}, Volker and {Bunker}, Andrew J. and others},
title = "{The search for Population III: Confirmation of a HeII emitter with no metal lines at z=10.6}",
journal = {\mnras, submitted},
year = 2026,
month = mar,
eid = {arXiv:2603.20362},
pages = {arXiv:2603.20362},
doi = {10.48550/arXiv.2603.20362},
archivePrefix = {arXiv},
eprint = {2603.20362},
primaryClass = {astro-ph.GA},
adsurl = {https://ui.adsabs.harvard.edu/abs/2026arXiv260320362M}
}

@ARTICLE{Ubler2026,
author = {{{"U}bler}, Hannah and {Maiolino}, Roberto and {P{'e}rez-Gonz{'a}lez}, Pablo G. and {Isobe}, Yuki and {Jones}, Gareth C. and {Kumari}, Nimisha and {Charlot}, St{'e}phane and {Rusta}, Elka and {Salvadori}, Stefania and {Nakajima}, Kimihiko and {Perna}, Michele and {Arribas}, Santiago and {Bunker}, Andrew J. and {Carniani}, Stefano and {D'Eugenio}, Francesco and {Rodr{'i}guez Del Pino}, Bruno and {Bertola}, Elena and {B{"o}ker}, Torsten and {Chevallard}, Jacopo and {Circosta}, Chiara and {Cresci}, Giovanni and {Curti}, Mirko and {Curtis-Lake}, Emma and {Eisenstein}, Daniel J. and {Hainline}, Kevin and {Johnson}, Benjamin D. and {Parlanti}, Eleonora and {Rinaldi}, Pierluigi and {Robertson}, Brant and {Scholtz}, Jan and {Tacchella}, Sandro and {Venturi}, Giacomo and {Witstok}, Joris and {Zamora}, Sandra},
title = "{GA-NIFS & JADES: Confirmation of pristine gas near GN-z11}",
journal = {\aap, submitted},
year = 2026,
month = mar,
eid = {arXiv:2603.20360},
pages = {arXiv:2603.20360},
doi = {10.48550/arXiv.2603.20360},
archivePrefix = {arXiv},
eprint = {2603.20360},
primaryClass = {astro-ph.GA},
adsurl = {https://ui.adsabs.harvard.edu/abs/2026arXiv260320360U}
}

@ARTICLE{Cai2025,
       author = {{Cai}, Sijia and {Li}, Mingyu and {Cai}, Zheng and {Wu}, Yunjing and {Yu}, Fujiang and {Dickinson}, Mark and {Sun}, Fengwu and {Fan}, Xiaohui and {Wang}, Ben and {Cullen}, Fergus and {Bian}, Fuyan and {Lin}, Xiaojing and {Zou}, Jiaqi},
        title = "{A Metal-free Galaxy at z = 3.19? Evidence of Late Population III Star Formation at Cosmic Noon}",
      journal = {\apjl},
         year = 2025,
        month = nov,
       volume = {993},
       number = {2},
          eid = {L52},
        pages = {L52},
          doi = {10.3847/2041-8213/ae1608},
archivePrefix = {arXiv},
       eprint = {2507.17820},
 primaryClass = {astro-ph.GA},
       adsurl = {https://ui.adsabs.harvard.edu/abs/2025ApJ...993L..52C}
}

@ARTICLE{Jeong2026,
author = {{Jeong}, Tae Bong and {Venditti}, Alessandra and {Bromm}, Volker and {Jeon}, Myoungwon and {Hsiao}, Tiger Yu-Yang and {Finkelstein}, Steven L. and {Chisholm}, John},
title = "{How Massive Can a Population III Starburst Be? Simulating the First Galaxies with High Lyman-Werner Background}",
journal = {\apj, submitted},
year = 2026,
month = mar,
eid = {arXiv:2603.23209},
pages = {arXiv:2603.23209},
doi = {10.48550/arXiv.2603.23209},
archivePrefix = {arXiv},
eprint = {2603.23209},
primaryClass = {astro-ph.GA},
adsurl = {https://ui.adsabs.harvard.edu/abs/2026arXiv260323209J}
}

@ARTICLE{Visbal2025,
author = {{Visbal}, Eli and {Hazlett}, Ryan and {Bryan}, Greg L.},
title = "{LAP1-B is the First Observed System Consistent with Theoretical Predictions for Population III Stars}",
journal = {\apjl},
year = 2025,
month = nov,
volume = {993},
number = {1},
eid = {L17},
pages = {L17},
doi = {10.3847/2041-8213/ae122f},
archivePrefix = {arXiv},
eprint = {2508.03842},
primaryClass = {astro-ph.GA},
adsurl = {https://ui.adsabs.harvard.edu/abs/2025ApJ...993L..17V}
}

@ARTICLE{Morishita2025,
       author = {{Morishita}, Takahiro and {Liu}, Zhaoran and {Stiavelli}, Massimo and {Treu}, Tommaso and {Bergamini}, Pietro and {Zhang}, Yechi},
        title = "{Pristine Massive Star Formation Caught at the Break of Cosmic Dawn}",
      journal = {arXiv e-prints},
         year = 2025,
        month = jul,
          eid = {arXiv:2507.10521},
        pages = {arXiv:2507.10521},
          doi = {10.48550/arXiv.2507.10521},
archivePrefix = {arXiv},
       eprint = {2507.10521},
 primaryClass = {astro-ph.CO},
       adsurl = {https://ui.adsabs.harvard.edu/abs/2025arXiv250710521M}
}

@ARTICLE{Liu2025,
       author = {{Liu}, Boyuan and {Mapelli}, Michela and {Bromm}, Volker and {Klessen}, Ralf S. and {Boco}, Lumen and {Hartwig}, Tilman and {Glover}, Simon C.~O. and {Lipatova}, Veronika and {Costa}, Guglielmo and {Dall'Amico}, Marco and {Iorio}, Giuliano and {Shepherd}, Kendall and {Bressan}, Alessandro},
        title = "{Impact of initial mass function on the chemical evolution of high-redshift galaxies}",
      journal = {arXiv e-prints},
         year = 2025,
        month = jun,
          eid = {arXiv:2506.06139},
        pages = {arXiv:2506.06139},
          doi = {10.48550/arXiv.2506.06139},
archivePrefix = {arXiv},
       eprint = {2506.06139},
 primaryClass = {astro-ph.GA},
       adsurl = {https://ui.adsabs.harvard.edu/abs/2025arXiv250606139L}
}

@ARTICLE{Liu2020,
       author = {{Liu}, Boyuan and {Bromm}, Volker},
        title = "{When did Population III star formation end?}",
      journal = {\mnras},
         year = 2020,
        month = sep,
       volume = {497},
       number = {3},
        pages = {2839-2854},
          doi = {10.1093/mnras/staa2143},
archivePrefix = {arXiv},
       eprint = {2006.15260},
 primaryClass = {astro-ph.GA},
       adsurl = {https://ui.adsabs.harvard.edu/abs/2020MNRAS.497.2839L}
}

@ARTICLE{Storck2026,
       author = {{Storck}, Anatole and {Katz}, Harley and {Devriendt}, Julien and {Slyz}, Adrianne and {Cadiou}, Corentin and {Choustikov}, Nicholas and {Rey}, Martin P. and {Saxena}, Aayush and {Agertz}, Oscar and {Kimm}, Taysun},
        title = "{MEGATRON: the environments of Population III stars at Cosmic Dawn and their connection to present-day galaxies}",
      journal = {\mnras},
         year = 2026,
        month = may,
       volume = {548},
       number = {1},
          eid = {stag529},
        pages = {stag529},
          doi = {10.1093/mnras/stag529},
archivePrefix = {arXiv},
       eprint = {2510.06853},
 primaryClass = {astro-ph.GA},
       adsurl = {https://ui.adsabs.harvard.edu/abs/2026MNRAS.548ag529S}
}

@ARTICLE{Jeon2026,
author = {{Jeon}, Junehyoung and {Jeong}, Tae Bong and {Zhang}, Saiyang and {Bromm}, Volker},
title = "{What is Powering the Enigmatic He II Emitter Hebe: The First Stars or Black Holes?}",
journal = {\apj, submitted},
year = 2026,
month = apr,
eid = {arXiv:2604.19075},
pages = {arXiv:2604.19075},
doi = {10.48550/arXiv.2604.19075},
archivePrefix = {arXiv},
eprint = {2604.19075},
primaryClass = {astro-ph.GA},
adsurl = {https://ui.adsabs.harvard.edu/abs/2026arXiv260419075J}
}

\begin{appendix}

\section{Detectability criteria}\label{appendix:obs}

\begin{figure}
    \begin{center}
    \includegraphics[width=.35\textwidth]{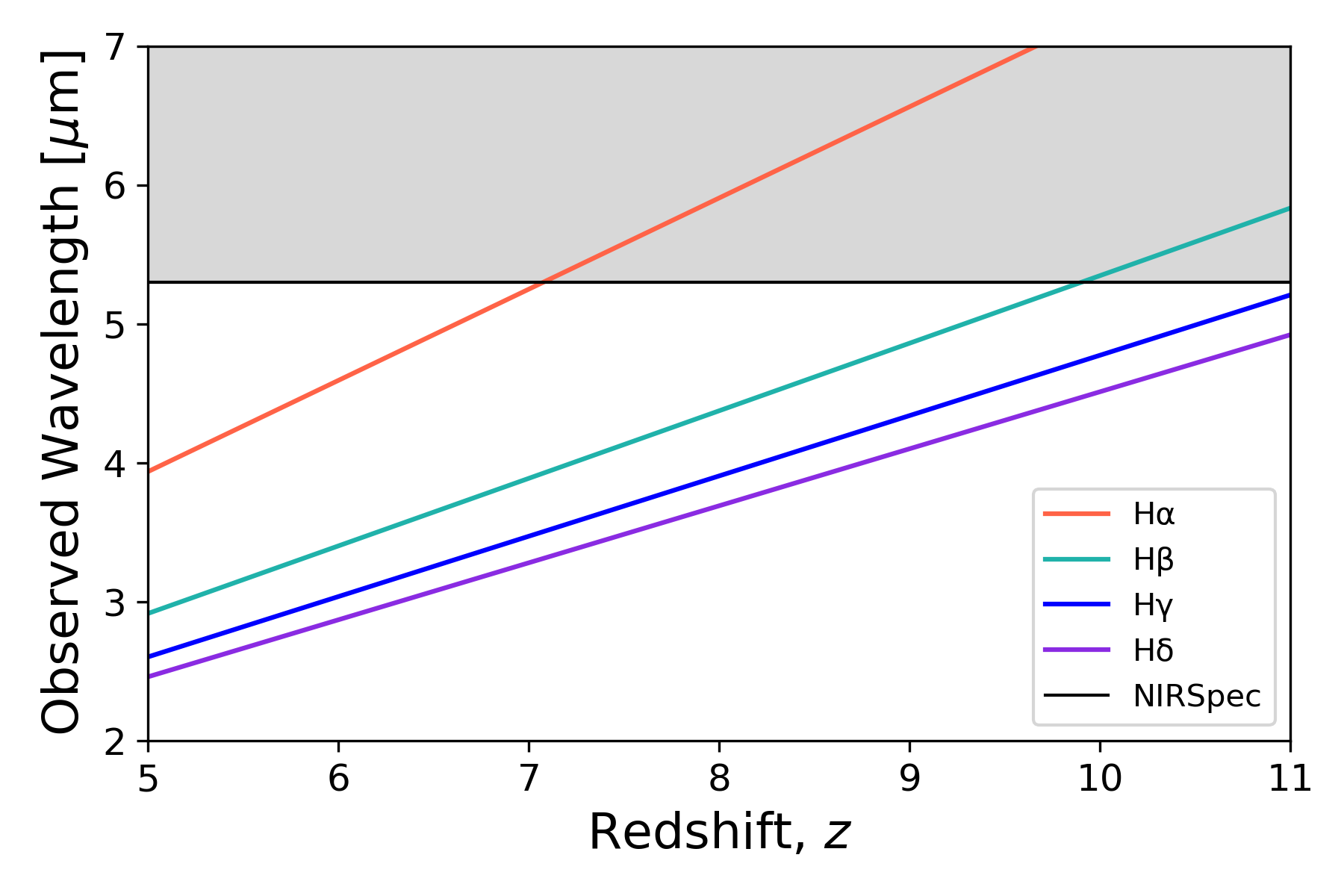}
    \caption{Observed wavelengths of the first few Balmer lines (H$\alpha$,  H$\beta$,  H$\gamma$,  H$\delta$) as a function of redshift, compared with the wavelength range accessible using the PRISM mode of JWST's NIRSpec instrument \citep{jakobsen2022nirspec}. H$\alpha$ is only detectable using NIRSpec out to $z \sim 7$, while the higher Balmer transitions remain potentially detectable out to $z \sim 10-11$.} 
    \label{fig:balmer_vs_JWST}
    \end{center}
\end{figure}
        
In this study, we focus on the use of the \textit{Near-Infrared Spectrograph} (NIRSpec) aboard JWST, which is specifically designed for medium- and high-resolution spectroscopic observations in the near-infrared regime. NIRSpec is particularly well-suited for detecting redshifted Balmer lines from high-$z$ sources, as illustrated in Fig.~\ref{fig:balmer_vs_JWST}, due to its wavelength coverage ($\sim$0.6 -- 5.3\,$\mu$m) and high sensitivity to faint emission features \citep{jakobsen2022nirspec}.

Alternative instruments such as NIRCam and MIRI are less suitable for this analysis \citep{rieke2005nircam, rieke2015miri}. NIRCam, while highly sensitive and capable of low-resolution slitless spectroscopy, is primarily optimised for imaging and broadband photometry rather than detailed line spectroscopy. The MIRI medium resolution spectrometer extends coverage into the mid-infrared (5-28\,$\mu$m), but suffers from significantly reduced sensitivity compared to NIRSpec, especially for weak emission lines from distant and low-mass halos. Thus, NIRSpec offers the optimal balance of spectral resolution, sensitivity, and wavelength coverage for probing redshifted hydrogen recombination lines from Population~III star-forming regions.

For the ETC simulations, we construct a synthetic source representing the nebular emission in the first four Balmer recombination lines (H$\alpha$, H$\beta$, H$\gamma$, H$\delta$) produced by a high-redshift halo containing Pop.\ III and/or Pop.\ II stars. At the redshifts of interest in this paper, the NIRSpec pixel scale of 0.1 arcsecond corresponds to a physical size of 400--600~pc.

We assume that the emission is produced on scales much smaller than this, so that we can treat the emitting sources as point sources. For each line, we input the corresponding redshifted wavelength into the ETC, define appropriate observational apertures, and set a target signal-to-noise ratio (S/N) of 5 to assess detectability.

\section{Parameter studies} \label{appendix:param}

   \begin{figure}[h]
        \begin{center}
        \includegraphics[width=.23\textwidth]{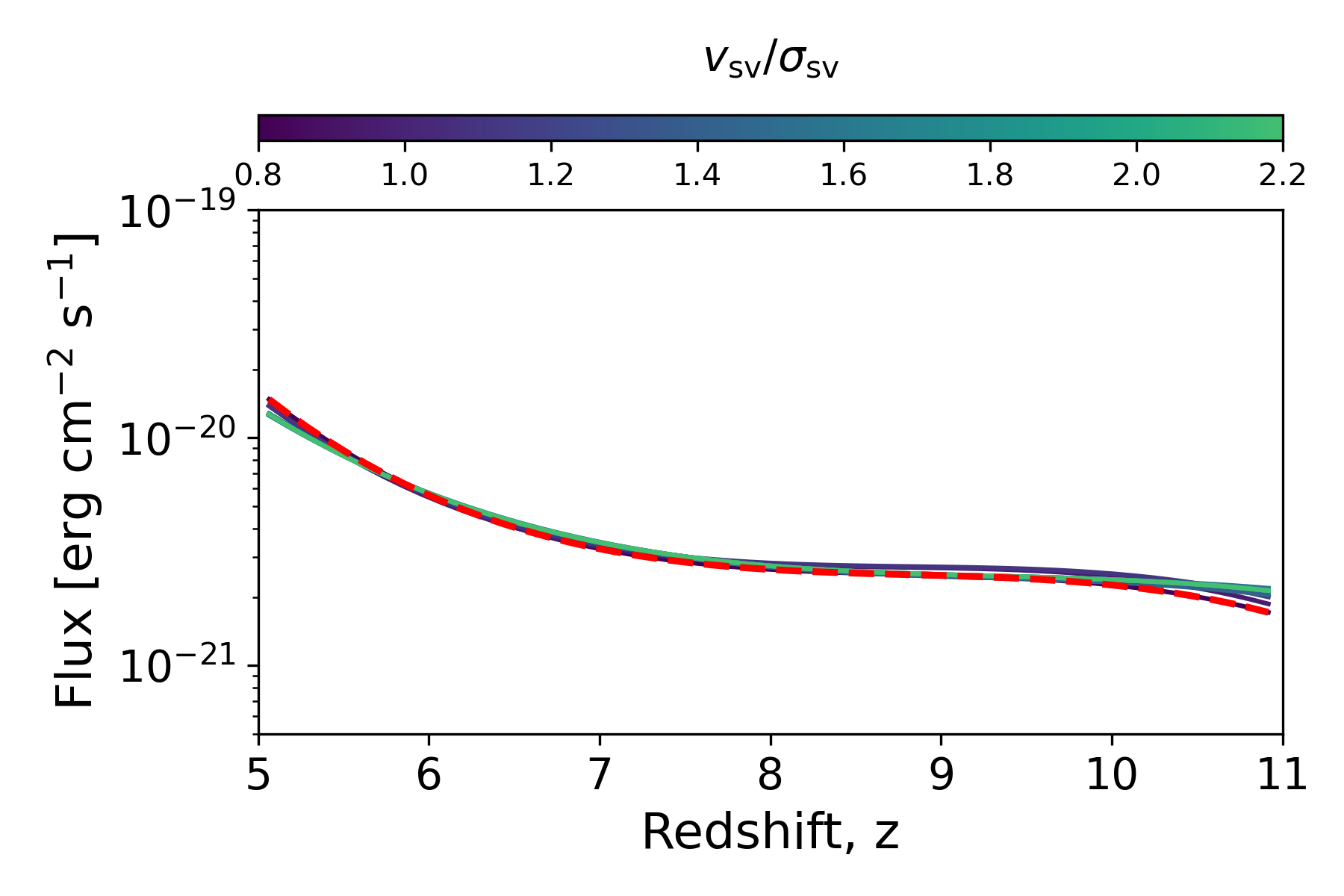}
        \includegraphics[width=.23\textwidth]{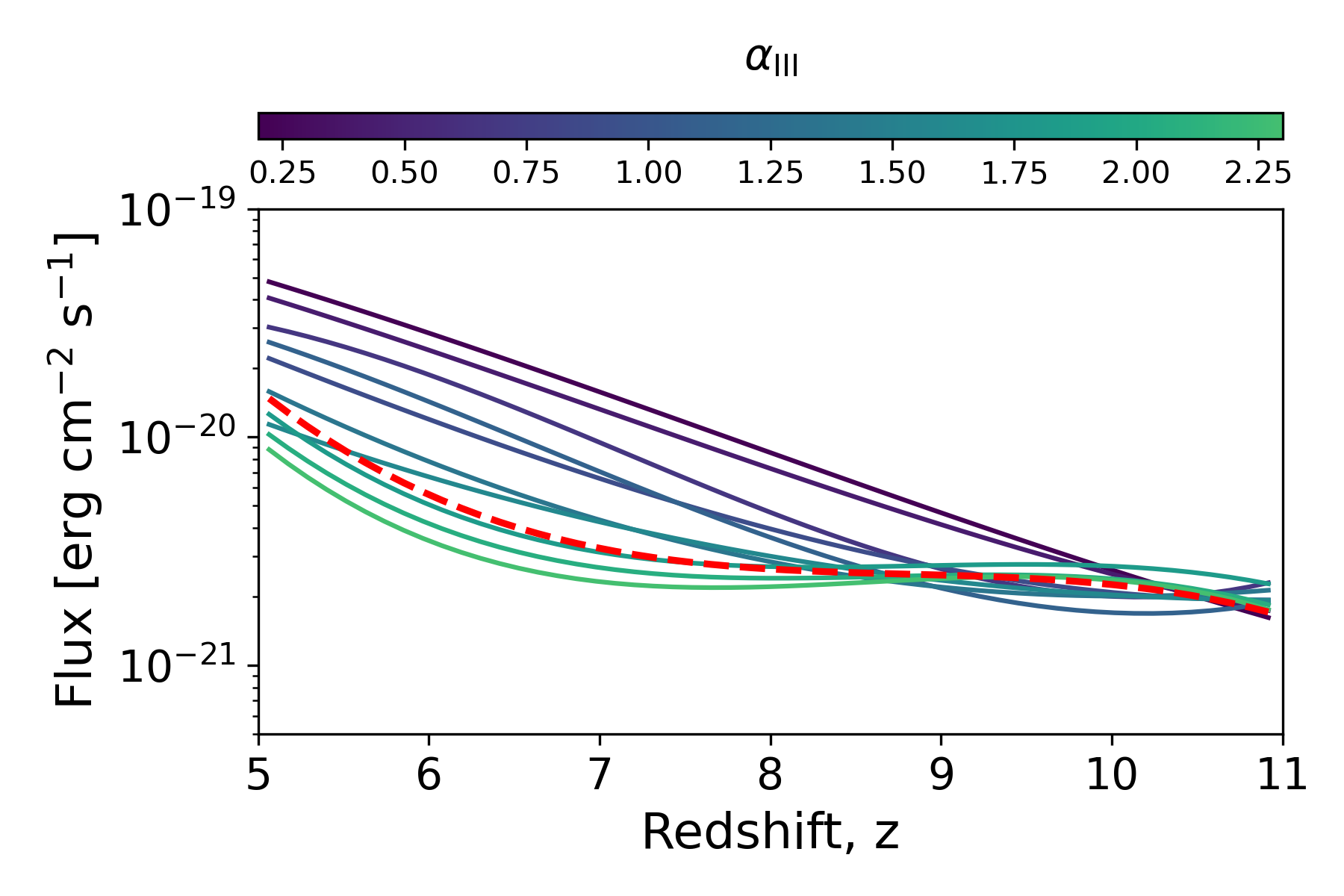}
        \includegraphics[width=.23\textwidth]{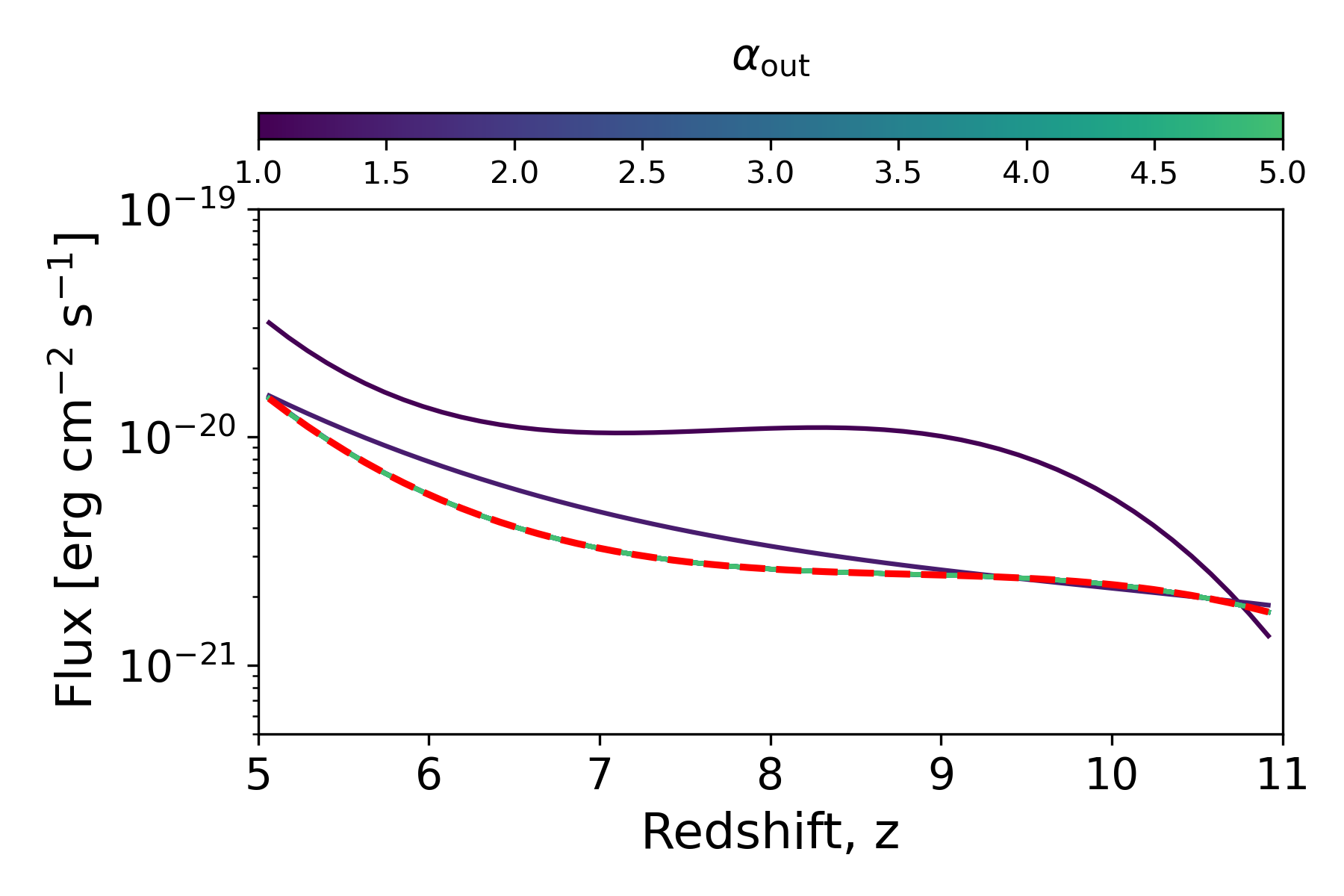}
        \includegraphics[width=.23\textwidth]{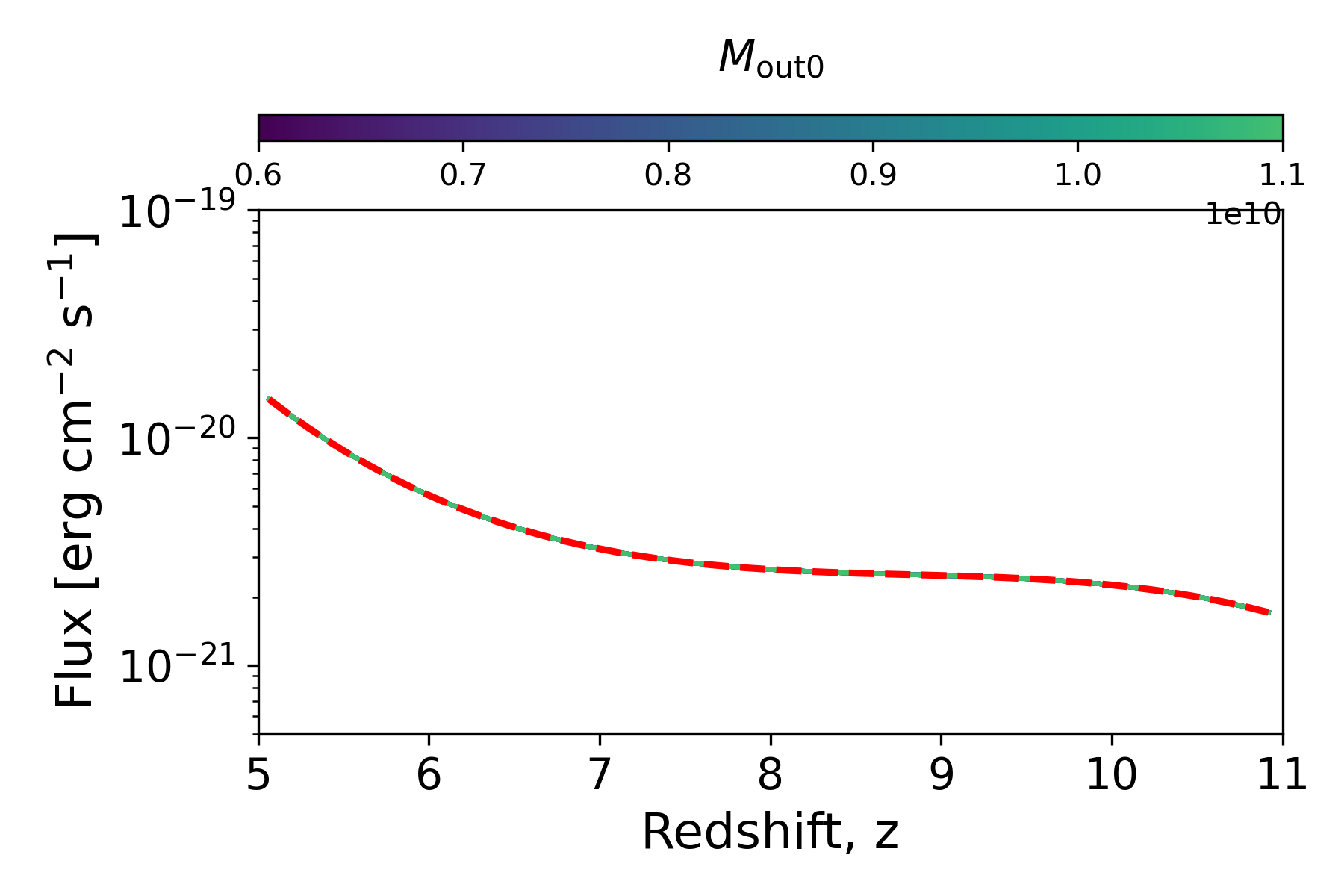}
        \includegraphics[width=.23\textwidth]{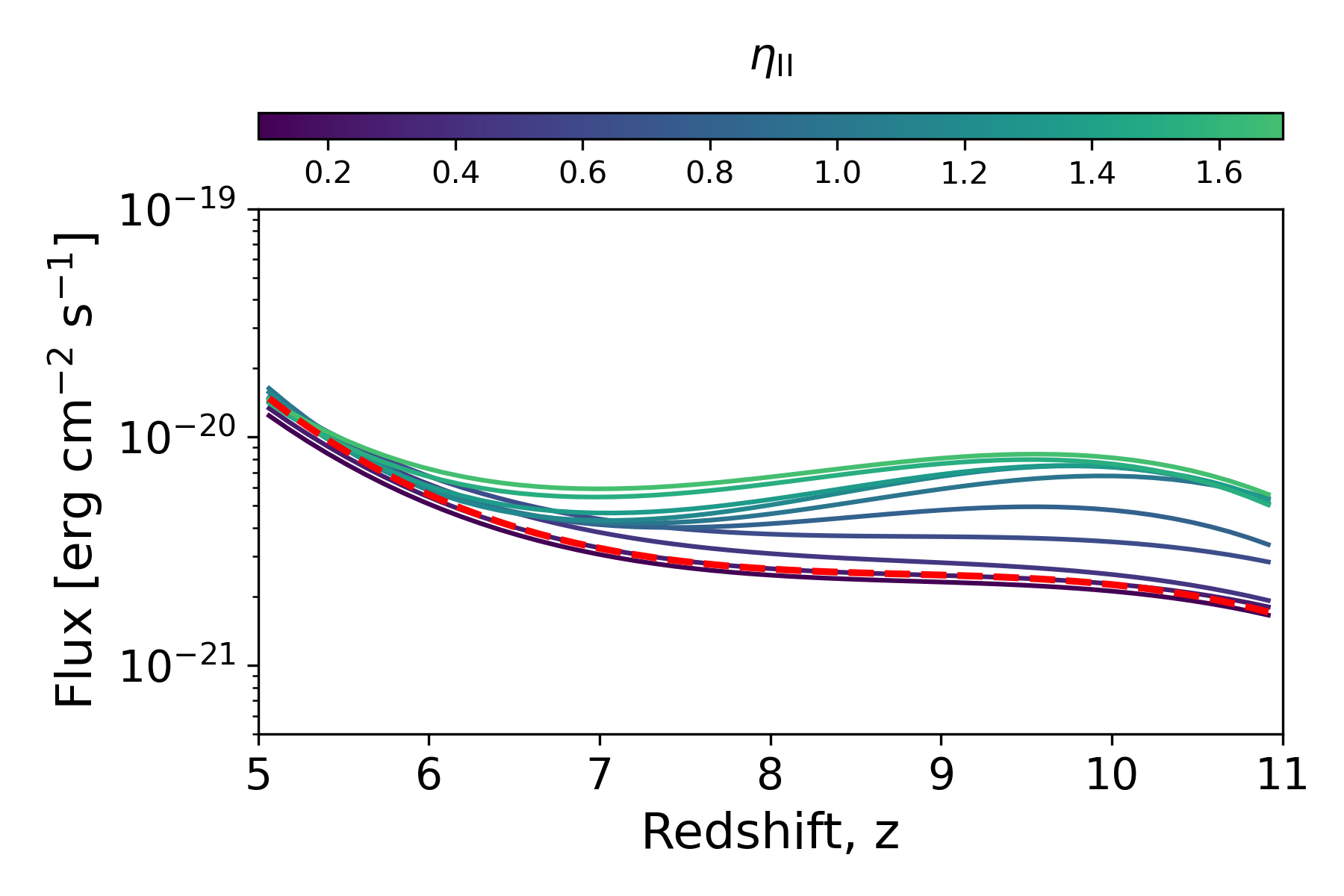}
        \includegraphics[width=.23\textwidth]{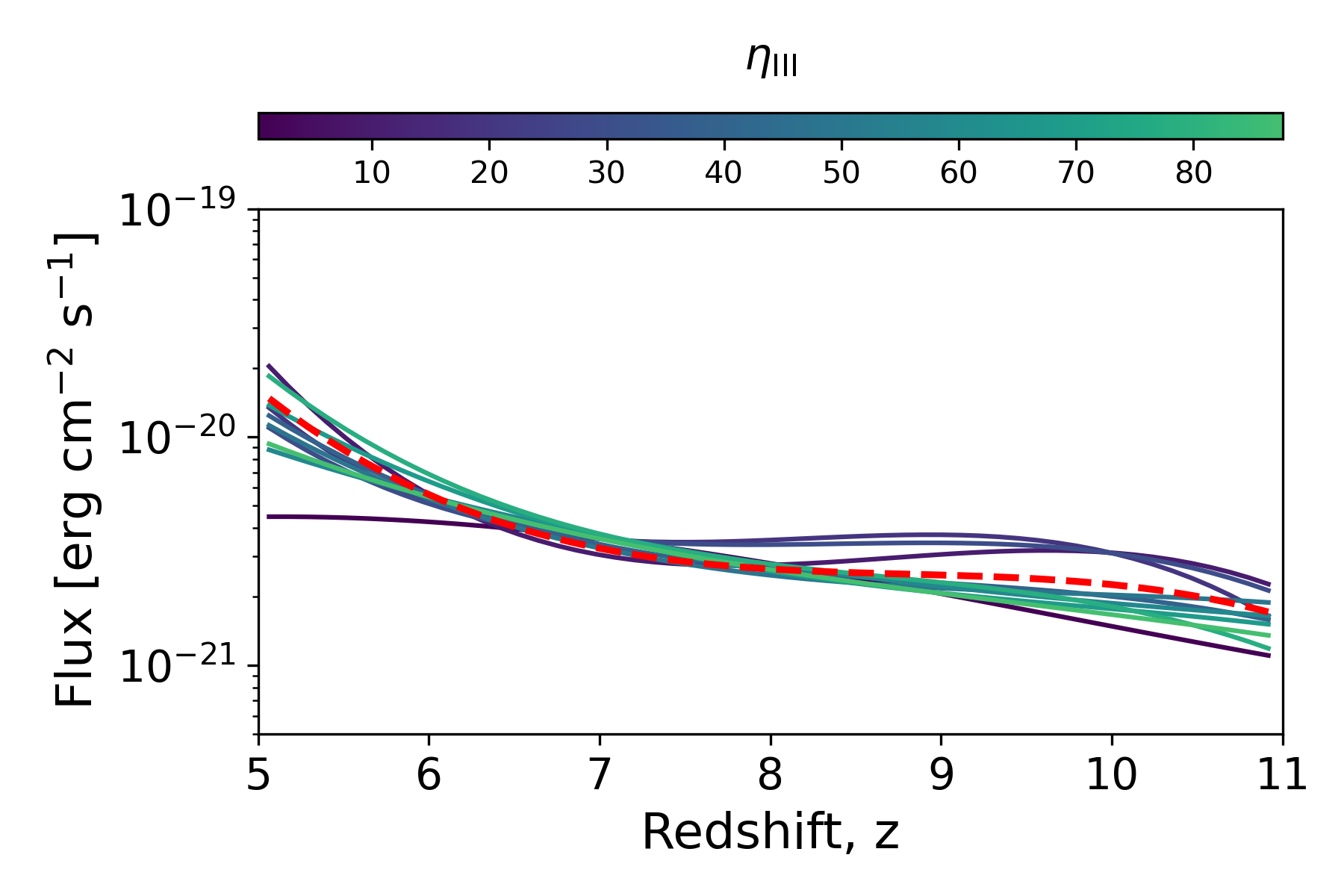}
        \includegraphics[width=.23\textwidth]{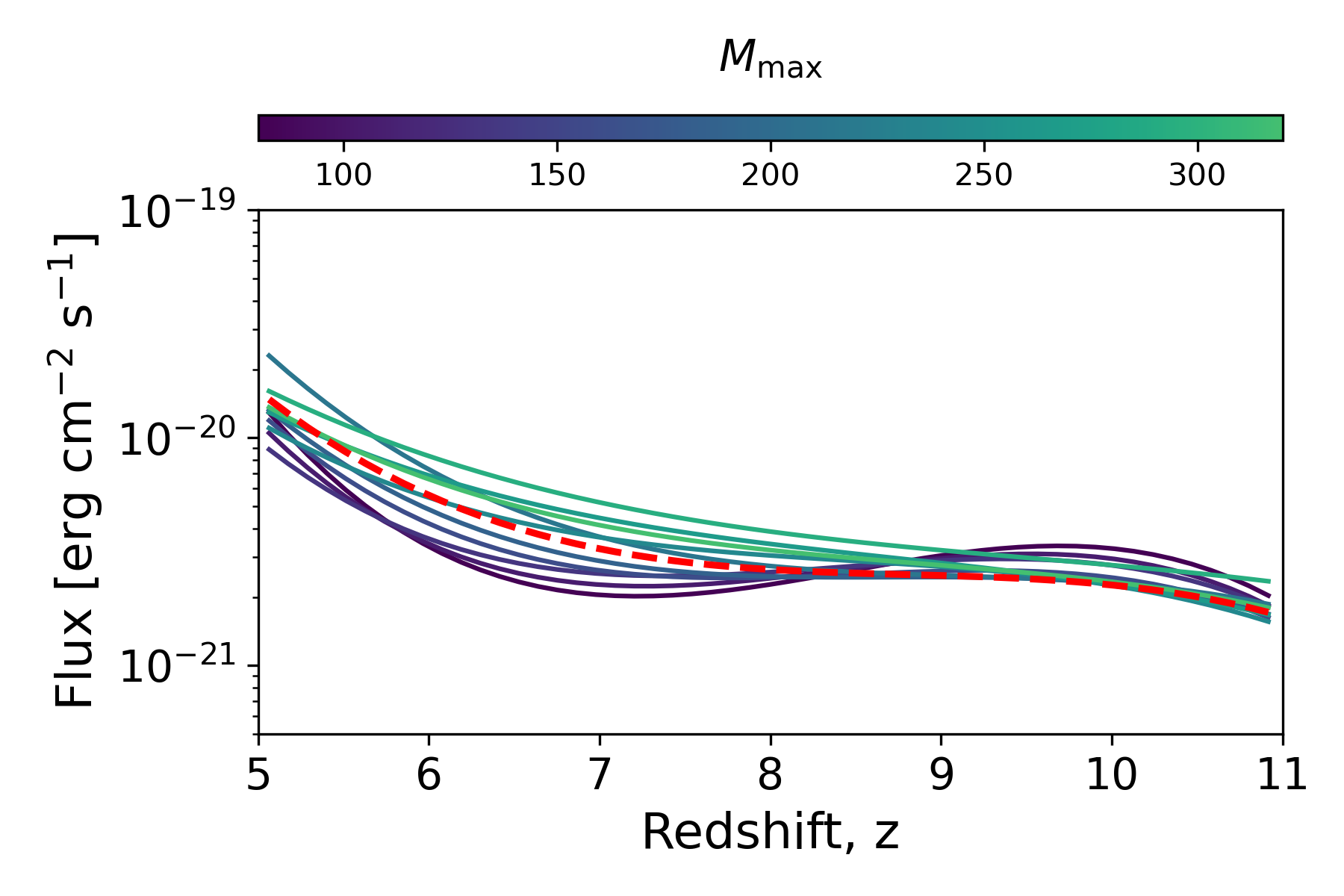}
        \includegraphics[width=.23\textwidth]{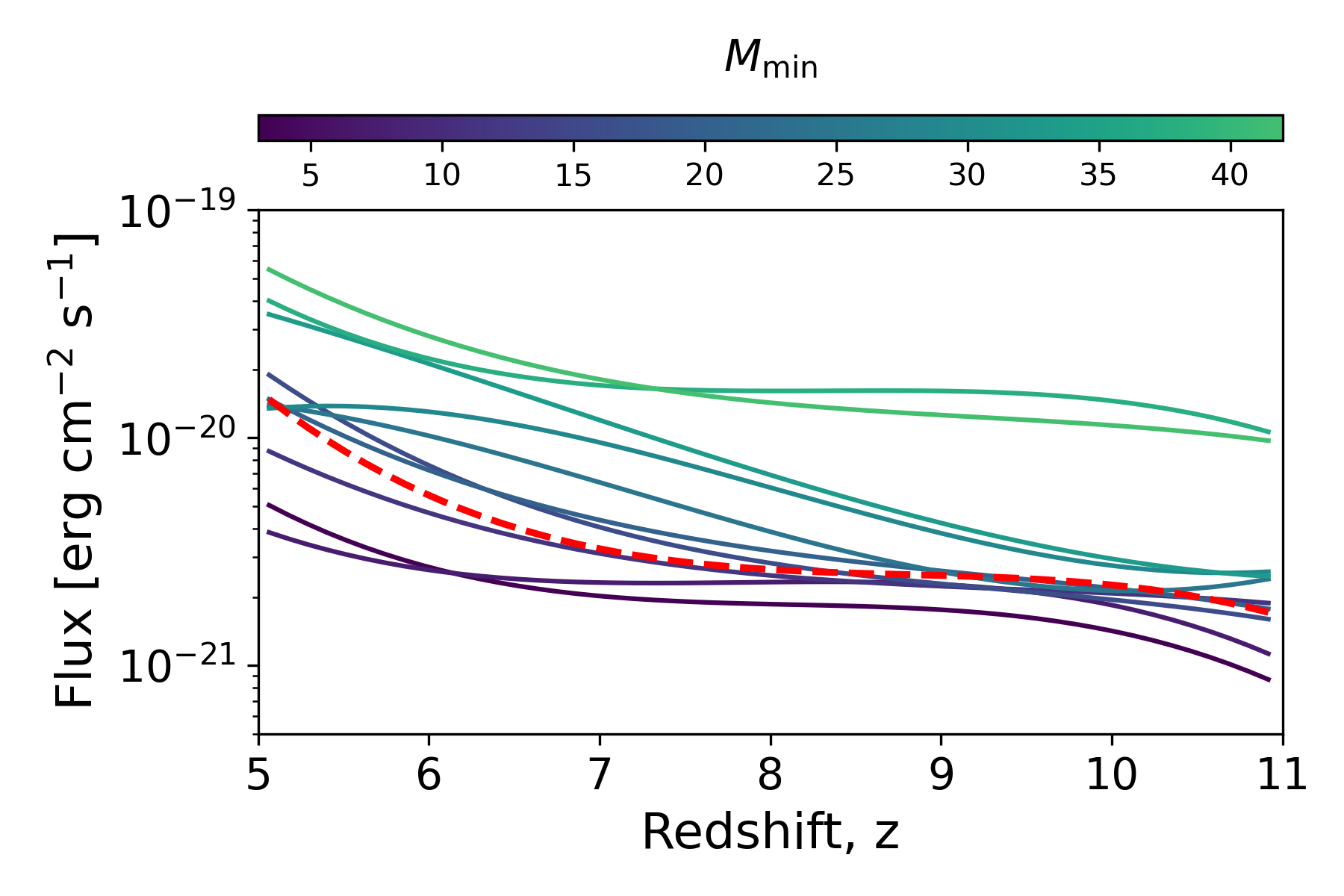}
        \includegraphics[width=.23\textwidth]{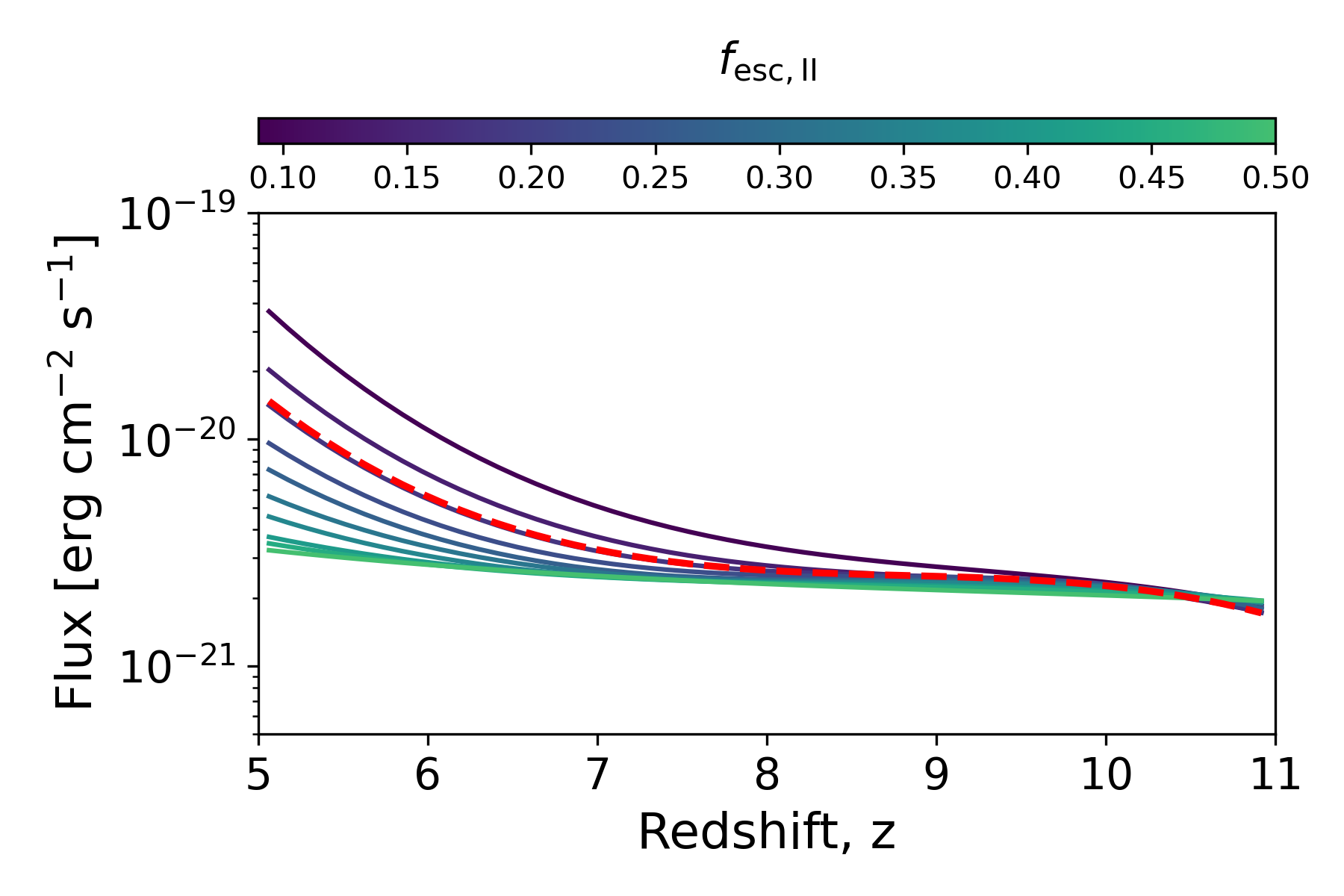}
        \includegraphics[width=.23\textwidth]{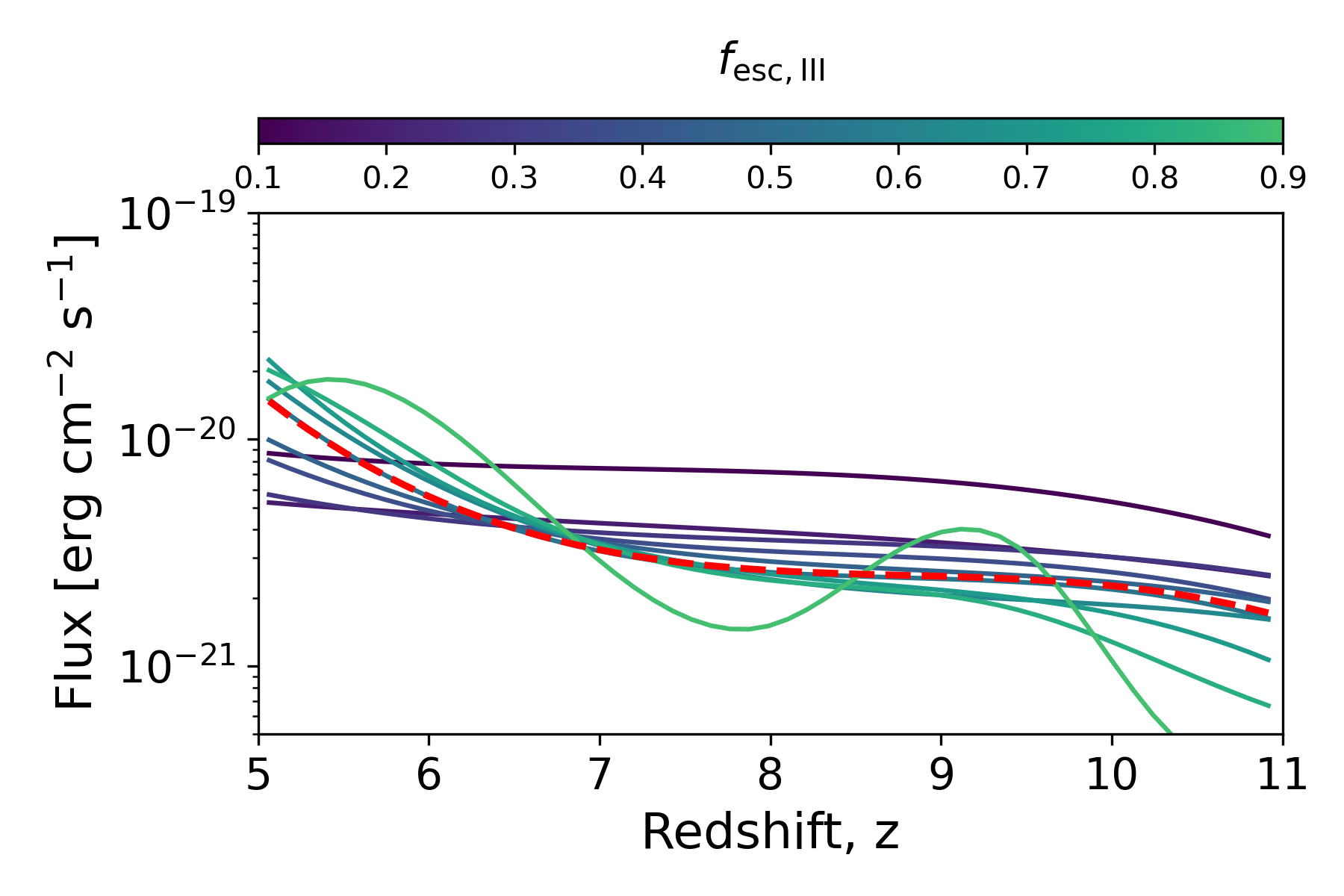}
        \includegraphics[width=.23\textwidth]{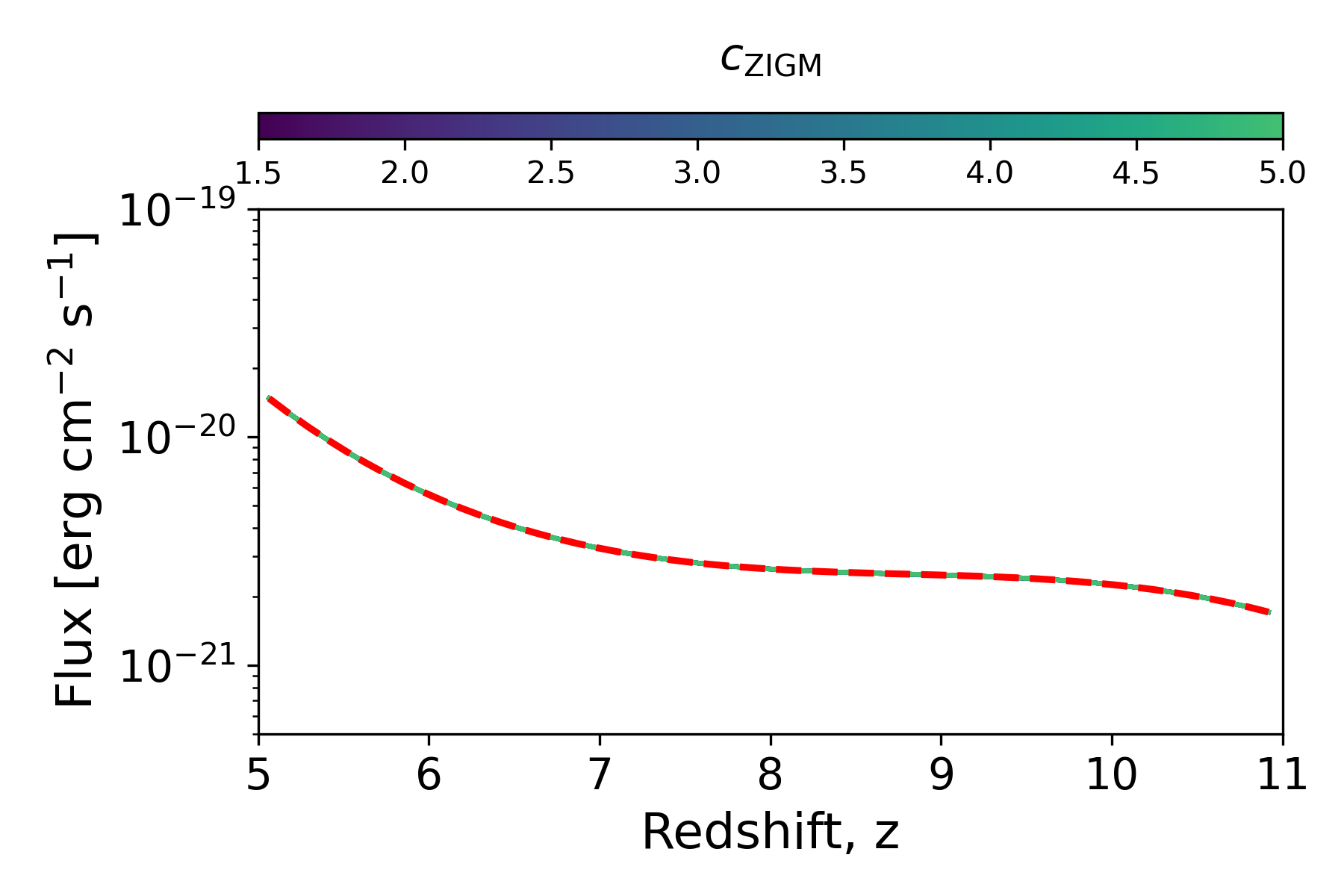}
        \caption{Upper envelope of simulated H$\alpha$ fluxes from Pop.\ III halos derived from the \asloth model using EPS merger tree as a function of redshift for variations in each free model parameter. Each curve corresponds to a different value of the varied parameter and shows the 97th percentile of Pop.\ III fluxes for the following value. This definition of the upper envelope is consistent with that used in Fig.~\ref{fig:alpha_flux_vs_redshift}. Line colour encodes the parameter value as indicated by the colourbar above each panel due to Table \ref{tab:parameters} ranges for each parameter that was chosen in accordance with the central 68\% confidence interval in the same Table. The red line indicates the results obtained with the default set of parameter values, for reference.} 
        \label{fig:full_param_variation}
        \end{center}
    \end{figure}

Here we present more details on the dependence of the fluxes on variations of key physics parameters in the semi-analytic model \asloth using the EPS approach. We illustrate in Fig.\ \ref{fig:full_param_variation} the predicted 97th percentile upper envelope of H$\alpha$ fluxes from Pop.\ III star-forming halos as a function of redshift, across the full physical range of each of the 11 model parameters listed in Table~\ref{tab:parameters}. Each subplot isolates the impact of varying a single parameter, holding all others fixed to their best-fit values obtained by \citet{hartwig24}. The spread in flux upper envelope values at each redshift thus directly reflects the sensitivity of the observable signal to the chosen parameter: 

\begin{itemize}

    \item {${v}_\mathrm{sv}/{\sigma}_\mathrm{sv}$:} The baryonic streaming velocity has only a modest effect on the flux envelope. The dependence of the upper flux envelope on the baryonic streaming velocity is weak and non-monotonic. At $z \gtrsim 9$, stronger streaming motion appears to reduce the number of low-mass halos contributing to the flux distribution, which can influence the 97th percentile in a complex, non-linear way. Since \asloth implements streaming effects via a raised star formation threshold and does not directly model gas accretion suppression, we refrain from attributing a clear physical trend and instead note the statistical origin of this weak dependence.

    \item{${\alpha}_\mathrm{III}$:} The Pop.\ III IMF slope strongly affects the envelope at the low-redshift end ($z\sim5$-7), with more top-heavy IMFs (smaller $\alpha_\mathrm{III}$) producing systematically higher H$\alpha$ fluxes. Toward higher redshift, the curves converge, and the dependence on $\alpha_\mathrm{III}$ becomes much weaker by $z\gtrsim9$-11.

    \item{${\alpha}_\mathrm{out}$:} The slope of outflow efficiency shows minimal influence on the upper envelope. Although it regulates baryon loss, the most extreme Pop.\ III halos remain luminous across a broad range of feedback scaling laws.

    \item{${M}_{\mathrm{out},0}$:} The outflow normalisation mass has negligible influence on the H$\alpha$ flux, suggesting that large-scale baryon ejection is less critical for the luminous Pop.~III star-forming halos producing the highest emissions.

    \item{${\eta}_\mathrm{II}$:} The Pop.\,II star-formation efficiency has a modest but systematic effect on the Pop.\,III H$\alpha$ flux envelope. Larger $\eta_\mathrm{II}$ shifts the envelope upward, most noticeably at $z\gtrsim9-10$, indicating that Pop.\,II-regulated halo conditions (and associated feedback/escape) can indirectly influence the maximum Pop.\,III recombination output.

    \item{${\eta}_\mathrm{III}$:} Pop.\ III star formation efficiency is one of the most impactful parameters. Increasing $\eta_\mathrm{III}$ boosts stellar mass and H$\alpha$ luminosity, raising the upper flux limit. However, the effect saturates at high values, as the total amount of stars that can form before feedback blows away the remaining halo gas is limited by the availability of cold gas. Beyond a certain point, further increasing $\eta_\mathrm{III}$ merely corresponds to converting this cold gas more rapidly into stars and does not result in a larger total mass in Pop.\ III stars. Consequently, the impact of varying $\eta_\mathrm{III}$ on the flux is limited: even substantial changes in its value do not translate into orders-of-magnitude changes in the maximum observable H$\alpha$ flux. 

    \item{${M}_\mathrm{max}$:} Increasing the maximum Pop.\,III stellar mass generally increases the upper envelope at $z\sim5$-7, consistent with a higher ionising-photon yield per unit stellar mass when the IMF extends further into the very massive regime. At $z\gtrsim8$-11 the dependence weakens and the curves largely converge.

    \item{${M}_\mathrm{min}$:} The minimum Pop.\,III stellar mass has a strong and monotonic effect: larger $M_\mathrm{min}$ (a more top-heavy IMF) produces a substantially higher H$\alpha$ flux envelope across the full redshift range. This parameter is therefore among the most effective IMF controls on the predicted maximum Pop.\,III Balmer emission.

    \item{${f}_{\mathrm{esc,II}}$:} The Pop.\,II escape fraction affects the envelope primarily at the low-redshift end, where varying $f_{\mathrm{esc,II}}$ changes the upper-envelope flux by a noticeable amount: smaller escape fractions yield brighter recombination emission. At $z\gtrsim9$-11 the curves converge, indicating that the brightest Pop.\,III emitters at the highest redshifts are less sensitive to the Pop.\,II escape fraction.

    \item {${f}_{\mathrm{esc,III}}$:} The Pop.\,III escape fraction slightly controlling the Pop.\,III H$\alpha$ envelope. Lower $f_{\mathrm{esc,III}}$ systematically increases the maximum flux (more ionising photons are retained in the halo and reprocessed into recombination emission), while higher $f_{\mathrm{esc,III}}$ suppresses the flux envelope.

    \item {${c_{\rm ZIGM}}$:} The IGM clumping factor does not affect predicted fluxes, confirming that the metallicity-dependent feedback primarily shapes global star formation trends (dominated by Pop.~II stars) rather than peak emission of Pop.~III stars from individual halos. 
\end{itemize}

To test whether the predicted fluxes could be significantly increased within the allowed parameter space, we performed an exploratory run in which several parameters were simultaneously pushed toward values expected to maximise the ionising output. In particular, we adopted a much more top-heavy Pop.\,III IMF by increasing the minimum stellar mass to $M_{\rm min}=21.1\,\Msun$, extending the upper limit to $M_{\rm max}=313\,\Msun$, and reducing the IMF slope to $\alpha_{\rm III}=0.23$. At the same time, we increased both star-formation efficiencies ($\eta_{\rm III}=8.15$, $\eta_{\rm II}=1.64$), reduced the escape fractions ($f_{\rm esc,III}=0.196$, $f_{\rm esc,II}=0.093$), and adopted a lower streaming velocity ($v_{\rm sv}/\sigma_{\rm sv}=0.8$), all of which should favour larger nebular fluxes. 

Despite these coordinated changes, the resulting flux–mass distribution shows only a modest upward shift relative to the fiducial model and remains well below the level required for detectable Balmer emission. This behaviour indicates a substantial degeneracy between the IMF parameters and the global star-formation efficiencies. Although a more top-heavy IMF increases the ionising photon production per unit stellar mass, the total stellar mass formed in individual halos remains limited by feedback and gas supply. Consequently, even when multiple parameters are tuned simultaneously to maximise the signal, the predicted Pop.\,III H$\alpha$ flux increases only moderately.

\section{Why some halos appear above the analytic ZAMS relations} \label{appendix:overline}

In Fig.~\ref{fig:mass_cross}, a small number of halos lie above the analytic ZAMS relations. 
These relations represent the expected H$\alpha$ flux from a stellar population whose ionising photon production follows the IMF-averaged ZAMS value used in our analytic estimate.

The apparent excess is primarily caused by the escape-fraction prescription used in the \asloth\ model. 
The analytic lines in Fig.~\ref{fig:mass_cross} are computed assuming the Pop.\,III escape fraction $f_{\mathrm{esc,III}}$. 
However, in halos where Pop.\,II stars have already formed, we adopt the Pop.\,II escape fraction $f_{\mathrm{esc,II}}$ for all stellar populations. 
For the best-fit parameters ($f_{\mathrm{esc,III}}=0.525$, $f_{\mathrm{esc,II}}=0.175$), this corresponds to an increase of a factor of $\simeq5$. 
As a result, halos that contain Pop.\,II stars can appear above the analytic Pop.\,III lines even though their intrinsic ionising photon production is consistent with the IMF-averaged expectation.

When the analytic relations are recomputed using $f_{\mathrm{esc,II}}$, the apparent discrepancy largely disappears. 
Similarly, if the sample is restricted to Pop.\,III-only halos ($M_{\star,\mathrm{II}}=0$), almost all systems lie below the analytic ZAMS relations.

The opposite trend of most halos lying below the analytic lines is also expected. The analytic relations assume an idealised stellar population with the IMF-averaged ZAMS ionising efficiency, whereas the simulated halos typically have a lower effective peak ionising efficiency due to insufficient sampling of the high-mass end of the IMF. The analytic relations should therefore be interpreted as an approximate upper envelope.

However, sometimes halos whose peak ionising efficiency significantly exceeds the IMF-averaged value can occur. These cases correspond to stochastic IMF sampling realisations in which a single very massive star dominates the ionising luminosity during the peak phase. Such events are expected in low-mass stellar systems and do not indicate a systematic inconsistency between the analytic model and the simulation results.

\end{appendix}

\end{document}